\documentclass[11pt,a4paper]{article}
\usepackage{jheppubnew}

\newcommand\nn{{\nonumber}}

\newcommand{\I}{i}

\newcommand{\C}{\mathbb{C}}

\newcommand{\id}{\mathbb{I}}
\renewcommand{\H}{\mathcal{H}}
\newcommand{\<}{\langle}
\renewcommand{\>}{\rangle}

\newcommand{\dbl}{\text{eff}}

\newcommand{\M}{\mathcal{M}}
\newcommand{\m}{\mu}

\newcommand\mat[4]{{\left( \begin{array}{cc} {#1} & {#2} \\ {#3} & {#4} \end{array} \right)}}

\def\be{\begin{equation}}
\def\ee{\end{equation}}
\def\beq{\begin{eqnarray}}
\def\eeq{\end{eqnarray}}
\def\a{\alpha}
\def\b{\beta}

\def\e{\epsilon}
\def\EE{\Delta}

\def\({\left(}
\def\){\right)}
\def\[{\left[}
\def\]{\right]}

\def\ra{\rightarrow}

\DeclareMathOperator{\Tr}{Tr}
\DeclareMathOperator{\Span}{span}
\DeclareMathOperator*{\wlim}{wlim}
\DeclareMathOperator{\EllipticK}{K}
\DeclareMathOperator{\EllipticE}{E}

\title{Interactions resolve state-dependence in a toy-model of AdS black holes}
\author[a,b]{Adam Bzowski,}
\author[c]{Alessandra Gnecchi,}
\author[a]{Thomas Hertog}
\affiliation[a]{Institute for Theoretical Physics, KU Leuven, Celestijnenlaan 200D, Leuven, Belgium}
\affiliation[b]{Institut de Physique Th\'{e}orique, CEA Saclay, Gif-sur-Yvette, France}
\affiliation[c]{Theoretical Physics Department, CERN, Geneva, Switzerland}
\emailAdd{adam.bzowski@ipht.fr}
\emailAdd{alessandra.gnecchi@cern.ch}
\emailAdd{thomas.hertog@kuleuven.be}
\abstract{We show that the holographic description of a class of AdS black holes with scalar hair involves dual field theories with a double well effective potential. Black hole microstates have significant support around both vacua in the dual, which correspond to perturbative degrees of freedom on opposite sides of the horizon. A solvable toy-model version of this dual is given by a quantum mechanical particle in a double well potential. In this we show explicitly that the interactions replace the state-dependence that is needed to describe black hole microstates in a low energy effective model involving the tensor product of two decoupled harmonic oscillators. A naive number operator signals the presence of a firewall but a careful construction of perturbative states and operators extinguishes this. 
}
\cern{CERN-TH-2018-027}

\begin{document}

\maketitle

\section{Introduction}

Holography has provided a fruitful new perspective on the black hole information paradox first formulated by Hawking \cite{Hawking:1976ra}. Specifically it has enabled an expression of the essence of the paradox in dual quantum field theoretic terms. This has further sharpened some of the underlying assumptions, and it has led to novel suggestions for its resolution. Key elements that have emerged in this recent discussion include the following;

\begin{itemize}
\item {\it Firewalls:} According to \cite{Almheiri:2012rt,Almheiri:2013hfa,Bousso:2012as,Marolf:2013dba} a generic black hole microstate exhibits a firewall, \textit{i.e.} that its horizon is not a smooth surface. In terms of field theory data the paradox is stated as an incompatibility of the following four conditions: \textit{i}) unitarity, \textit{ii}) validity of the semiclassical approximation for an asymptotic observer, \textit{iii}) existence of black hole microstates visible to an asymptotic observer as states with exponentially small energy differences, and \textit{iv}) existence of a firewall-free number operator for an infalling observer.
\item {\it State-dependence:} In contrast with this \cite{Papadodimas:2013jku,Papadodimas:2013wnh,Papadodimas:2015jra,Papadodimas:2015xma} proposed a dual description of the black hole interior by introducing state-dependent operators. The black hole horizon remains smooth, but such operators depend on the specific black hole microstate and hence go beyond the standard paradigm of quantum field theory. In this way one avoids many pitfalls presented by the firewall arguments such as existence of creation-annihilation operators satisfying the `wrong sign' commutation relations.
\item {\it Vacuum structure:} Hawking's original argument has been recast as a no-go theorem \cite{Mathur:2008wi,Mathur:2009hf} that states that quantum gravity effects cannot prevent information loss if they are confined to within a given scale and if the vacuum of the theory is assumed to be unique. This is a particularly sharp puzzle in the context of holography given the apparent uniqueness of the vacuum of dual CFTs. It can be seen as one motivation for recent investigations into a possible non-trivial vacuum structure as in \cite{Hawking:2016msc}.
\end{itemize}

The reformulation of the information paradox in terms of purely field theoretic data elucidates the nature of the underlying issues. Devoid of a geometrical interpretation one can ask whether there exists any tractable well-defined quantum theories modeling black holes and satisfying basic properties required by the aforementioned papers.

In this paper we put forward a new holographic toy-model for AdS black holes that incorporates in a toy-model fashion a specific proposal for the nature of non-perturbative quantum gravity corrections to black hole physics. The model consists of a quantum mechanical particle in a double well potential. 

Our motivation to advance this as a toy-model for black holes in AdS is twofold. First, there is a class of single-sided black hole solutions in global AdS with scalar hair outside the horizon whose dual description involves a field theory with a double well effective potential \cite{Hertog:2004dr,Hertog:2004ns,Hertog:2005hu,Hertog:2006rr}. We review these black holes, which are solutions in truncations of AdS supergravity with so-called designer gravity boundary conditions, in Section \ref{hairy}. The potential barrier in their dual description separates the perturbative degrees of freedom on both sides of the horizon, but they are coupled through multi-trace interactions. We argue that black hole microstates are states with significant support around both perturbative vacua. The quantum mechanical model we put forward can be viewed as a toy-model for systems of this kind since it amounts to two harmonic oscillators coupled through a `non-perturbative' interaction modeled as a double well potential. Outside the context of holography Giddings \cite{Giddings:2012gc,Giddings:2017mym} has studied how novel, non-local interactions (in the bulk) can resolve the information paradox. 

Secondly, recent work in the context of two-sided black holes in AdS has advocated that not only entanglement but also interactions between the two boundary CFTs are needed to describe the bulk \cite{Chowdhury:2013tza,Maldacena:2013xja,Susskind:2014moa,Balasubramanian:2014gla}. Some implications of adding a specific example of such interactions were explored in nearly $AdS_2$ in \cite{Shenker:2013pqa,Shenker:2013yza,Balasubramanian:2014gla,Gao:2016bin,Maldacena:2017axo,vanBreukelen:2017dul}. These studies yield a different motivation for our toy-model in which the two perturbative vacua are thought of as being dual to the two asymptotic regions on both sides of the horizon. The potential barrier in the dual toy-model amounts to a proposal for a specific interaction between a single pair of left and right modes in the bulk, $|n_k\>_L$ and $| n_k \>_R$, with fixed frequency $\omega$ and fixed wave vector $k$ and related to boundary states by some form of the HKLL construction \cite{Hamilton:2005ju,Hamilton:2006az}. For small values of $n_k$ the left and right modes are essentially separated by a potential barrier. By contrast, for sufficiently large occupation numbers left and right modes interact strongly and the semiclassical approximation breaks down. In the context of holography \cite{Fitzpatrick:2016ive} argued that such non-perturbative effects can be sufficient to resolve the information paradox.

Motivated by these developments we consider a quantum mechanical particle with the following potential
\begin{equation} \label{introV}
V(x) = \frac{1}{32 \lambda} - \frac{1}{4} \omega^2 x^2 + \frac{\lambda}{2} \omega^4 x^4
\end{equation}
governed by the standard Hamiltonian
\begin{equation} \label{H}
H = \frac{1}{2} p^2 + V(x)
\end{equation}
This system is characterized by a dimensionless coupling constant $\lambda$ and a frequency $\omega$ of the approximate harmonic oscillators around the semiclassical vacuum states $\varphi_0^L$ and $\varphi_0^R$ at $x_{R,L} = \pm 1/(2 \omega \sqrt{\lambda})$. 
The vacua in our model correspond to the left and right, or interior and exterior, semiclassical vacua in the bulk. Excited states from the standpoint of observers in one of these vacua then naturally correspond to perturbative states $\varphi_n^L$ and $\varphi_n^R$. Black hole microstates finally are linear combinations of perturbative states in both vacua. We will argue that typical microstates correspond to states with significant support around both perturbative vacua. By contrast, bulk spacetimes without a black hole correspond to states with support around one of the vacua only. 

To make contact with the usual perturbative expansion in semiclassical gravity we introduce a dimensionless parameter $N$ as 
\begin{equation}
\lambda = \frac{1}{N^2}.
\end{equation}
Hence perturbation theory in $\lambda$ models the usual large $N$ expansion. The height of the potential barrier equals $V_{\ast} = V(0) = 1/(32 \lambda)$. In the limit $\lambda\to 0$ the barrier grows, and the two minima move apart. In the exact $\lambda = 0$ limit the excitations around both perturbative vacua decouple completely and the system reduces to two decoupled harmonic oscillators with frequency $\omega$. In the bulk, with designer gravity boundary conditions, this decoupling limit corresponds precisely to a limit in which the horizon of the hairy black holes becomes singular.

In this paper we carefully study how states and operators in the full interacting toy-model relate to quantities in a low energy effective theory involving the tensor product of two decoupled harmonic oscillators. At the effective theory level our model captures many of the usual paradoxes associated with the semiclassical approximation of black hole physics in a remarkably precise manner. A major advantage of our model is that it is solvable. This enables us to sharpen the limitations of the perturbative description of black holes, to explore dynamical processes, and to understand in this concrete toy-model setup how non-perturbative interactions resolve the paradoxes. In particular we show explicitly that the interactions eliminate the state-dependence that is needed to describe black hole microstates in the effective low energy dual. We also find that a naive number operator signals the presence of a firewall, but that a careful construction of perturbative states and operators in the full model extinguishes this. Finally, when it comes to dynamical processes, we point out that tunneling near the potential maximum corresponds to Hawking radiation in the bulk, and that the scattering of classical waves nicely captures the behavior of shock waves in the bulk.

We conclude this introduction with an important caveat. Evidently our model is not suitable for the analysis of properties of black holes that depend on a collection of modes. This in particular encompasses all thermodynamical properties that rely on the existence of an ensemble of modes. In this context it would be necessary to consider more complicated models such as matrix or tensor models.

\section{Dual description of hairy AdS black holes}\label{hairy}

In this section we review the dual description in terms of a field theory with a double well effective potential of a class of single-sided asymptotically $AdS_4$ static black hole solutions with scalar hair. The perturbative degrees of freedom on both sides of the horizon correspond in the dual description to excitations around two distinct perturbative vacua. However, multi-trace interactions in the dual imply a non-perturbative coupling between both sides. As such this setup motivates the quantum mechanical particle in a double well potential as a toy-model for black holes in AdS. 

The black hole solutions we construct are variations of the solutions found in \cite{Hertog:2004dr,Hertog:2004ns,Hertog:2005hu,Hertog:2006rr,Martinez:2004nb,Acena:2013jya,Anabalon:2012ta} and recently in \cite{Faedo:2017veq,Anabalon:2017yhv}.
Consider the low energy limit of M theory with $AdS_4 \times S^7$ boundary conditions. The massless sector of the compactification of $D=11$ supergravity on $S^7$ is ${\cal N}=8$ gauged supergravity in four dimensions. It is possible to consistently truncate this theory to include only gravity and a single scalar with action 
\be\label{4-action}
S=\int d^4x\sqrt{-g}\left[\frac{1}{2}R
-\frac{1}{2}(\nabla\phi)^2 +2+\cosh(\sqrt{2}\phi) \right]
\ee
where we have set $8\pi G=1$ and chosen the gauge coupling so that the AdS radius is one. The potential has a maximum at $\phi =0$ corresponding to an $AdS_4$ solution with unit radius. It is unbounded from below, but small fluctuations have
$m^2 = -2$, which is above the Breitenlohner-Freedman bound $m_{BF}^2 = -9/4$ so with the usual boundary conditions $AdS_4$ is stable. Consider global coordinates in which the $AdS_4$ metric takes the form
\be \label{adsmetric}
ds^2_0 = \bar g_{\mu \nu} dx^{\mu} dx^{\nu}=
-(1+r^2 )dt^2 + \frac{dr^2}{1+r^2} + r^2 d\Omega_{2}
\ee
In all asymptotically AdS solutions, the scalar $\phi$ decays at large radius as
\be\label{hair4d}
\phi(r) \sim\frac{\alpha}{r}+\frac{\beta}{r^2}\ ,\ \qquad r\to\infty
\ee
where $\a$ and $\b$ are functions of $t$ and the angles. To have a well-defined theory one must specify boundary conditions at spacelike infinity. The standard choice of boundary condition corresponds to taking $\a=0$. However one can consider more general `designer gravity' boundary conditions \cite{Hertog:2004ns} with $\a \neq 0$ that are specified by a functional relation $\b(\a)$ in (\ref{hair4d}). The backreaction of the $\a$-branch of the scalar field and its self-interaction modify the asymptotic behavior of the gravitational fields. Writing the metric as $g_{\mu \nu}=\bar g_{\mu \nu} +h_{\mu \nu}$ the corresponding asymptotic behavior of the metric components is given by
\be \label{4-grr}
h_{rr}=  -\frac{(1+\alpha^2/2)}{r^4} + {\cal O} (1/r^5), \quad h_{rm} = {\cal O}(1/r^2), \quad  h_{mn}={\cal O}(1/r)
\ee
Nevertheless, the Hamiltonian generators of the asymptotic symmetries remain well-defined and finite when $\a \neq 0$ \cite{Hertog:2004dr,Henneaux:2004zi,Hertog:2004bb}. They acquire however an explicit contribution from the scalar field. For instance, the conserved mass of spherical solutions is given by
\be\label{mass}
M = {\rm Vol}(S^{2}) \left[ M_0 +\a\b +W \right]
\ee
where $M_0$ is the coefficient of the $1/r^{5}$ term in the asymptotic expansion of $g_{rr}$, and where we have defined the function
\be \label{bc}
W(\a) = \int_0^\a \b(\tilde \a) d \tilde \a \ ,
\ee
which defines the choice of boundary conditions. 

Consider now a specific class of boundary conditions defined by the following relation
\be \label{bccurve}
\b_{bc}(\a) = - c_1 \a^2 +c_2 \a^3
\ee
where $c_1$ and $c_2$ are constants. With these, the conserved mass (\ref{mass}) is given by
\be \label{bhmass2}
M = 4\pi \(M_0 - \frac{4}{3} c_1 \a^3 + \frac{1}{6} c_2 \a^4\)
\ee

Both the vacuum and the dynamical properties of the theory -- as well as the possible black hole endstates of gravitational collapse -- depend significantly on $W$ \cite{Hertog:2004ns,Papadimitriou:2007sj}. In the context of the AdS/CFT correspondence, adopting designer gravity boundary conditions defined by a function $W \neq 0$ corresponds to adding a potential term $\int W ({\cal O})$ to the dual CFT action, where ${\cal O}$ is the field theory operator that is dual to the bulk scalar \cite{Witten:2001ua,Berkooz:2002ug}. This is generally a complicated multi-trace interaction. Certain deformations $W$, including those corresponding to boundary conditions of the form \eqref{bccurve}, give rise to field theories with additional, possibly metastable vacua. 

\begin{figure}[htb]
\includegraphics[width=0.45\textwidth]{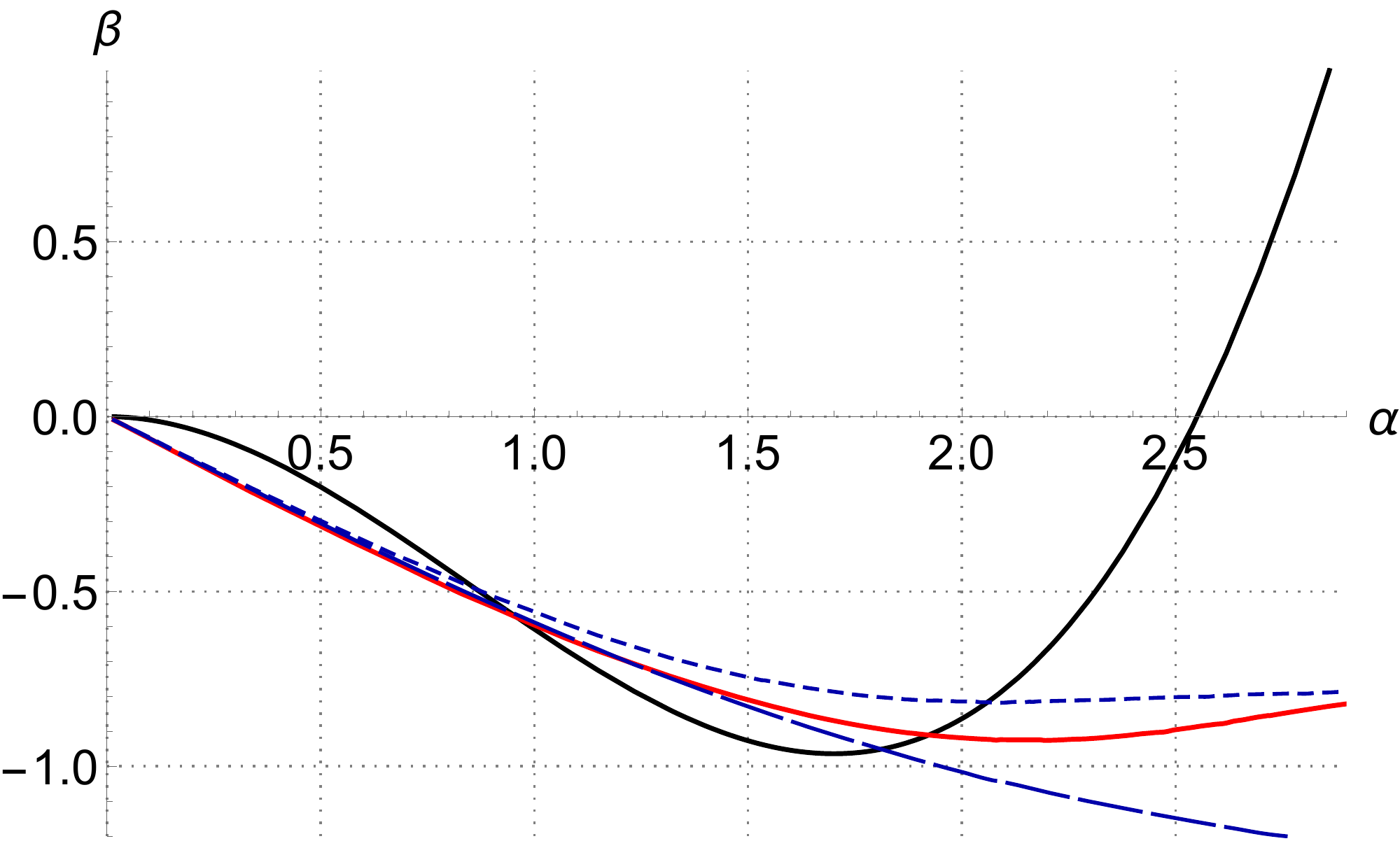}
\qquad
\includegraphics[width=0.45\textwidth]{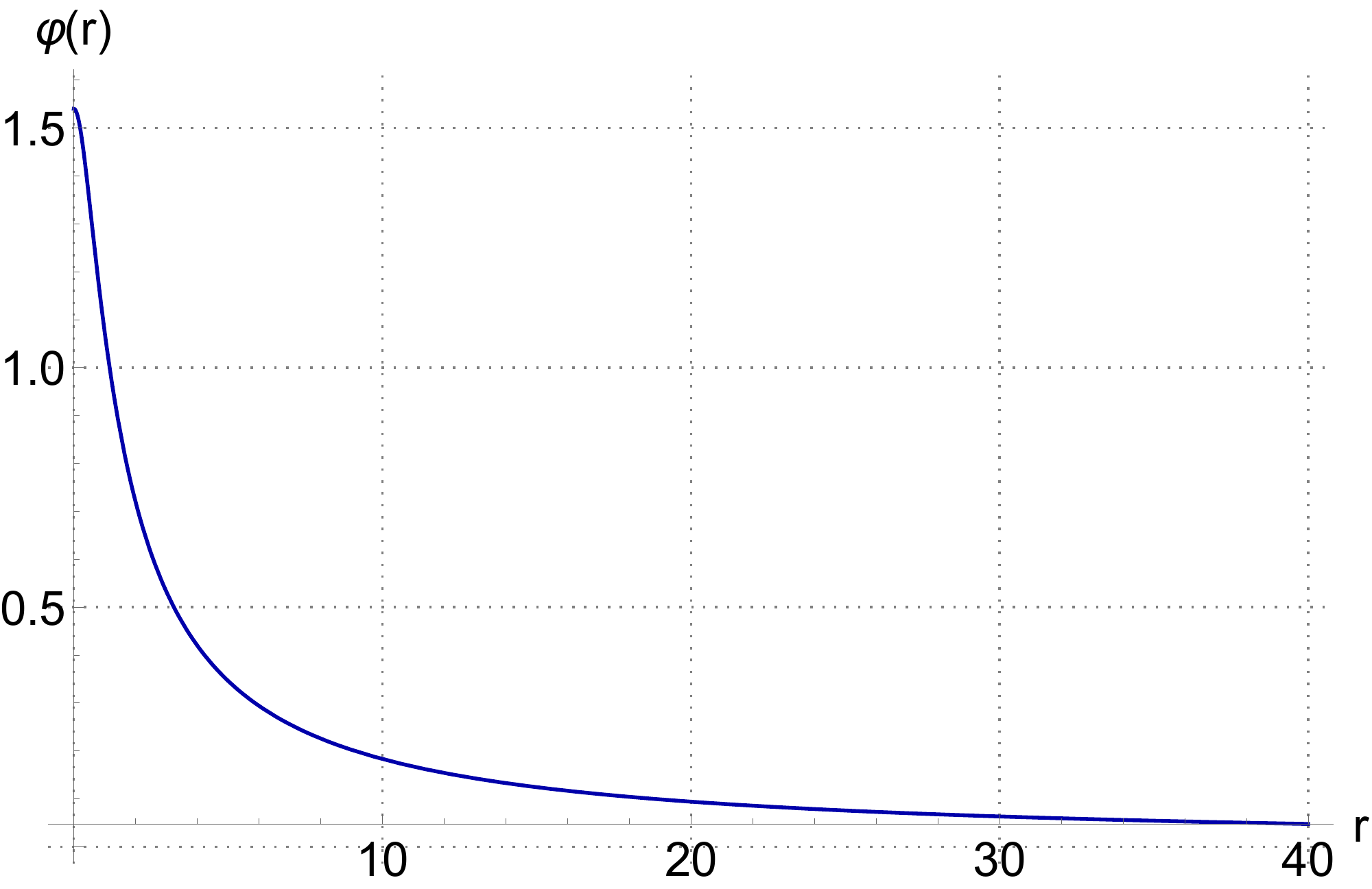}
\centering
\caption{Left: The functions $\b(\a)$ that characterize the set of solitons (red) in the theory, and hairy black holes with horizon size $R_e=.2$ (blue dotted) and $R_e=1$ (blue dashed). The black curve shows the boundary condition function $\b_{bc}(\a)=- \a^2 + 0.392 \a^3$ which, together with the characteristics of the solitons, leads to the dual double well effective potential shown in Figure \ref{2}(a). Right: The scalar radial profile of the soliton associated with the second intersection point of $\b_{bc}(\a)$ with the soliton curve $\beta_s(\alpha)$.}
\label{1}
\end{figure}

The AdS/CFT correspondence relates the expectation values $\langle {\cal O} \rangle$ in different field theory vacua to the asymptotic scalar profile of regular static solitons in the bulk. The precise correspondence between solitons and field theory vacua is given by the following function \cite{Hertog:2004bb},
\be \label{effpot}
{\cal V}(\alpha) = -\int_{0}^{\a} \b_{s} (\tilde \a) d\tilde \a + W(\alpha)
\ee
where $\b_{s}(\a)$ is obtained from the asymptotic scalar profiles of spherical soliton solutions with different values $\phi(0)$ at the origin $r=0$. This curve was first obtained in \cite{Hertog:2004ns} for the theory \eqref{4-action} and is plotted in Figure \ref{1}(a). 

Given a choice of boundary condition $\b(\a)$, the allowed solitons are simply given by the points where the soliton curve intersects the boundary condition curve: $\b_s(\a) = \b(\a)$. For any $W$ the location of the extrema of ${\cal V}$ in \eqref{effpot} yield the vacuum expectation values $ \langle {\cal O} \rangle = \a$, and the value of ${\cal V}$ at each extremum yields the energy of the corresponding soliton. Hence ${\cal V}(\a)$ can be interpreted as an effective potential for $\langle {\cal O} \rangle$. This led \cite{Hertog:2004ns} to conjecture that (a) there is a lower bound on the gravitational energy in those designer gravity theories where ${\cal V}(\a)$ is bounded from below, and that (b) the solutions locally minimizing the energy are given by the spherically symmetric, static soliton configurations. 

\begin{figure}[htb]
\includegraphics[width=0.45\textwidth]{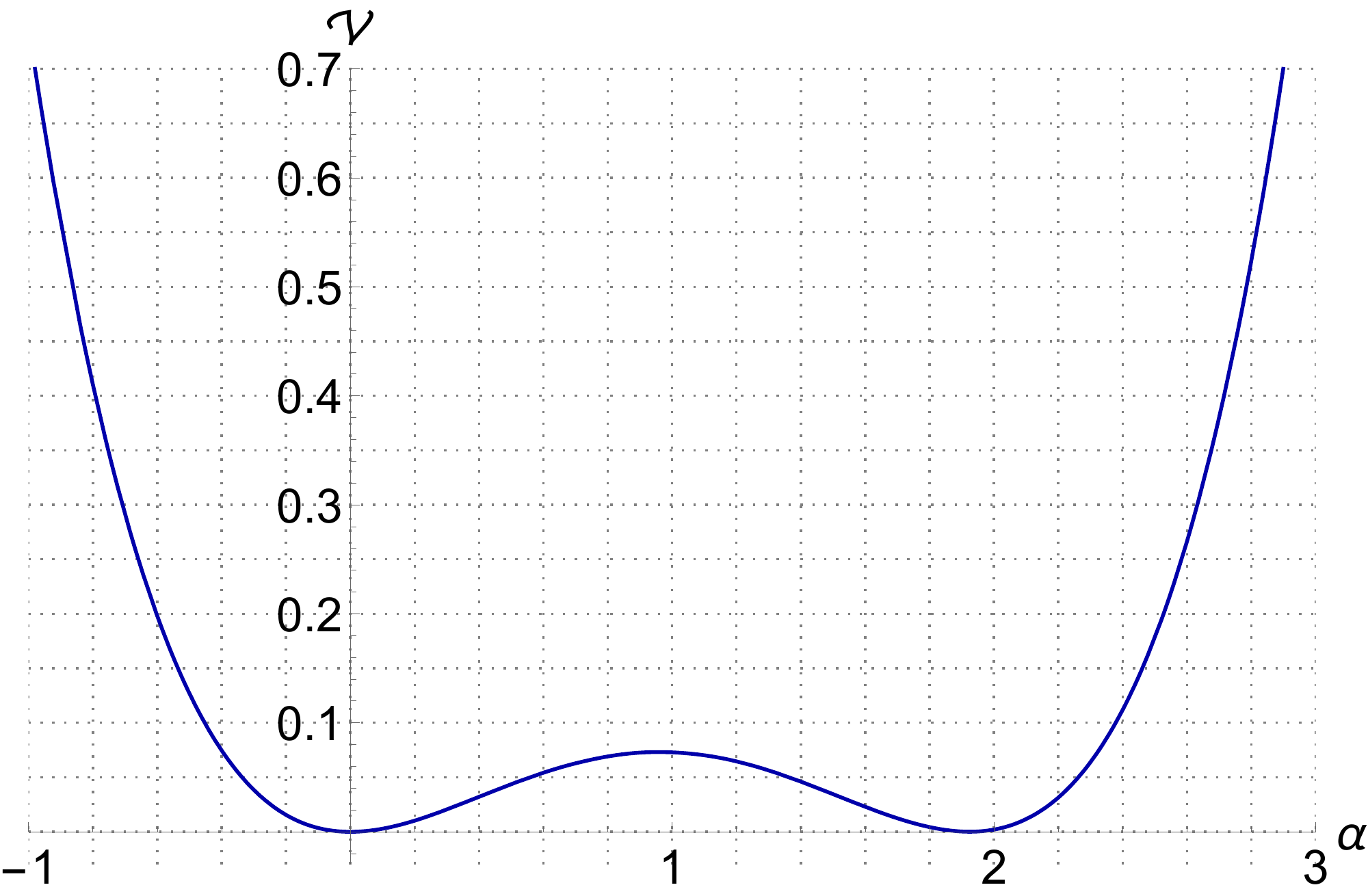}
\qquad
\includegraphics[width=0.45\textwidth]{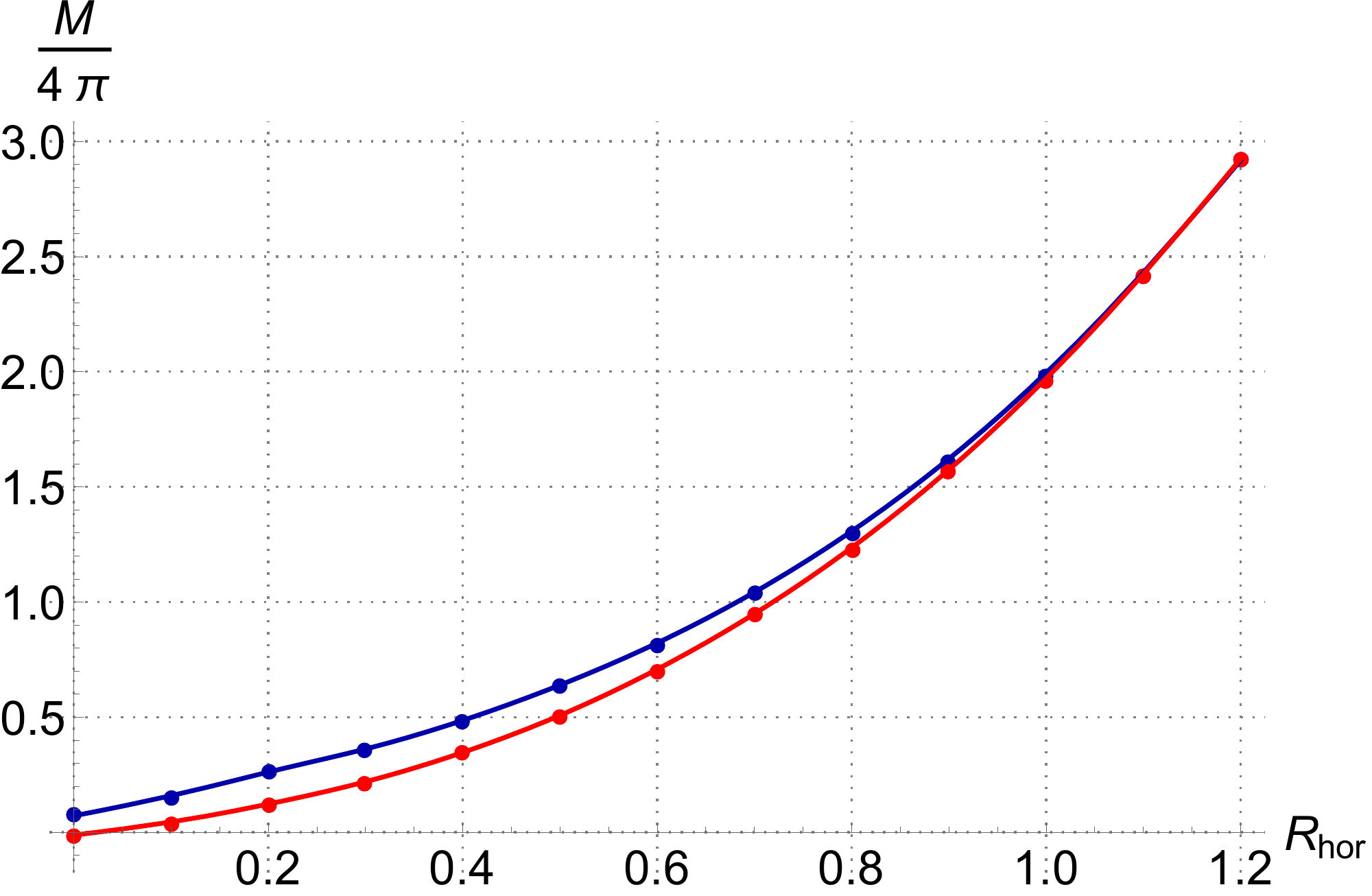}
\centering
\caption{Left: The effective potential ${\cal V}(\a)$ for the vacuum expectation values $\langle {\cal O} \rangle$ in the dual field theory with deformation $W = - \frac{1}{3} \a^3 + 0.098 \a^4$. The parameters are chosen in such a way that the potential exhibits two minima of equal depth. Right: The mass of the hairy black holes that obey these boundary conditions. The red line gives the masses of the second (perturbatively stable) branch of solutions, which are associated with the second intersection point of the curves $\b_{R_e}(\a)$ with $\b_{bc}(\a)$, and hence have more hair.}
\label{2}
\end{figure}

For the boundary conditions \eqref{bccurve} the effective potential is generally of the form shown in Figure \ref{2}, indicating the emergence of a second, metastable vacuum\footnote{In the bulk this new vacuum corresponds to the second intersection point of $\b_s(\a) = \b(\a)$ in Figure \ref{1}(a). The first intersection point corresponds to unstable solitons associated with the local maximum of ${\cal V}(\a)$.}. The AdS/CFT correspondence then suggests that the bulk theory \eqref{4-action} with such boundary conditions satisfies the Positive Mass Theorem, and that empty AdS remains the true ground state\footnote{See \cite{Hertog:2005hm} for a stability analysis of this theory (with more stringent conditions on $W$) using purely gravitational arguments.}. However the constants in \eqref{bccurve} can be tuned so that the new vacuum has precisely the same energy as the AdS vacuum. For this choice of boundary conditions the effective potential in the dual takes the form of a double well potential with two vacua at equal energy, separated by a barrier.

One can also consider excitations around each of these perturbative vacua. A particular class of excitations corresponds to `adding' a black hole at the centre of the soliton. When non-linear backreaction is included, these are spherical static black hole solutions with scalar hair. Black holes of this kind were found numerically in \cite{Hertog:2004dr,Hertog:2005hu} for boundary conditions similar to \eqref{bccurve}. Regularity of the event horizon $R_e$ implies the relation 
\be \label{horcon}
\phi'(R_{e}) =  \frac{R_e V_{,\phi_{e}}}{1-R_e^2 V(\phi_e)}
\ee
where $\phi_{e}\equiv \phi(R_e)$. The usual Schwarzschild-AdS black holes with $\phi=0$ everywhere are still valid solutions of the theory with boundary conditions (\ref{bccurve}), since the curve $\b(\a)$ intersects the origin. However in addition the theory admits black holes with scalar hair at and outside the horizon. The scalar asymptotically behaves again as \eqref{hair4d}, so we obtain a point in the $(\a,\b)$ plane for each combination $(R_e,\phi_e)$. Repeating for all $\phi_e$ gives a curve $\b_{R_e}(\a)$. In Figure \ref{1}(a) we show this curve for hairy black holes of two different sizes. As one increases $R_e$, the curve decreases faster and reaches larger (negative) values of $\b$. Given a choice of boundary conditions $\b(\a)$, the allowed black hole solutions are given by the points where the black hole curves intersect the boundary condition curve: $\b_{R_e}(\a) = \b(\a)$. It follows immediately that for boundary conditions \eqref{bccurve} there are two hairy black holes of a given horizon size provided $R_e$ is sufficiently small. Each branch of hairy black holes tends to one of the two spherical static solitons in the limit $R_e \ra 0$. The mass (\ref{bhmass2}) of both branches of black holes is shown in Figure \ref{2}(b). 

The hairy black holes reviewed here are solutions where a normal Schwarzschild-AdS black hole interior solution is smoothly glued at the horizon onto a scalar soliton solution outside, slightly modified by the non-linear backreaction of the black hole. Hence the usual AdS vacuum outside the black hole is essentially replaced by a solitonic vacuum\footnote{The upper branch of more massive hairy black holes in Figure \ref{2}(b) corresponds to black holes glued onto the unstable soliton associated with the maximum of ${\cal V}(\a)$. Those black holes are, like the soliton, unstable.}. In this way one separates to first approximation the excitations that make up the black hole interior from the degrees of freedom outside the horizon, but without introducing a second boundary. This separation is clearly manifest in the dual description of the black holes which involve a double well effective potential of the form shown in Figure \ref{2}(a). In this description, the soliton corresponds to the ground state wave function around the new vacuum whereas the black hole degrees of freedom correspond to excitations around the original vacuum.

However, there is evidently also an important coupling between both vacua. On the field theory side the vacuum structure emerges from a complicated set of multi-trace interactions. In the bulk the coupling is at the semiclassical level encoded in the regularity condition at the horizon. Note furthermore that one can consider a one-parameter family of boundary conditions of the form (\ref{bccurve}) for which the second vacuum is always at zero energy, but is gradually taken further away. In the limit of large separation in which both vacua decouple, the hairy black holes become singular on the horizon. 

These features of the dual description of hairy black holes form the basic motivation to put forward a quantum mechanical particle in a double well potential as an extremely simplified -- but solvable --  toy-model for (this class of) black holes in AdS. In the remainder of this paper we study this toy-model, and its connection to black hole physics.

\section{Quantum mechanics in a double well and black hole microstates}

\subsection{Canonical quantization and Hilbert spaces}

Consider the 1D quantum mechanical system of a particle in the double well potential\footnote{For simplicity we set $\omega = 1$ from now on and restore $\omega$ only where it is illuminating.} \eqref{introV}. One can carry out the procedure of canonical quantization either around $x_L$ or around $x_R$ by ignoring all interaction terms. This leads to two separate Fock spaces $\mathcal{F}_L$ and $\mathcal{F}_R$ which come equipped with two pairs of creation and annihilation operators $b_L, b_L^+$ and $b_R, b_R^+$. However, while formally independent, these two Fock spaces must be related since they arise from the same system. 

We first discuss the perturbative structure around the minumum at $x_R$. When expanded around $x_R$, the Hamiltonian can be written as $H = H_R^{(0)} + H_R^{(1)}$, with $H_R^{(0)}$ the Hamiltonian of a harmonic oscillator and $H_R^{(1)}$ its (perturbative) correction,
\begin{align}
\label{HR0}
H_R^{(0)} & = \frac{1}{2} p^2 + \frac{1}{2} y_R^2, \\
\label{HR1}
H_R^{(1)} & = \sqrt{\lambda} y_R^3 + \frac{\lambda}{2} y_R^4,
\end{align}
where $y_R = x - x_R$. Standard canonical quantization based on $H_R^{(0)}$ around $x_R$ yields a Fock space $\mathcal{F}_R$ with a set of basis states $|n\>_R$, together with a pair of creation-annihilation operators $b_R, b_R^+$ satisfying
\begin{equation}
b_R|0\rangle_R=0, \qquad\qquad |n\rangle_R = \frac{1}{\sqrt{n!}}(b^+_R)^n|0\rangle_R \ .
\end{equation}
A similar analysis is applicable to the left minimum at $x_L$\footnote{In particular, we can define analogous Hamiltonians $H_L^{(0)}$ and  $H_L^{(1)}$ which differ from their $x_R$ counterparts but satisfy by definition $H_L^{(0)} + H_L^{(1)}= H_R^{(0)} + H_R^{(1)}=H$. It was argued in \cite{Harlow:2014yoa} that, when modelling an eternal black hole, $H_R$ itself should be regarded as the total Hamiltonian of the theory, instead of $H_R \pm H_L$. Our model realizes this intuition since there is only a single Hamiltonian $H$. This Hamiltonian can be split into its free and interacting part as in \eqref{HR0} and \eqref{HR1} to better describe an experience of the right and the left observer. But nevertheless $H_R = H_L = H$ is always the same operator.} 
and gives another Fock space $\mathcal{F}_L$ spanned by the states $|n\>_L$ and with creation and annihilation operators $b_L, b_L^+$. One can associate to these Fock spaces two observers, left and right. The right observer perceives the state $| 0 \>_R$ as the natural semiclassical vacuum and the excited state $|n\>_R$ as an $n$-particle state. Similary, the left observer regards $| 0 \>_L$ as the semiclassical vacuum and $| n \>_L$ as an $n$-particle state.

The above canonical quantizations eliminate any relation between both Fock spaces. In particular expressions such as $[b_L, b_R]$ make no sense as the operators involved act on different Hilbert spaces. To relate $\mathcal{F}_L$ and $\mathcal{F}_R$ we have to embed these into the total Hilbert space $\mathcal{H} = L^2(\mathbb{R}; \C)$ of complex-valued, square-integrable wave functions. Consider the set $\{\varphi_n\}_{n\in \mathbb N}$ of normalized eigenfunctions of the Hamiltonian of a harmonic oscillator,
\begin{equation}\label{phiRL}
\varphi_n(x) = \frac{1}{\pi^{1/4} \sqrt{2^n n!}} H_n(x) e^{-\frac{x^2}{2}}\ ,
\qquad x\in \mathbb R\ .
\end{equation}
We can define two morphisms $F_L$ and $F_R$ between the Fock spaces $\mathcal{F}_L$, $\mathcal{F}_R$ and $\mathcal{H}$ as
\begin{align}
F_R \ : \ \mathcal{F}_R \ni |n_{R} \> & \mapsto \varphi_n^R \in \mathcal{H}, \ \hspace{1cm} \varphi_n^R(x) =  \varphi_n(x - x_{R}) \ , \label{isoR} \\[2mm]
F_L \ : \ \mathcal{F}_L \ni |n_{L} \> & \mapsto \varphi_n^L \in \mathcal{H}, \ \hspace{1cm} \varphi_n^L(x) = (\Theta \varphi^R_n)(x) = (-1)^n \varphi_n(x - x_{L})\ . \label{isoL}
\end{align}
where the CPT operator $\Theta$ acts on elements $\psi \in \mathcal{H}$ as $(\Theta \psi)(x) = \psi^{\ast}(-x)$, and the asterisk denotes complex conjugation. We have introduced an additional factor $(-1)^n$ in the definition of $\varphi_n^L$ which allows us to relate left and right modes as CPT conjugates of each other. The maps $F_R$ and $F_L$ are obviously isomorphisms that can be thought of as two different choices of basis of $\mathcal{H}$ associated with harmonic oscillator eigenstates around either $x_L$ or $x_R$. Hence the total Hilbert space $\mathcal{H}$ is isomorphic to each Fock space $\mathcal{F}_L$ and $\mathcal{F}_R$ \emph{separately},
\begin{equation}\label{Hlambda}
\mathcal{H}_{\lambda} =\mathcal{H} \cong \mathcal{F}_R \cong \mathcal{F}_L \ ;
\end{equation}
There is no tensor product. The interactions provide a non-trivial identification of the two Fock spaces within a single $\mathcal{H}$.

On the other hand, the Fock spaces $\mathcal{F}_R$ and $\mathcal{F}_L$ have distinct sets of creation and annihilation operators. This means that an expression such as $b_L | n \>_R$ makes \textit{a priori} no sense, since $b_L$ acts on $\mathcal{F}_L$, whereas $| n \>_R$ is a state in $\mathcal{F}_R$. However the isomorphisms \eqref{isoR} and \eqref{isoL} can be used to define new annihilation operators $a_R, a_L$ constructed from $b_R, b_L$ that do have a well defined action in $\mathcal{H}$. In particular, defining the operators
\begin{equation} \label{defaRaL}
a_R = F_R b_R F_R^{-1}, \qquad\qquad a_L = F_L b_L F_L^{-1},
\end{equation}
together with their conjugates $a_R^+$ and $a_L^+$, we get the following actions,
\begin{align}\label{aRphiR}
& a_R \varphi_n^R = \sqrt{n} \varphi_{n-1}^R, && a_R^{+} \varphi_n^R = \sqrt{n+1} \varphi_{n+1}^R \ , \\
\label{aLphiL}
& a_L \varphi_n^L =  \sqrt{n} \varphi_{n-1}^L, && a_L^{+} \varphi_n^L = \sqrt{n+1} \varphi_{n+1}^L \ ,
\end{align}
In particular, the action of $a_L$ is related to the action of $a_R$ by the parity operator,
\begin{equation}\label{aLR-rel-1}
a_L = \Theta a_R \Theta
\end{equation}
which is the standard relation between left and right creation-annihilation operators featuring in black hole physics, \textit{e.g.}, \cite{Harlow:2014yka}. Expressions such as $a_L \varphi_n^R$ are now meaningful because both pairs of operators $a_L, a_L^+$ and $a_R, a_R^+$ act on the same Hilbert space $\mathcal{H}$. Their relation to the fundamental field operator $x$ is simply that of a shifted harmonic oscillator,
\begin{align}
y_R & = x - x_R = \frac{1}{\sqrt{2}}(a_R + a_R^+), \\
y_L & = \Theta y_R \Theta = x_L - x = \frac{1}{\sqrt{2}}(a_L + a_L^+).
\end{align}
Since the two oscillators are related by a displacement (up to a sign), it follows that the creation-annihilation operators are related as well,
\begin{equation} \label{aLofaR}
a_L = - \frac{1}{\sqrt{2 \lambda}} \mathbb{I} - a_R = - \frac{N}{\sqrt{2}} \mathbb{I} - a_R.
\end{equation}
A striking feature of this expression is that it does not possess a finite decoupling limit $N \rightarrow \infty$ as an operator statement. Instead of approximating the free field creation-annihilation operators $b_L, b_L^+, b_R, b_R^+$, these operators diverge in the decoupling limit.\footnote{Another indication that the natural creation-annihilation operators $a_L, a_L^+, a_R, a_R^+$ are not to be identified with the perturbative creation-annihilation operators are their commutation relations. Using \eqref{aLofaR} we can find that
\begin{equation} \label{commaLaR}
[a_L, a_R] = [a_L^{+}, a_R^{+}] = 0, \qquad\qquad [a_L, a_R^{+}] = [a_R, a_L^{+}] = -1 \ .
\end{equation}
Hence $a_L$ and $a_R^{+}$ do not commute, in sharp contrast with the usual situation in black hole physics, where one expects the left and right creation-annihilation operators to commute as a consequence of the locality of semiclassical physics near the horizon.}
 
As we can see, the $\lambda \rightarrow 0$ limit should be taken with care. The low occupancy modes $\varphi_n^R$ and $\varphi_m^L$, $n,m, \ll N$ nearly decouple for small $\lambda$ and in the limit $\lambda \rightarrow 0$, the two sets of modes decouple completely. We end up with two separate harmonic oscillators with the tensor product Hilbert space  $\mathcal{H}_0 \cong \mathcal{F}_L \otimes \mathcal{F}_R$. On the other hand the interaction Hamiltonian $H_R^{(1)}$ vanishes for $\lambda = 0$. 
From the point of view of the vacuum at $x_R$, the second vacuum moves away and a single harmonic oscillator Hilbert space $\mathcal{F}_R$ remains. Hence from this viewpoint every state $\varphi_n^L$ disappears as $\lambda \rightarrow 0$. In the language of mathematical perturbation theory \cite{Reed/Simon} this limit is singular.

This is also the case in the bulk for the black holes with scalar hair described in the Introduction and reviewed in Section \ref{hairy}. The distance between the vacua in the dual is related to the boundary conditions in the bulk, which in turn determine the amount of scalar hair. Increasing the distance between both vacua, keeping the black hole mass constant, also increases the value of the scalar hair on the horizon. In the limit in which the vacua in the dual theory decouple, a curvature singularity at the horizon develops, effectively dividing the inside and outside regions in two separate spacetimes.

In black hole physics one usually assumes at the outset that the Hilbert space splits in a tensor product of Fock spaces associated with modes inside and outside the horizon. In particular this is a fundamental assumption behind much of the discussion of the information paradox (see \textit{e.g.}, \cite{Almheiri:2012rt,Almheiri:2013hfa,Bousso:2012as,Marolf:2013dba,Maldacena:2001kr,Mathur:2008wi,Mathur:2009hf,Czech:2012be,VanRaamsdonk:2013sza}). Our toy-model shows that small, non-perturbative interactions can drastically change this structure.\footnote{Interactions in the form of shock waves between the left and right Hilbert spaces of two-sided eternal black holes in AdS were recently analyzed in \cite{Maldacena:2017axo,vanBreukelen:2017dul}.}

\subsection{Black Hole Microstates}

Another important characteristic of our toy-model is that it is solvable. The energy eigenstates and the corresponding energies can be computed numerically to arbitrary precision by various methods, most notably the standard Ritz-Rayleigh method\footnote{One cannot rely on perturbation theory around a single minimum, since perturbative series diverge and lead to asymptotic expansions around $\lambda = 0$.}.

\begin{figure}[ht]
	\includegraphics[width=0.45\textwidth]{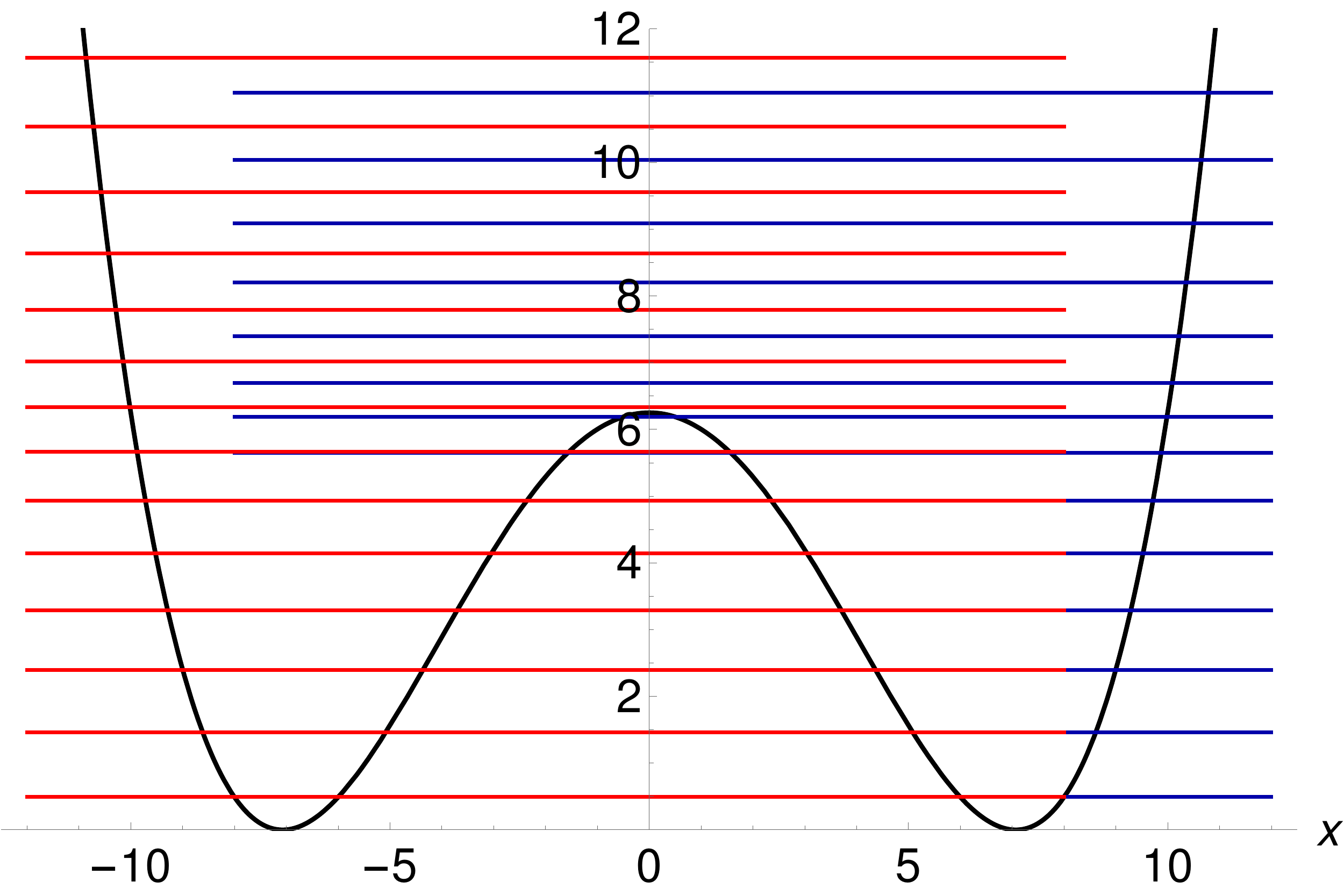}
	\qquad
	\includegraphics[width=0.45\textwidth]{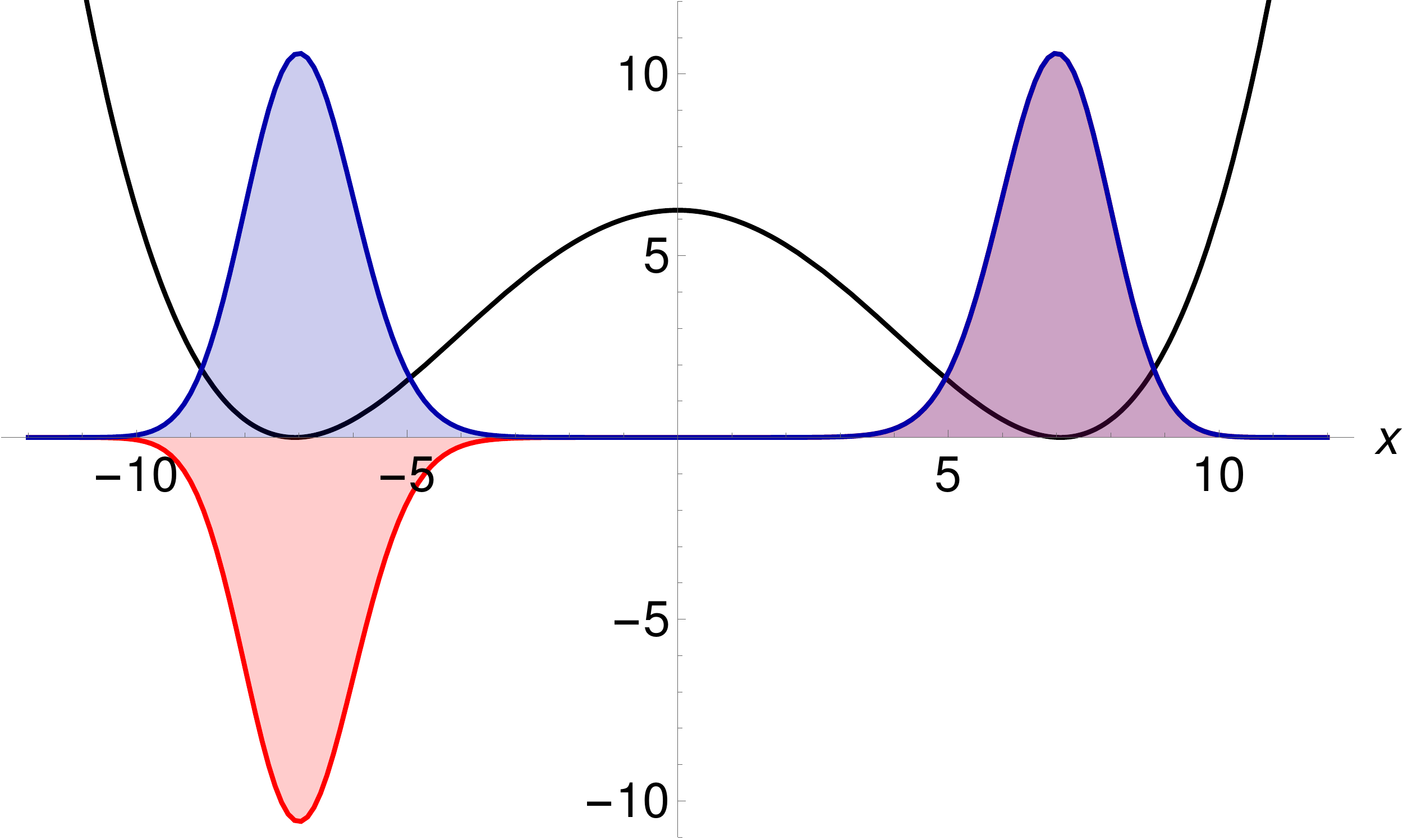}
	\centering
	\caption{The structure of energy levels (left) and the profile of the two lowest energy eigenstates $\Psi_0^+$ and $\Psi_0^-$ (right). For energies lower than the top of the barrier $V_{\ast}$ odd eigenstates $\Psi_n^{-}$ (red) have slightly larger energy than even eigenstates $\Psi_n^{+}$ (blue).\label{fig:en}}
\end{figure}

We will say that $f(\lambda)$ is \emph{non-perturbatively small} if $f \sim 0$ as $\lambda \rightarrow 0^{+}$, where $\sim$ denotes the asymptotic expansion. Equivalently, $f(\lambda)$ is non-perturbatively small if $f(\lambda) = o(\lambda^n)$ for all $n \geq 0$ as $\lambda \rightarrow 0^{+}$. We will denote any non-perturbatively small terms by $o(\lambda^\infty)$. Finally we say that two quantities are \emph{equal in perturbation theory} or \emph{equal up to non-perturbative terms} if they have identical asymptotic expansions. These definitions are needed when considering, for example, the energy eigenstates of the full system, to which we now turn.

Since the double well potential is invariant under $x \rightarrow -x$ the operator $\Theta$ commutes with the Hamiltonian. Hence every energy eigenstate has definite parity, and the Hilbert space can be decomposed as $\mathcal{H} = \mathcal{H}^{+} \oplus \mathcal{H}^{-}$, where $\Theta \mathcal{H}^{+} = \mathcal{H}^{+}$ and $\Theta \mathcal{H}^{-} = - \mathcal{H}^{-}$. We denote energy eigenstates by
\begin{equation}
H\Psi_n^{\pm} = E_n^{\pm} \Psi_n^{\pm}\ ,
\end{equation}
where $\Psi_n^{\pm}$ are even/odd eigenstates.\footnote{We choose all eigenstates $\Psi_n^{\pm}$ to be real and normed to one. Overall signs are such that for $x \rightarrow \infty$, $\Psi_n^{\pm} \approx 2^{-1/2} \varphi_n^R$. The two lowest energy eigenstates $\Psi_0^{\pm}$ can be seen in Figure \ref{fig:en}.} The corresponding energies satisfy $E_n^{+} < E_n^{-}$ and their difference $\Delta E_n = E_n^- - E_n^+$ is exponentially small.
In particular, the energy difference between the ground state and the first excited state is dominated by the 1-instanton effect \cite{ZinnJustin:2004ib} and satisfies
\begin{equation} \label{split}
E_0^{-} - E_0^{+} = \frac{2}{\sqrt{\pi \lambda}} e^{-\frac{1}{6 \lambda}} \left[ 1 + O(\lambda) \right].
\end{equation}
The vacuum energy $E_0^{+}$ is always smaller than $1/2$ -- the energy of an unperturbed ground state of the harmonic oscillator.

\begin{figure}[ht]
	\includegraphics[width=0.45\textwidth]{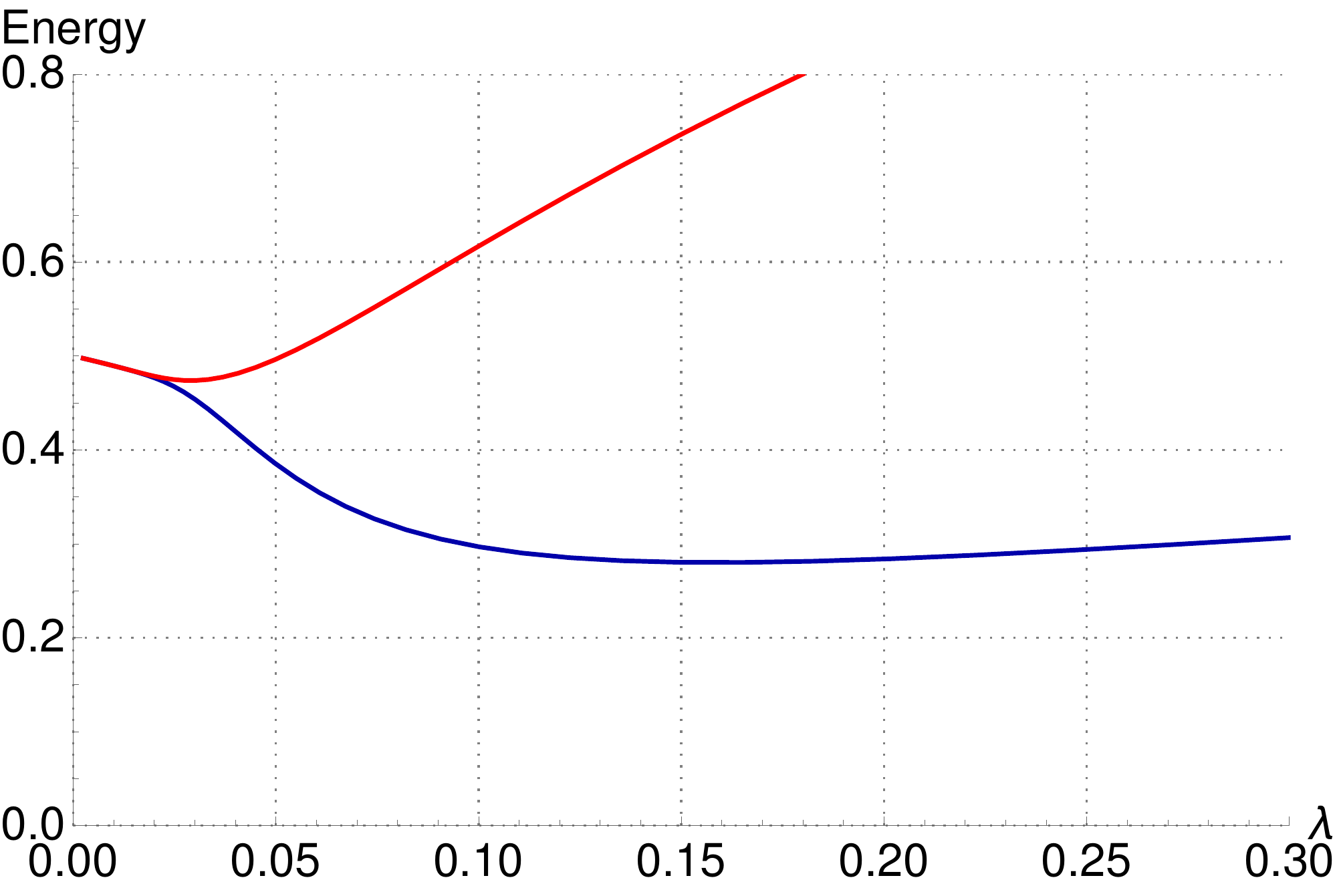}
	\qquad
	\includegraphics[width=0.45\textwidth]{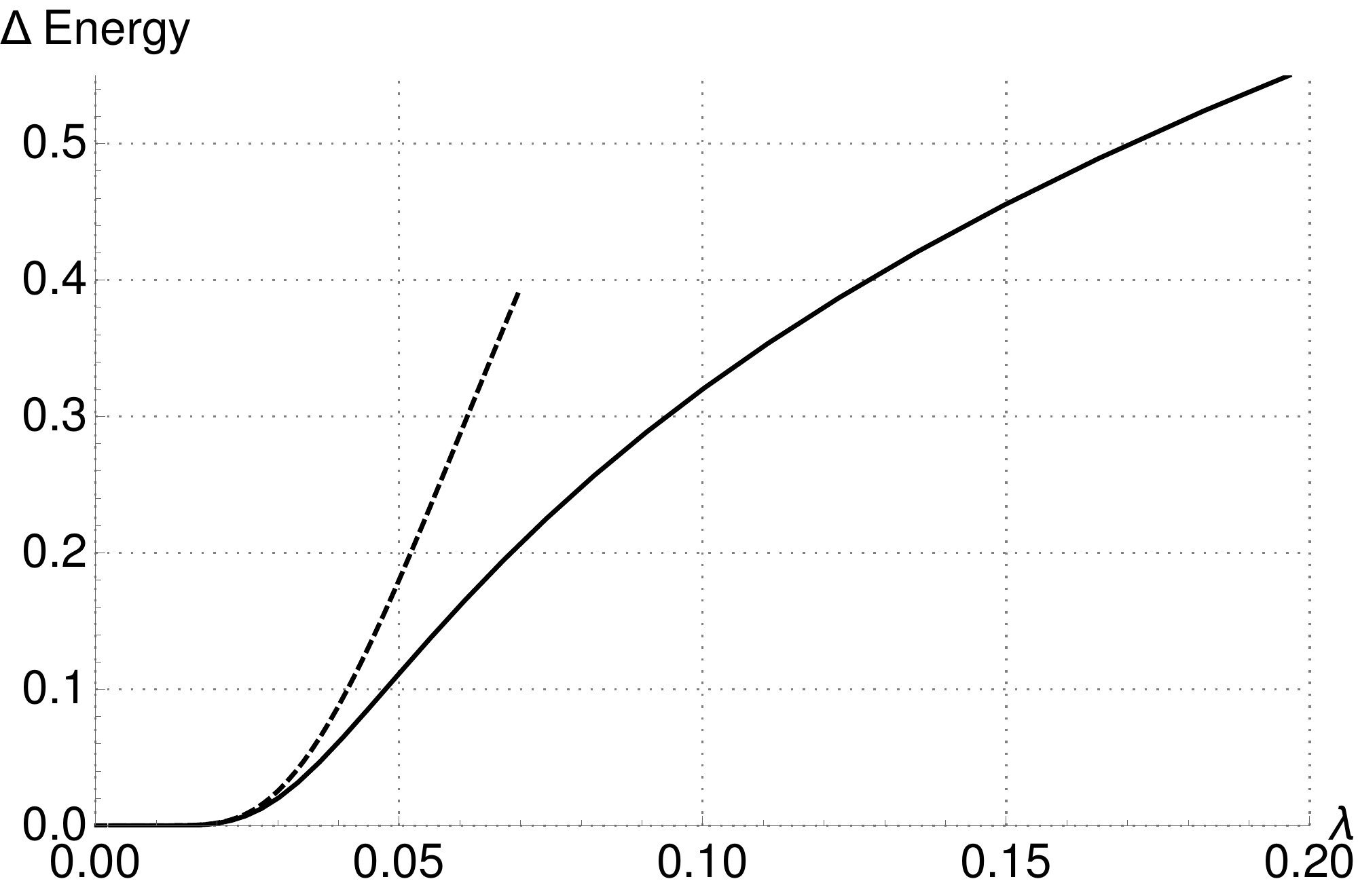}
	\centering
	\caption{Left: energy of the vacuum (blue) and the first excited state (red) as a function of the coupling $\lambda$. Right: the difference between the two energies as a function of $\lambda$. The dashed line represents the leading 1-instanton approximation given by \eqref{split}.\label{fig:split}}
\end{figure}

The difference $\Delta E_n \sim 0$ in general is a non-perturbative effect and hence exponentially small for energies well below the maximum of the potential, $E_n^{\pm} < V_{\ast} = 1/(32 \lambda)$. For energies larger than this, non-perturbative effects are numerically large and the difference $\Delta E_n \simeq 1/2$, as one can observe in Figure \ref{fig:en}. In this sense every pair of energy eigenstates $\Psi_n^{\pm}$ corresponds to two microstates with exponentially small energy splitting due to non-perturbative effects.
	
Motivated by the dual description of the black holes with scalar hair in Section \ref{hairy}, where the (perturbative) degrees of freedom on both sides of the horizon correspond to excitations around two different perturbative vacua, we interpret states with significant support around both minima of the potential as the dual description in our toy-model of a \emph{black hole microstate}. By contrast, semiclassical states centered around one of the two vacua only are interpreted as spacetimes without a black hole\footnote{At first sight this description of microstates is at odds with the intuition that black holes should be high-energy states $E \sim N^2$, compared to the vacuum energy, as perceived by an asymptotic observer. However for an interacting system a number operator and the Hamiltonian are in general very different; whereas the Hamiltonian of the system is a unique operator once the time direction is chosen, a number operator is an inherently semiclassical object that depends on a choice of a semiclassical vacuum. We return to this point below.}. We consider microstates of any energy\footnote{This restriction on the energies is needed for a perturbative description to be meaingful and resonates with the bulk where there are no hairy black holes above a certain mass (cf. Figure \ref{2}(b)).} $E \ll V_{\ast}$, but the lowest energy states are of a particular interest. Denote
\begin{equation} \label{defM}
\M = \{ \alpha_+ \Psi_0^+ + \alpha_- \Psi_0^- \ : \ \alpha_{\pm} \in \C \}.
\end{equation}
and consider any normalized state $\mu \in \M$. The energy of any such state is non-perturbatively close to the ground state energy, $\< \mu | H | \mu \> = E_0^+ + o(\lambda^\infty)$. Hence, in perturbation theory the ground state is degenerate. For this reason we will refer to $\M$ as the subspace of \emph{perturbative vacua} and each element $\mu \in \M$ will be called a \emph{perturbative vacuum}. Roughly we have in mind a correspondence between the (degenerate) energy $E_n$ of the states and the mass of the black holes.

A relation between microstates and macrostates can be twofold. We first discuss this from the viewpoint of a single, say right, asymptotic observer with easy access to the right portion of the wave function only. This is the natural perspective if we consider our model to be a dual toy-model description of the single-sided hairy black holes discussed earlier. In this context the right portion of the wave function specifies a macrostate, and a set of microstates differing in the shape of the wave function around the left minimum can be considered. 

A second characterisation of macrostates follows from considerations of a pair of observers in two distinct asymptotic regions as \textit{e.g.}, in the case of eternal black holes. From the point of view of two such perturbative observers a macrostate is given by two independent pieces of the wave function: the left portion, $\psi_L$, and the right portion, $\psi_R$. We can regard a macrostate as being represented by a tensor product $\psi_L \otimes \psi_R$, while a set of corresponding microstates is given by all states of the form $\alpha_L \psi_L + \alpha_R \psi_R$ for $\alpha_L, \alpha_R \in \C$. Each microstate is a specific continuation through the potential barrier that eludes both observers. We will discuss the relation between such macrostates and microstates in detail in Section \ref{sec:eff}.

This interpretation also fits in the black hole paradigm of \cite{Papadodimas:2012aq} and with a more general quantum perspective on black holes \cite{Mathur:2008wi,Hartle:2015bna,Hertog:2017vod} according to which, from the point of view of a single asymptotic observer, the Hilbert space $\mathcal{H}$ factors as $\mathcal{H} \cong \mathcal{H}_{\text{coarse}} \otimes \mathcal{H}_{\text{fine}}$. The coarse degrees of freedom $\mathcal{H}_{\text{coarse}}$ are clearly distinguishable by the asymptotic observer within perturbation theory whereas $\mathcal{H}_{\text{fine}}$ contains fine degrees of freedom that require non-perturbative effects to identify. In our model $\mathcal{H}_{\text{fine}} \cong \C^2$ is a two-dimensional space, which can be identified with the space of perturbative vacua $\M$. The energy difference of any two microstates is then non-perturbatively small, and hence our model satisfies postulate 3 of \cite{Susskind:1993if,Almheiri:2012rt}.

Finally the fact that in perturbation theory various microstates cannot be distinguished also leads to the Boltzmann entropy,
\begin{equation}\label{SBoltz}
S_B = \log \dim \mathcal{H}_{\text{fine}} = \log 2.
\end{equation}
This is the Bolzmann entropy associated with a single pair of harmonic oscillators\footnote{To get an area factor as in black holes, one should consider an ensemble of oscillators with different frequencies \cite{Harlow:2014yka}.}.

\subsection{Firewalls?} \label{sec:fire}

We now explore further the implications of the above identification of black hole microstates in our toy-model.
The model has two natural number operators that describe (perturbative) excitations from one or the other asymptotic viewpoint,
\begin{equation}
N_L = H_L^{(0)} = a_L^+ a_L, \qquad\qquad N_R = H_R^{(0)} = a_R^+ a_R
\end{equation}
However we are also interested to describe observations from the viewpoint of an infalling observer with easy access to perturbative physics in both asymptotic regions, inside and outside the horizon of the black hole. A first guess for this is to consider the following number operator  
\begin{equation} \label{defNA}
N_A = N_L + N_R + O(\sqrt{\lambda}) = a_L^+ a_L + a_R^+ a_R + O(\sqrt{\lambda}),
\end{equation}
possibly up to small corrections in $\lambda$.\footnote{Formally, the number operator for the infalling observer is of this form. In fact, for a fixed mode, the number operator for the infalling observer is non-perturbatively close to the number operator for the asymptotic observer, since $\< N_A \> \sim e^{-8 \pi \omega M}$ with $M$ of order $N$.} Indeed, if $\lambda = 0$ this operator counts a sum of excitations of two decoupled harmonic oscillators. One would expect that with a small coupling $\lambda$, the sum $N_L + N_R$ recieves corrections of order $O(\sqrt{\lambda})$ in such a way that $N_A$ vanishes (or at least is small) in the new vacuum $\Psi_0^+$. However, to the contrary, the expectation value of $N_A$ in a generic state turns out to be very large,
\begin{equation} \label{firewallA}
\< \psi | N_A | \psi \> \gtrsim \frac{1}{2} N^2 \left[ 1 + O \left( \frac{1}{N} \right) \right].
\end{equation}
Since this diverges as $N$ grows, in the language of \cite{Almheiri:2012rt} the microstate appears to exhibit a \emph{firewall}. In particular even very low energy states including semiclassical vacua exhibit firewalls. Indeed, while
\begin{equation}
\< \varphi_0^L | N_L | \varphi_0^L \> = \< \varphi_0^R | N_R | \varphi_0^R \> = 0\ ,
\end{equation}
as expected, relation \eqref{aLofaR} leads to a large expectation value
\begin{equation} \label{Nvarphi}
\< \varphi_0^L | N_R | \varphi_0^L \> = \< \varphi_0^R | N_L | \varphi_0^R \> = \frac{1}{2 \lambda} = \frac{1}{2} N^2.
\end{equation}
By contrast, the energy of the state remains small,
\begin{equation} \label{en_0}
\< \varphi_0^L | H | \varphi_0^L \> = \< \varphi_0^R | H | \varphi_0^R \> = \frac{1}{2} + \frac{3}{8} \lambda.
\end{equation}

\begin{figure}[ht]
	\includegraphics[width=0.45\textwidth]{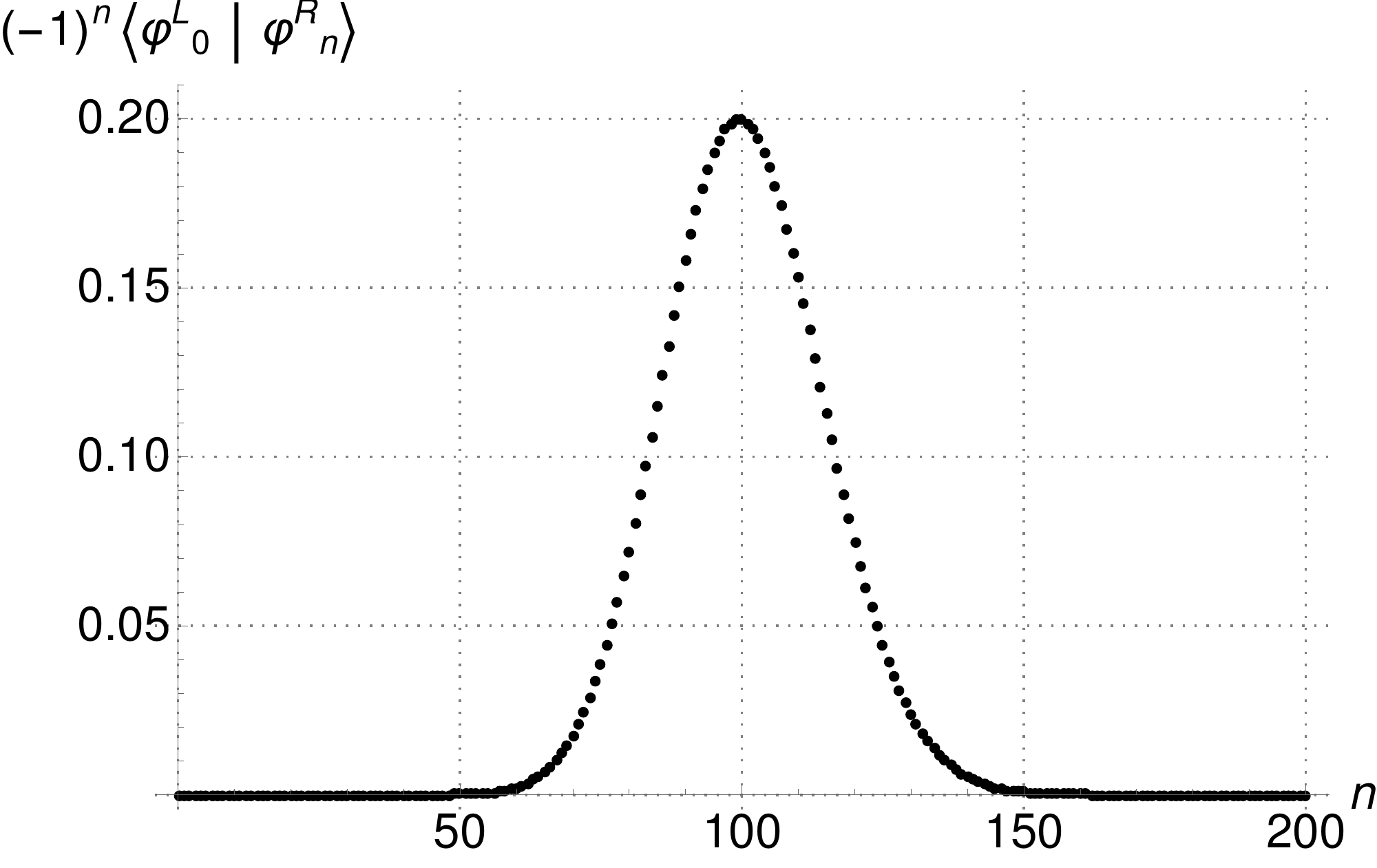}
	\qquad
	\includegraphics[width=0.45\textwidth]{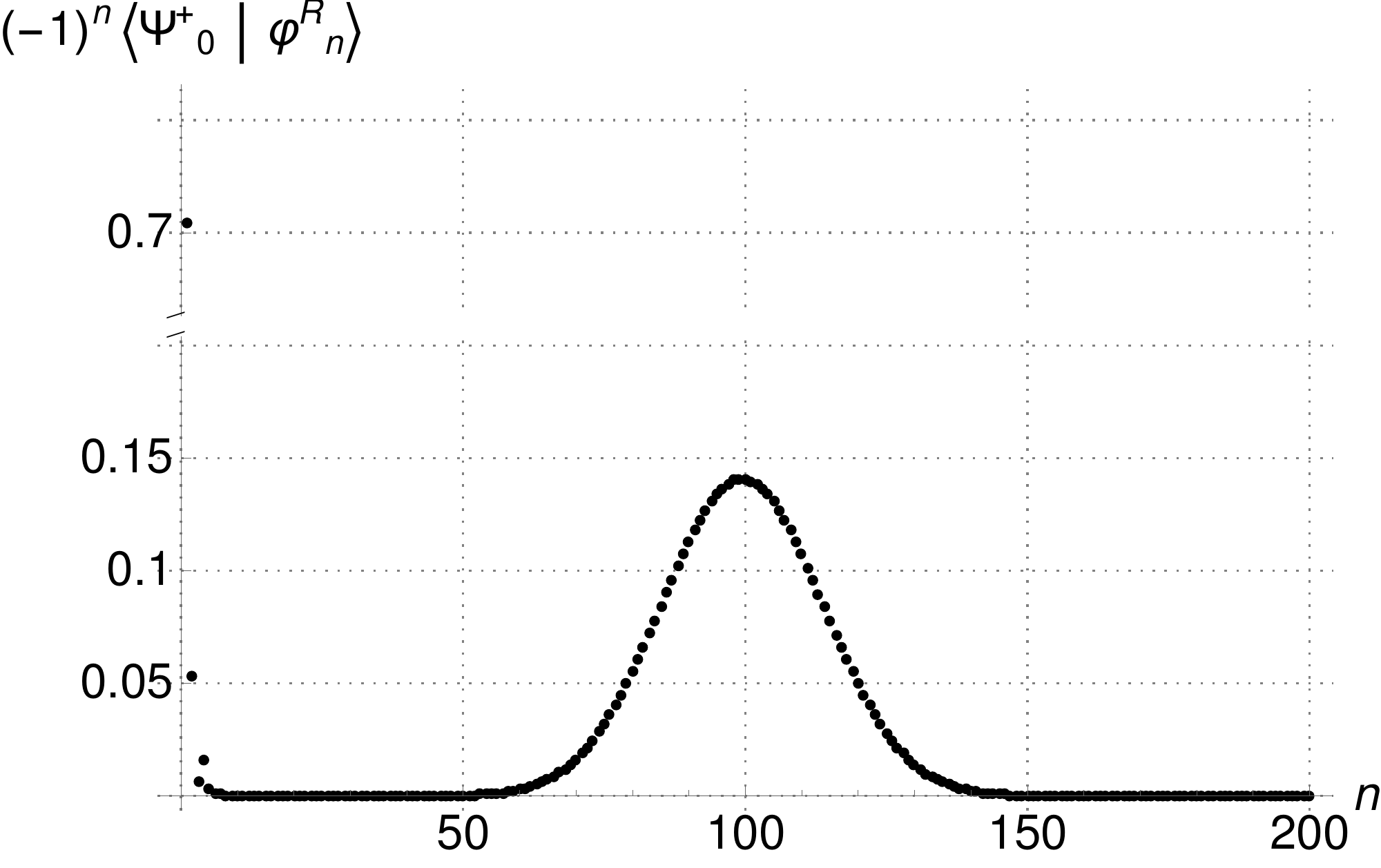}
	\centering
	\caption{Left: matrix coefficients \eqref{mat:n0} between $\varphi_0^L$ and $\varphi_n^R$ as a function of $n$. The position of the maximum is always around $n \sim 1/(2\lambda)$. Right: matrix coefficients between the true ground state $\Psi_0^+$ and $\varphi_n^R$ as a function of $n$. Recall that in the leading order the ground state is given by \eqref{ground_approx}. Hence the plot shows excitations of a few initial modes $\varphi_n^R$ for $n \sim 0$ as well as a wide peak around $n \sim 1/(2 \lambda)$. In both figures $\lambda = 1/200$.\label{fig:cfs}}
\end{figure}

The large expectation values \eqref{Nvarphi} are a manifestation of the fact that from the viewpoint of, say, the right minimum, the state $\varphi_0^L$ is a highly excited state. Indeed, since the full Hilbert space $\mathcal{H}$ is isomorphic to $\mathcal{F}_R$, both sets $\{ \varphi_n^R \}_n$ and $\{ \varphi_n^L \}_n$ span the entire Hilbert space $\mathcal{H}$ separately. We can decompose the left modes $\varphi_n^L$ in terms of right modes $\varphi_n^R$ as,
\begin{equation} \label{ambiga}
\varphi_m^L = \sum_{n=0}^{\infty} \< \varphi_n^R | \varphi_m^L \> \varphi_n^R.
\end{equation}
The value of the matrix element $\< \varphi_n^R | \varphi_m^L \>$ can be calculated by noticing that the left and right modes are related by a displacement, up to a sign,
\begin{equation}
\varphi_n^L(x) = (-1)^n \varphi_n^R(x + a), \qquad a = 2 x_R = \lambda^{-1/2}.
\end{equation} 
The matrix element then reads
\begin{align} \label{LR}
\< \varphi^L_m | \varphi^R_n \> & = \< \varphi^L_n | \varphi^R_m \> = \< \varphi^R_n | \Theta | \varphi^R_m \> = \< \varphi^L_m | \Theta | \varphi^L_n \> \nn\\
& = (-1)^{n} \sqrt{\frac{m!}{n!}} e^{-\frac{1}{4 \lambda}} (2 \lambda)^{\frac{1}{2}(m - n)} L_m^{(n-m)}\left( \frac{1}{2 \lambda} \right),
\end{align}
where $L_m^{(\alpha)}$ denotes Laguerre polynomials. This expression simplifies for the semiclassical vacuum state $m = 0$, for which $L_0^{(n)} = 1$,
\begin{equation} \label{mat:n0}
\< \varphi^L_0 | \varphi^R_n \> = \frac{(-1)^n e^{-\frac{1}{4 \lambda}}}{\sqrt{2^n \lambda^n n!}}.
\end{equation}

Figure \ref{fig:cfs} shows numerical values of these matrix elements as a function of $n$.
For small $\lambda$ the distance between the minima is large and one can use Stirling's formula to find that \eqref{mat:n0} attains its maximum at $n = 1/(2 \lambda)$. This shows that, in order to write a semiclassical vacuum state $\varphi^L_0$ as a superposition of the semiclassical states around the right minimum, $\varphi^R_n$, one needs to excite highly energetic states, namely those with $n$ of order $N^2$. Low energy states of the left asymptotic observer are detected as highly excited states by the right asymptotic observer and \textit{vice versa}.

\begin{figure}[ht]
	\includegraphics[width=0.65\textwidth]{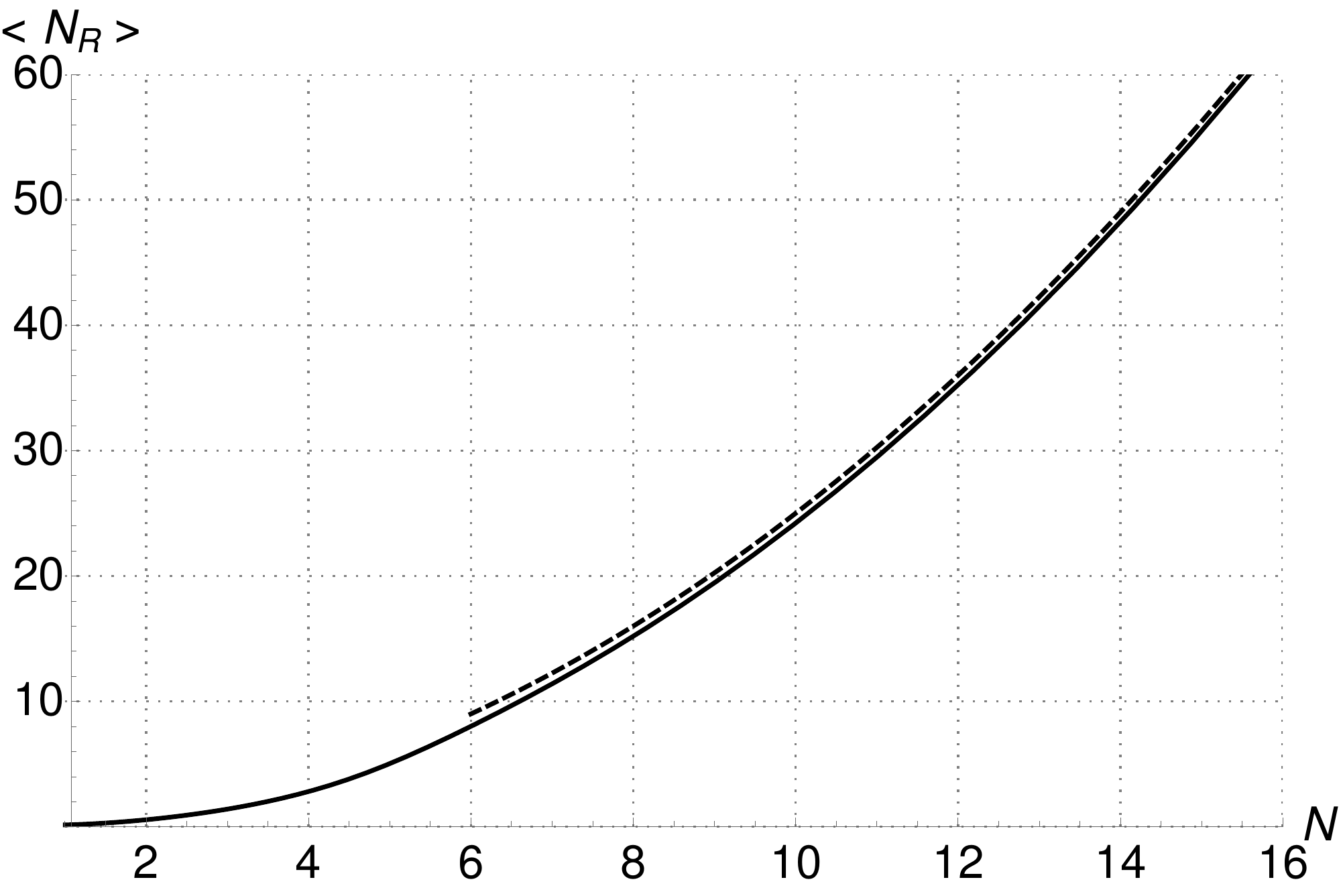}
	\centering
	\caption{The expectation value of the naive number operator $N_R$ in the ground state $\Psi_0^+$ as function of $N = \lambda^{-1/2}$. The dashed line represents the leading term $N^2/4$ in equation \eqref{firewall}.\label{fig:NaiveNA}}
\end{figure}

The above conclusions are directly applicable to the lowest energy eigenstates. It follows from perturbation theory that to leading order in the coupling $\lambda$ we have
\begin{equation} \label{ground_approx}
\Psi_0^{\pm} = \frac{1}{\sqrt{2}} ( \varphi_0^R \pm \varphi_0^L ) + O(\sqrt{\lambda}).
\end{equation}
Hence these states are all highly populated both with respect to $N_L$ and $N_R$. In particular equation \eqref{Nvarphi} implies that
\begin{equation} \label{firewall}
\< \Psi_0^{\pm} | N_R | \Psi_0^{\pm} \> = \< \Psi_0^{\pm} | N_L | \Psi_0^{\pm} \> = \frac{1}{4} N^2 \left[ 1 + O \left( \frac{1}{N} \right) \right].
\end{equation}
By considering a general microstate of the form $\m \in \M$ the definition \eqref{defNA} leads to a firewall expressed by equation \eqref{firewallA}.

Thus we find that even the ground state $\Psi_0^{+}$ is a highly populated state of very small energy as measured by the total Hamiltonian \eqref{H}. This seems paradoxical but is in fact just a consequence of the interacting nature of the system and related to non-perturbative effects. Actual numerical values of the matrix elements $\< \Psi_0^+ | \varphi_n^R \>$ as a function of $n$ are shown in Figure \ref{fig:cfs}, while the expectation value of the right number operator $N_R$ in the ground state $\Psi_0^R$ is presented in Figure \ref{fig:NaiveNA}.

To conclude, we have identified black hole microstates in our dual toy-model as relatively low energy states in the dual theory that are nevertheless heavily populated from the point of view of an asymptotic observer. The microstates are indistinguishable by an asymptotic observer with access to the perturbative physics only. A microstate structure emerges in our model as a consequence of non-perturbative level splitting \eqref{split} in the presence of interactions. This splitting can then be regarded as a source for the entropy \eqref{SBoltz}.

At first sight the results of this section would seem to support the conclusions of \cite{Almheiri:2012rt,Almheiri:2013hfa}, \textit{i.e.}, the presence of firewalls. This, however, will turn out to be false. A caveat is that in the decoupling limit $\lambda \rightarrow 0$ the creation-annihilation operators $a_L, a_L^+, a_R, a_R^+$ do not approach the perturbative operators $b_L, b_L^+, b_R, b_R^+$ in any sense. In fact, as indicated by equation \eqref{aLofaR} such a limit is ill-defined. In Section \ref{sec:eff}, starting from the interacting model, we carefully identify the perturbative degrees of freedom from the point of view of both observers, and we construct well-defined perturbative operators. In particular we will construct another set of creation-annihilation operators on $\mathcal{H}$, which will have a well-defined decoupling limit.

\subsection{Limitations of perturbation theory}

Before we proceed we pause to formulate precisely the limitations of the validity of perturbation theory around one of the minima in our model. This will be important in what follows. 

\begin{enumerate}
	\item[a.] \textit{Perturbation theory breaks down when the overlap between the left and right semiclassical modes, $\< \varphi^L_m | \varphi^R_n \>$ becomes significant when $n \sim m \sim N^2$}. 
\\
Notice that all matrix elements \eqref{LR} are exponentially damped by a factor of $e^{-1/(4\lambda)}$. In other words one could write $\< \varphi^L_m | \varphi^R_n \> = O(e^{-1/(4\lambda)}) = o(\lambda^\infty)$. This is the correct behavior for an amplitude associated with the tunneling process, but it does not imply that the amplitude remains small for all states. Indeed, a degree of a Laguerre polynomial $L_n^{(\alpha)}(z)$ is equal to $n$, and the leading term is $(-1)^n z^n/n!$. Hence, for $n = m$ the matrix element becomes
\begin{equation}
\< \varphi^L_n | \varphi^R_n \> = \frac{e^{-\frac{1}{4 \lambda}}}{(2 \lambda)^n n!} \left[1 + O(\lambda) \right].
\end{equation}
For $n \sim m \sim 1/(4 \lambda)$ the Stirling's formula indicates that the denominator vanishes faster than the numerator. Therefore the non-perturbative terms become numerically large.

\item[b.] \textit{Time-independent perturbation theory breaks down for states with occupancy numbers $m, n \sim N^2$.}
	\\
Since the potential \eqref{introV} is quartic, $\< \varphi_m^R | H | \varphi_n^R \> = 0$ if $|n - m| > 4$. Furthermore, $\< \varphi_m^R | H | \varphi_n^R \> = O(\sqrt{\lambda})$ if $m \neq n$. Hence we can concentrate on the energy of the $n$-th perturbative state $\varphi_n^R$,
\begin{equation} \label{en_nR}
\< \varphi_n^R | H | \varphi_n^R \> = \frac{1}{2} + n + \frac{3}{8} \lambda (2 n^2 + 2 n + 1).
\end{equation}
Clearly, if $n$ is of order $1/\lambda$ the correction is of the same order than the unperturbed part. This is one of the many indications that the perturbative methods break down for states with occupancy numbers of order $1/\lambda$.

	\item[c.] \textit{The subleading terms in the commutation relation
\begin{equation} \label{commHaR}
[H, a_R] = - a_R - \frac{3}{\sqrt{2}} \sqrt{\lambda} y_R^2 - \lambda \sqrt{2} y_R^3
\end{equation}
	become relevant whenever applied to states with occupancy numbers $n \sim N^2$.}
\\
In the commutation relation \eqref{commHaR} the correction is formally of order $\sqrt{\lambda}$. However, when applied to the state $\varphi_n^R$ with $n \sim 1/\lambda$ the unperturbed part is of the same order than the perturbation.
	\item[d.] \textit{Time-dependent perturbation theory breaks down for times $t \sim N$ for any state}.
\\
Breakdown of perturbation theory can also be seen in time evolution. For example, time-dependent perturbation gives a matrix element
\begin{equation}
\< \varphi_n^R | e^{- \I t H} | \varphi_m^R \> = \delta_{nm} - \I t \< \varphi_n^R | H_R^{(1)} | \varphi_m^R \> + \ldots
\end{equation}
When $|n-m|>4$, the matrix element in the second term is identically zero.
Hence such term becomes relevant in the expansion in $\lambda$, when either $n$, $m$ are of order $1/\sqrt{\lambda}$. Even for low occupancy states with $m,n$ of order $1$ in $\lambda$, the perturbation theory breaks down after time $t \sim 1/\sqrt{\lambda} \sim N$. Notice that this is significantly shorter than the exponentially large tunneling time. In particular it agrees with the scrambling time of \cite{Sekino:2008he,Shenker:2013pqa} with entropy \eqref{SBoltz} and the mass $M \sim N$. We will discuss time-dependent processes in Section \ref{sec:time} in more detail, where we will also recover the relation $M \sim N$ for our toy-model.
\end{enumerate}

As we discussed the microstates $\Psi^{\pm}_0$ are highly excited. Hence, according to point {\it a} above, perturbation theory is expected to break down whenever `black hole microstate' effects are probed from the point of view of one of the semiclassical vacua. 
A very similar conclusion was reached in \cite{Ghosh:2016fvm,Ghosh:2017pel} on the basis of gravitational (bulk) arguments. However this does not invalidate perturbation theory in general, which remains valid for excitations close to the semiclassical vacuum. In this sense postulate 2 of \cite{Susskind:1993if,Almheiri:2012rt} holds in our model.

\section{Low energy excitations} \label{sec:eff}

In the previous section we identified several aspects of the holographic dictionary that relate our quantum mechanical toy-model to black hole physics in a dual bulk spacetime. In particular we established a notion of asymptotic observers, perturbative vacua, semiclassical states and their Fock spaces as well as dual black hole microstates. We have shown how non-perturbative effects enter in the picture leading to the breakdown of perturbation theory when it comes to the fine-grained features of microstates. We have also shown that the natural (naive) creation and annihilation operators \eqref{defaRaL} do not possess a well-defined decoupling limit $\lambda \rightarrow 0$, and therefore cannot represent creation-annihilation operators associated with the asymptotic regions. As a consequence a firewall \eqref{firewallA} emerged.

In this section we correctly identify perturbative degrees of freedom as perceived by the asymptotic observers. We are able to distinguish perturbative and non-perturbative physics and to define suitable creation and annihilation operators with a well-defined decoupling limit.

\subsection{Extinguishing firewalls} \label{sec:no_firewalls}

In Section \ref{sec:fire} we have shown that any microstate exhibits a firewall as measured by the naive number operator \eqref{defNA}. We have established that the source of the firewall is the fact that the creation and annihilation operators $a_L, a_L^+, a_R, a_R^+$ act on both left and right perturbative states $\varphi_n^L$ and $\varphi_n^R$. A natural resolution would seem to be to define a different set of creation-annihilation operators $\hat{a}_L, \hat{a}_L^+$ and $\hat{a}_R, \hat{a}_R^+$ such that $\hat{a}_R$ and $\hat{a}_R^+$ act only on $\varphi_n^R$, while $\hat{a}_L$ and $\hat{a}_L^+$ act only on $\varphi_n^L$,
\begin{align} \label{hata}
& \hat{a}_R \varphi_n^R \stackrel{?}{=} \sqrt{n} \varphi_{n-1}^R, && \hat{a}_R \varphi_n^L \stackrel{?}{=} 0, \\
& \hat{a}_R^+ \varphi_n^R \stackrel{?}{=} \sqrt{n + 1} \varphi_{n+1}^R, && \hat{a}_R^+ \varphi_n^L \stackrel{?}{=} 0 \label{hataa}
\end{align}
A new number operator $\hat{N}_A$ defined by means of the hatted operators
\begin{equation}
\hat{N}_A = \hat{a}_L^+ \hat{a}_L + \hat{a}_R^+ \hat{a}_R
\end{equation}
would then act on the energy eigenstates $\Psi_0^{\pm}$ according to \eqref{ground_approx} as
\begin{align}
\hat{N}_A | \Psi_0^{\pm} \> & = \frac{1}{\sqrt{2}} ( a_R^+ a_R \varphi^R_0 \pm a_L^+ a_L \varphi^L_0 ) + O(\sqrt{\lambda}) \nn\\
& = 0 + O(\sqrt{\lambda}).
\end{align}
Thus, no firewall!

The only problem with this reasoning is that the operators satisfying \eqref{hata} or \eqref{hataa} cannot exist. Since the full Hilbert space $\mathcal{H}$ is isomorphic to any single Fock space associated to a minimum, the set $\{ \varphi_n^L, \varphi_n^R \}_n$ constitutes an overcomplete basis. Given an action of $a_R$ on all right modes $\varphi_n^R$, its action on left modes $\varphi_n^L$ is fixed by means of \eqref{ambiga}. One could however hope to achieve relations \eqref{hata} and \eqref{hataa} approximately for low energy modes $\varphi_n^L$ and $\varphi_n^R$ with $n \ll N$. In fact, according to \eqref{LR}, the overlap between $\varphi_m^L$ and $\varphi_n^R$ for $m, n \ll N$ is exponentially small, and the subset of modes $\{\varphi_m^L, \varphi_n^R\}_{m, n \ll N}$ constitutes an `almost' orthonormal basis. Hence, we expect that the low energy physics should be well-approximated by the tensor product $\mathcal{F}_L \otimes \mathcal{F}_R$, where the excitations on the left and the right become independent.

We now give a specific proposal for `orthogonalizing' the overcomplete basis $\{ \varphi_n^L, \varphi_n^R \}_n$ in such a way that the hatted operators \eqref{hata} or \eqref{hataa} can be successfully defined. To be more precise, we will split the total Hilbert space $\mathcal{H}$ into two orthogonal components, $\mathcal{H} = \mathcal{H}_L \oplus \mathcal{H}_R$. The left and right hatted annihilation operators $\hat{a}_L, \hat{a}_R$ can then be defined as projections of the unhatted operators $a_L, a_R$ onto the appropriate subspaces. The `orthogonalization' is highly non-unique, but all ambiguities are non-perturbative and hence inaccessible in perturbation theory around any minimum. In the context of black hole physics the problem of overcompleteness of the basis has been pointed out in \cite{Jafferis:2017tiu}.

To resolve the overcompleteness of the set $\{ \varphi^L_n, \varphi^R_n \}_n$ consider symmetric and antisymmetric combinations of all energy eigenstates,
\begin{equation} \label{defLR}
\Psi_n^L = \frac{1}{\sqrt{2}} (\Psi_n^+ - \Psi_n^-), \qquad \Psi_n^R = \frac{1}{\sqrt{2}} (\Psi_n^+ + \Psi_n^-)
\end{equation}
and consider two Hilbert subspaces of $\mathcal{H}$, spanned by $\Psi^L_n$ and $\Psi^R_n$ respectively,
\begin{equation} \label{splitdef}
\mathcal{H}_L = \Span \{ \Psi_n^L \}_n, \qquad\qquad \mathcal{H}_R = \Span \{ \Psi_n^R \}_n.
\end{equation}
From \eqref{defLR} we have that $\< \Psi_m^L | \Psi_n^R \> = 0$ for all $n,m$ and hence $\mathcal{H}_L$ and $\mathcal{H}_R$ are orthogonal to each other. The full Hilbert space splits into a \emph{direct sum},
\begin{equation} \label{HRplusHL}
\mathcal{H} = \mathcal{H}_L \oplus \mathcal{H}_R, \qquad \mathcal{H}_L \perp \mathcal{H}_R, \qquad \Theta \mathcal{H}_L = \mathcal{H}_R, \qquad \Theta \mathcal{H}_R = \mathcal{H}_L.
\end{equation}
In other words $\Theta$ is a polarization of $\mathcal{H}$. We refer to $\mathcal{H}_L$ and $\mathcal{H}_R$ as \emph{left} and \emph{right perturbative Hilbert spaces} respectively.\footnote{We will discuss the precise meaning of the word \emph{perturbative} in the next section. For now notice that all low occupancy states $\Psi_n^R$ for $n \ll N$ are localized around the right minimum only, whereas $\Psi_n^L$ are localized around the left vacuum.} Furthermore by $P_L$ and $P_R$ we denote canonical orthogonal projections of $\mathcal{H}$ onto $\mathcal{H}_L$ and $\mathcal{H}_R$ respectively.

\begin{figure}[ht]
	\includegraphics[width=0.45\textwidth]{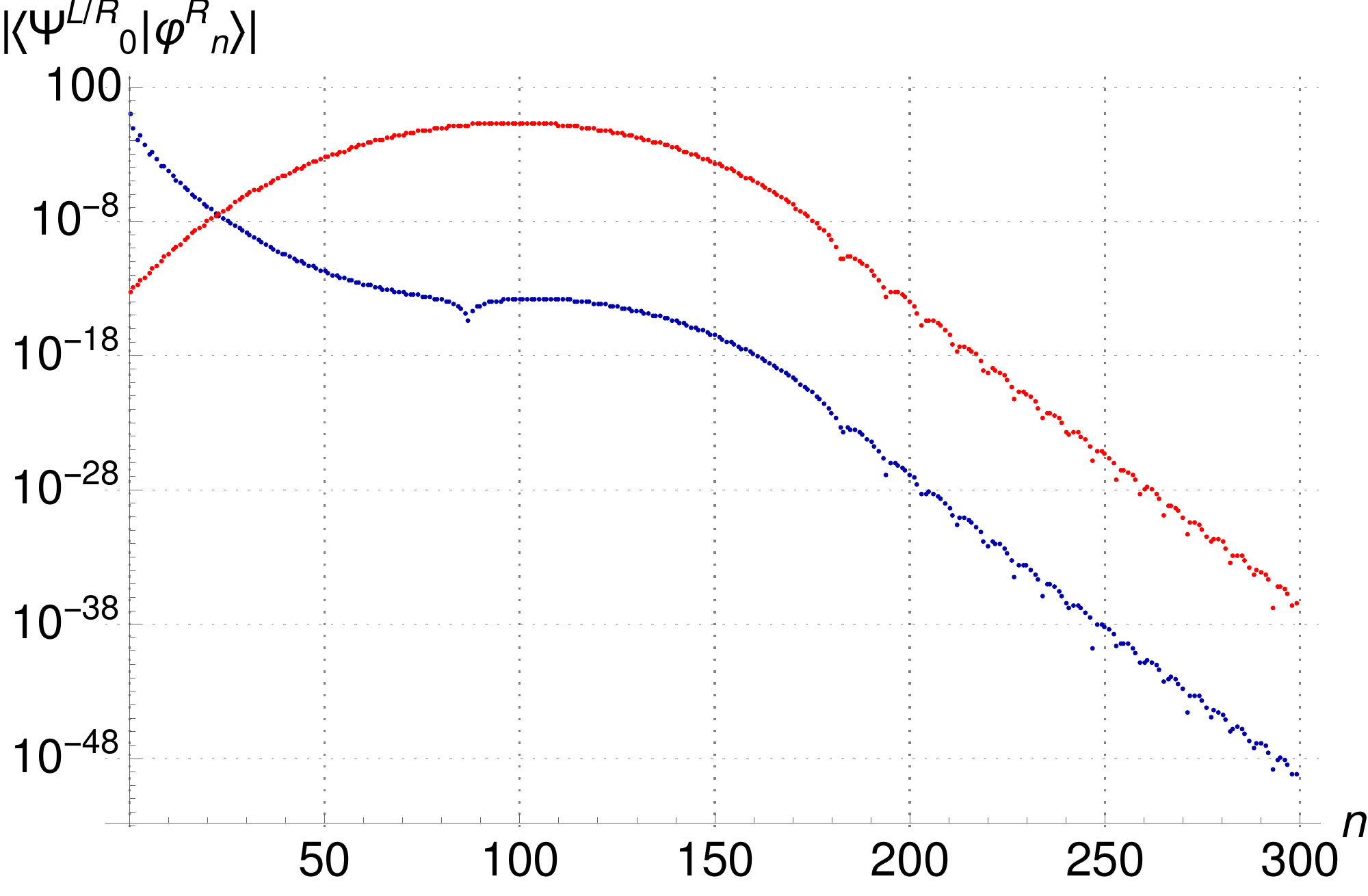}
	\qquad
	\includegraphics[width=0.45\textwidth]{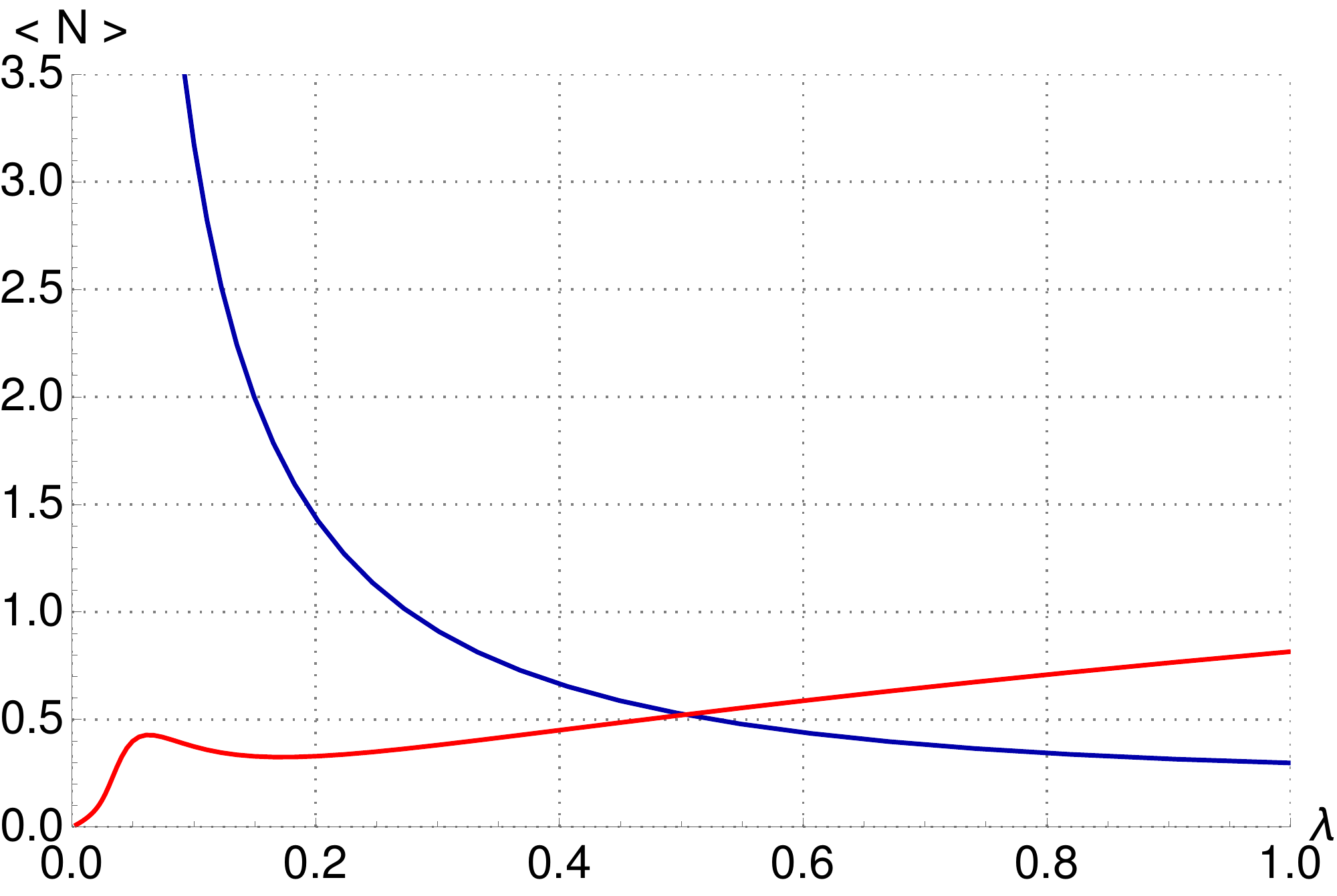}
	\centering
	\caption{On the left: a comparison between matrix elements $\< \Psi^R_0 | \varphi^R_n \>$ (blue) and $\< \Psi^L_0 | \varphi^R_n \>$ (red). On the right: The expectation value $\< \Psi_0^+ | N_A | \Psi_0^+ \>$ for number operators in the true vacuum state. The blue line shows the expectation value for the operator $N_A = a_L^+ a_L + a_R^+ a_R$. This operator leads to a firewall at $\lambda \rightarrow 0$. The red line shows the expectation value of the operator $\hat{N}_A = P_L a_L^+ a_L P_L + P_R a_R^+ a_R P_R$. The result approaches zero at $\lambda \rightarrow 0$ in a characteristic non-perturbative fashion and exhibits a small non-vanishing value for $\lambda > 0$ due to the interactions.\label{fig:NA}}
\end{figure}

Define $\hat{a}_R$ as $a_R$ restricted to $\mathcal{H}_R$ and similarly define $\hat{a}_L$ as $a_L$ restricted to $\mathcal{H}_L$,
\begin{equation} \label{defhata}
\hat{a}_L = P_L a_L P_L, \qquad\qquad \hat{a}_R = P_R a_R P_R.
\end{equation}
Define $\hat{a}_L^+$ and $\hat{a}_R^+$ as their Hermitian conjugates. We repeat this prescription to define a number operator $\hat{N}_A$, taking into account the fact that it should count excitations both in $\mathcal{H}_L$ and $\mathcal{H}_R$,
\begin{equation} \label{defhatNA}
\hat{N}_A = P_L N_L P_L \oplus P_R N_R P_R = P_L a_L^+ a_L P_L + P_R a_R^+ a_R P_R.
\end{equation}
This operator is now defined globally on the entire $\mathcal{H}$. It counts excitations on top of microstates from the point of view of both perturbative vacua. In particular, we argue that its expectation value in a state $\psi = \alpha_L \varphi_m^L + \alpha_R \varphi_n^R$ representing approximately $m$ particles on the left and $n$ particles on the right, is given by
\begin{equation} \label{expNA}
\< \psi | \hat{N}_A | \psi \> = | \alpha_L |^2 m + | \alpha_R |^2 n + O(\sqrt{\lambda}).
\end{equation} 
Hence $\hat{N}_A$ is a natural candidate for a global, firewall-free number operator\footnote{There are two slightly different choices here. One can define $\hat{a}_L$ and $\hat{a}_R$ with their images unrestricted or restricted to the corresponding subspaces $\mathcal{H}_L$ and $\mathcal{H}_R$. The latter definition is $\hat{a}_R = P_R a_R P_R$ as we have defined, the former means that $\hat{a}_R = a_R P_R$, and similarly for $\hat{a}_L$. The difference, $P_L a_R P_R$, is however non-perturbatively small, as we will argue in Section \ref{sec:pert_opers}, and hence invisible in perturbation theory. For the same reason one can consider another number operator built up with hatted creation-annihilation operators
\begin{equation}
\hat{N}'_A = \hat{a}_L^+ \hat{a}_L \oplus \hat{a}_R^+ \hat{a}_R = P_L a_L^+ P_L a_L P_L + P_R a_R^+ P_R a_R P_R.
\end{equation}
While formally different, the fact that $P_L a_R P_R = o(\lambda^\infty)$ as well as $P_R a_R P_L = o(\lambda^\infty)$ implies that the two number operators can differ by non-perturbative terms only, $\hat{N}_A = \hat{N}'_A + o(\lambda^\infty)$. Hence, in perturbation theory the two operators are indistinguishable. We will stick to the definition \eqref{defhatNA}, which is slightly more convenient for numerical calculations.}. Specifically in any perturbative vacuum $\mu \in \M$, $\< \mu | \hat{N}_A | \mu \> = O(\sqrt{\lambda})$\footnote{In \cite{Papadodimas:2013jku,Papadodimas:2013wnh,Papadodimas:2015jra,Papadodimas:2015xma} the Authors insist on a number operator $\hat{N}_A$ which satisfies $\hat{N}_A | \Psi_0^{\pm} \> = 0$ exactly. From the point of view of the QFT this seems unnecessarily strong, since interactions do create particles. Nevertheless, one can define the appropriate number operator $\hat{N}_A = \hat{A}_L^+ \hat{A}_L \oplus \hat{A}_R^+ \hat{A}_R$, where $\hat{A}_L$ is defined on $\mathcal{H}_L$ as $\hat{A}_L \Psi_n^L = \sqrt{n} \Psi_{n-1}^L$, and $\hat{A}_R$ is defined on $\mathcal{H}_R$ as $\hat{A}_R \Psi_n^R = \sqrt{n} \Psi_{n-1}^R$.}. The numerical plot of the expectation value of the number operator in the ground state $\Psi_0^+$ as a function of $\lambda$ is given in Figure \ref{fig:NA}.

While mathematically we defined hatted operators in \eqref{defhata} using projectors, it may be more physically accurate \emph{not} to specify the action of $\hat{a}_R$ on $\mathcal{H}_L$ nor $\hat{a}_L$ on $\mathcal{H}_R$. Similarly, we could define left and right number operators $N_L$ and $N_R$ on $\H_L$ and $\H_R$ only, as their physical meaning is associated with perturbative physics percieved by the corresponding observers. We can either refuse to act with perturbative operators on non-perturbative states or accept the fact that natural perturbative observables from the point of view of a given observer become non-perturbative from the point of view of the another observer. The total number operator \eqref{defNA}, however, remains globally defined.

In order to conclude the proof of \eqref{expNA} we have to study a relation between states $\varphi_n^R$ and $\Psi_n^R$, or equivalently between $\mathcal{H}_R$ and $\mathcal{F}_R$. The perturbation theory \cite{ZinnJustin:2004ib} implies that $\Psi_n^R = \varphi_n^R + O(\sqrt{\lambda})$, and hence 
\begin{equation} \label{PRphi}
P_R \varphi_n^R = P_R \left[ \Psi_n^R + O(\sqrt{\lambda})  \right] = \Psi_n^R + O(\sqrt{\lambda}) = \varphi_n^R + O(\sqrt{\lambda}).
\end{equation}
This implies that 
\begin{equation}	
\hat{a}_R^+ \hat{a}_R \varphi_n^R = P_R a_R^+ a_R P_R \varphi_n^R = n \varphi_n^R + O(\sqrt{\lambda})
\end{equation}
and \eqref{expNA} follows.

Equation \eqref{PRphi} suggests that the familiar semiclassical state $\varphi_n^R$ is not an element of the right perturbative space $\H_R$. Indeed, if some $\varphi_n^R$ was an element of $\mathcal{H}_R$, then $\Theta \varphi_n^R = \varphi_n^L$ would belong to $\mathcal{H}_L$. But the two states $\varphi_n^R$ and $\varphi_n^L$ are not orthogonal as their scalar product \eqref{LR} is non-vanishing. Hence $\varphi_n^R \notin \H_R$.

\begin{figure}[ht]
	\includegraphics[width=0.55\textwidth]{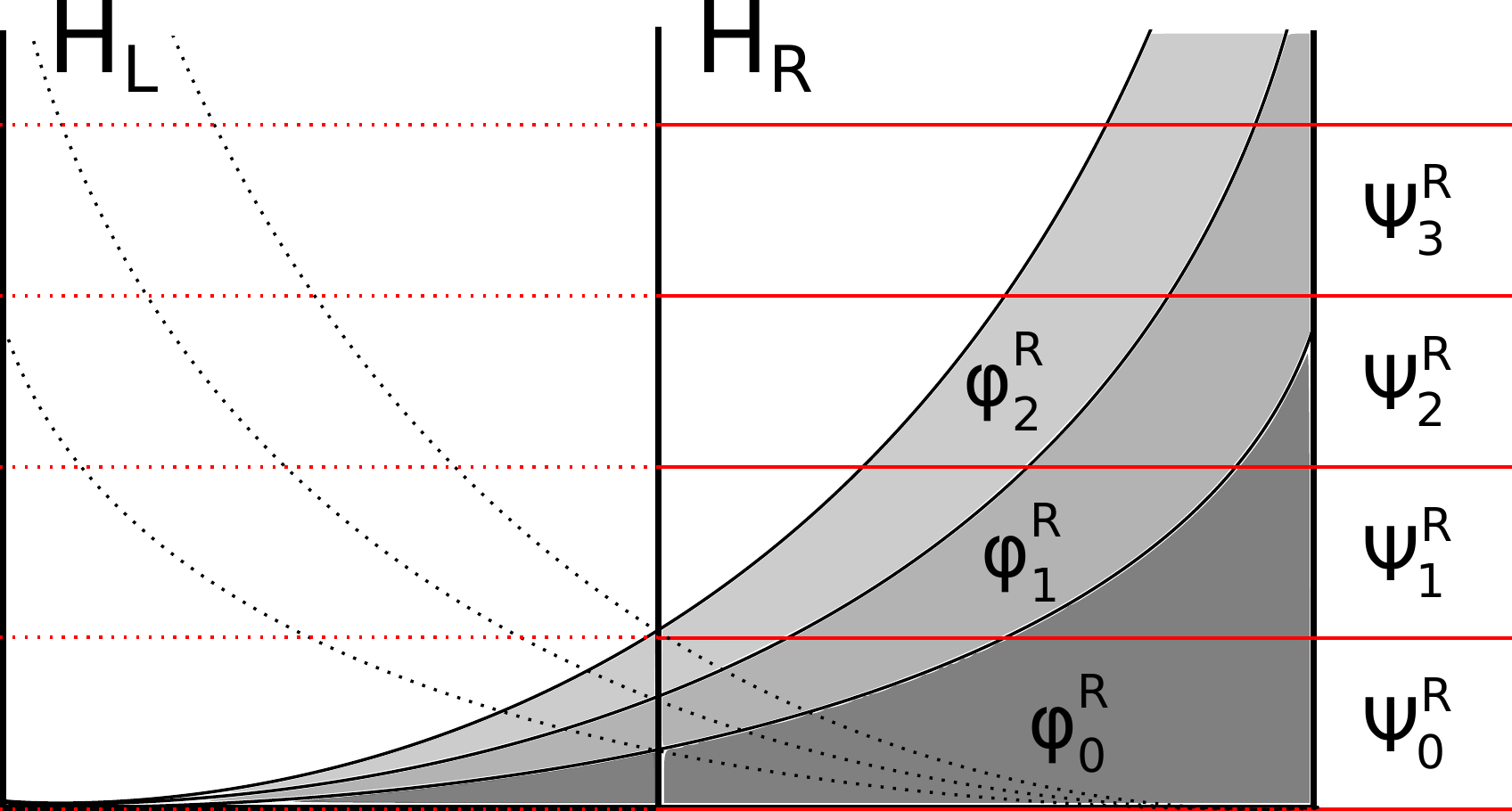}
	\centering
	\caption{A (very) schematic structure of the Hilbert space. The full Hilbert space $\H$ is split into a simple sum $\H \cong \H_L \oplus \H_R$. 1-dimensional subspaces spanned by $\Psi_n^R$ do not align exactly with semiclassical states $\varphi_n^R$. One the other hand, the states $\varphi_n^R$ are not entirely contained in $\H_R$, as they have a small non-perturbative overlap with states in $\H_L$. The structure of $\H_L$ and coresponding states $\Psi_n^L$ and $\varphi_n^L$ is symmetric and denoted by dotted lines.\label{fig:H}}
\end{figure}

While no $\varphi_n^R$ belongs to $\mathcal{H}_R$, for each $n$ a difference between the state $\varphi_n^R$ and its projection $P_R \varphi_n^R$ on $\mathcal{H}_R$ is non-perturbatively small. Equivalently, by following Example 6 of Section XII.3 of \cite{Reed/Simon} one can argue that $\| P_L \varphi_n^R \| = o(\lambda^\infty)$ and $\| P_R \varphi_n^L \| = o(\lambda^\infty)$ for any $n$. Hence, while $\varphi_n^R$ is not an element of the perturbative Hilbert space $\mathcal{H}_R$, its projection $P_R \varphi_n^R \in \mathcal{H}_R$ is non-perturbatively close to $\varphi_n^R$. In perturbation theory, one cannot distinguish the two states. A schematic relation between various states is presented in Figure \ref{fig:H}.

By defining left and right perturbative spaces $\H_L$ and $\H_R$ we effectively resolved the overcompleteness of the set $\{ \varphi_n^L, \varphi_n^R\}_n$. While stricktly speaking no semiclassical state $\varphi_n^R$ belongs to $\H_R$, there exist states $P_R \varphi_n^R$ non-perturbatively close to $\varphi_n^R$ lying in $\H_R$. We have found `approximate isomorphisms'
\begin{equation}
\mathcal{F}_L \cong \mathcal{H}_L + o(\lambda^\infty), \qquad\qquad \mathcal{F}_R \cong \mathcal{H}_R + o(\lambda^\infty)
\end{equation}
up to non-perturbative terms.

\subsection{Perturbative states}

A notion of a perturbative state is crucial for the discussion of the information paradox. An intuitive idea is that its support is concentrated around a single minimum. As an example consider basis states $\varphi_n^R$. Are all these states perturbative with respect to the right minimum or only those with $n \ll N^2$, so that their support is concentrated around the minimum? In the language of \cite{Papadodimas:2013jku,Papadodimas:2013wnh,Papadodimas:2015jra,Papadodimas:2015xma} one would call a state $\varphi_n^R$ perturbative only if $n \ll N^2$ is small. In this paper, however, we will introduce a weaker definition that allows for a wider range of perturbative states.

Our definition of a perturbative state deals with the behavior of the state as the coupling $\lambda$ approaches zero. Therefore, instead of a single state $\psi$, we consider a family $\{ \psi_\lambda\}_{\lambda > 0}$ labeled by the coupling. Essentially all states we consider depend on $\lambda$ in some implicit way. For example the perturbative states $\varphi_n^R$ and $\varphi_n^L$ are defined by (\ref{isoR}-\ref{isoL}) and hence they implicitly depend on $\lambda$. For that reason we will refer to the elements of the family $\{ \psi_\lambda\}_{\lambda > 0}$ as a state $\psi_\lambda \in \mathcal{H}$.

We say that the state $\psi_\lambda$ is \emph{perturbative with respect to the right minimum} if $F_R^{-1} \psi_\lambda$ converges in norm in $\mathcal{F}_R$ when $\lambda \rightarrow 0^+$. Here $F_R^{-1} : \mathcal{H} \stackrel{\cong}{\rightarrow} \mathcal{F}_R$ is the inverse of the isomorphism \eqref{isoR} between $\mathcal{H}$ and $\mathcal{F}_R$. Analogously we define states perturbative with respect to the left minimum.

First notice that all states $\varphi_n^R$ are mapped to $|n\>_R \in \mathcal{F}_R$, which are $\lambda$-independent in $\mathcal{F}_R$. Hence all $\varphi_n^R$ are trivially perturbative with respect to the right minimum. Consider now a state such as a ground state $\Psi_0^{+}$, which possesses two bumps around both left and right minimum. By going to $\mathcal{F}_R$ we may simply position ourselves at $x = x_R$ and send $\lambda$ to zero. The right portion of the wave function then concentrates around the right minimum and approaches $\varphi_0^R$. As the left minimum moves away to $-\infty$, the left portion of the wave function is lost in the decoupling limit $\lambda = 0$. Indeed, neither $\Psi_0^+$ nor $\Psi_0^-$ are perturbtive with respect to any minimum.

On the other hand, all states $\Psi_n^R$ and $\Psi_n^L$ as defined in \eqref{defLR} are perturbative with respect to right and left minima respectively and we have, \cite{Reed/Simon}, $\lim_{\lambda \rightarrow 0^{+}} F_R^{-1} \Psi_n^R = |n\>_R$ and $\lim_{\lambda \rightarrow 0^{+}} F_L^{-1} \Psi_n^L = | n \>_ L$ with the convergence in norm\footnote{Even if a state $\psi_\lambda$ is non-perturbative with respect to, say, right minimum, one can still define its decoupling limit. We will say that a state $| \psi_0 \> \in \mathcal{F}_R$ is a \emph{decoupling limit with respect to the right minimum} of $\psi_\lambda \in \mathcal{H}$ if $| \psi_0 \> = \wlim_{\lambda \rightarrow 0^{+}}F_R^{-1} \psi_\lambda$, where $\wlim$ denotes the weak limit. In this sense we have the following decoupling limits,
\begin{equation} \label{wlims}
\wlim_{\lambda \rightarrow 0^{+}} F_R^{-1} \Psi_n^{R} = | n \>_R, \qquad \wlim_{\lambda \rightarrow 0^{+}} F_R^{-1} \Psi_n^{L} = 0, \qquad \wlim_{\lambda \rightarrow 0^{+}} F_R^{-1} \Psi_n^{\pm} = \frac{1}{\sqrt{2}} | n \>_R.
\end{equation}
Numerical values of the matrix elements $\< \Psi_0^L | \varphi_n^R \>$ and $\< \Psi_0^R | \varphi_n^R \>$ as functions of $n$ can be seen in Figure \ref{fig:NA}.}. Hence every element of $\mathcal{H}_L$ is perturbative with respect to the left minimum and every element of $\mathcal{H}_R$ is perturbative with respect to the right minimum. This justifies their names as left and right \emph{perturbative} Hilbert spaces $\mathcal{H}_L$ and $\mathcal{H}_R$. We can also sharpen our definition of a \emph{right (left) observer} by declaring $\H_R$ ($\H_L$) as the Hilbert space available to the observer.

Notice that our definition of a perturbative state depends only on what happens with the state when $\lambda$ approaches to zero. For example, for a fixed value of $\lambda > 0$ the support of $\varphi_n^R$ is concentrated around the right minimum only for $n \ll \lambda^{-1} = N^2$. Nevertheless, according to our definition, all $\varphi_n^R$ are perturbative. As $\lambda$ approaches zero, each $\varphi_n^R$ concentrates around the right minimum, since for each $n$ there exists $\lambda$ so small that $n \ll \lambda^{-1}$.

In the language of \cite{Papadodimas:2013jku,Papadodimas:2013wnh,Papadodimas:2015jra,Papadodimas:2015xma} the space of perturbative states was finitely dimensional, as the condition $n \ll N^2$ on the occupancy numbers was imposed. In particular such a space was not generated by a genuine algebra acting on a cyclic vector: the issue that led the Authors of \cite{Papadodimas:2013wnh} to use a concept of `algebras with a cut-off'. In our approach such issues are completely avoided. The `small algebra' $\mathcal{A}_R$ associated with the right minimum is generated by $\hat{a}_R$ and $\hat{a}_R^+$ and the perturbative Hilbert space is then $\mathcal{H}_R = \mathcal{A}_R | \Psi_0^R \>$. No cut-offs of any sort are required and all states in $\mathcal{H}_R$ are perturbative with respect to the right minimum.

\subsection{Perturbative operators} \label{sec:pert_opers}

Having defined perturbative states, one can also define perturbative operators. These should be operators which: (i) preserve the decoupling of the potential wells up to non-perturbative effects; and (ii) have a well-defined decoupling limit. If we write an operator $A$ in the matrix form
\begin{equation}
A = \mat{A_{LL}}{A_{LR}}{A_{RL}}{A_{RR}} \ : \  \mathcal{H}_L \oplus \mathcal{H}_R \rightarrow \mathcal{H}_L \oplus \mathcal{H}_R
\end{equation}
with $A_{IJ}$ mapping $\mathcal{H}_I$ into $\mathcal{H}_J$, $I,J \in \{L,R\}$, then: (i) $A_{LR}$ and $A_{RL}$ must be non-perturbatively small, \textit{i.e.}, 
\begin{equation} \label{np-small}
A_{LR} = o(\lambda^\infty) \text{ and } A_{RL} = o(\lambda^\infty);
\end{equation}
and (ii) the decoupling limits $\lambda \rightarrow 0$ of $A_{LL}$ in $\H_L$ and $A_{RR}$ in $\H_R$ must exist. We will call such an operator $A$ as \emph{perturbative}.

\begin{figure}[ht]
	\includegraphics[width=0.45\textwidth]{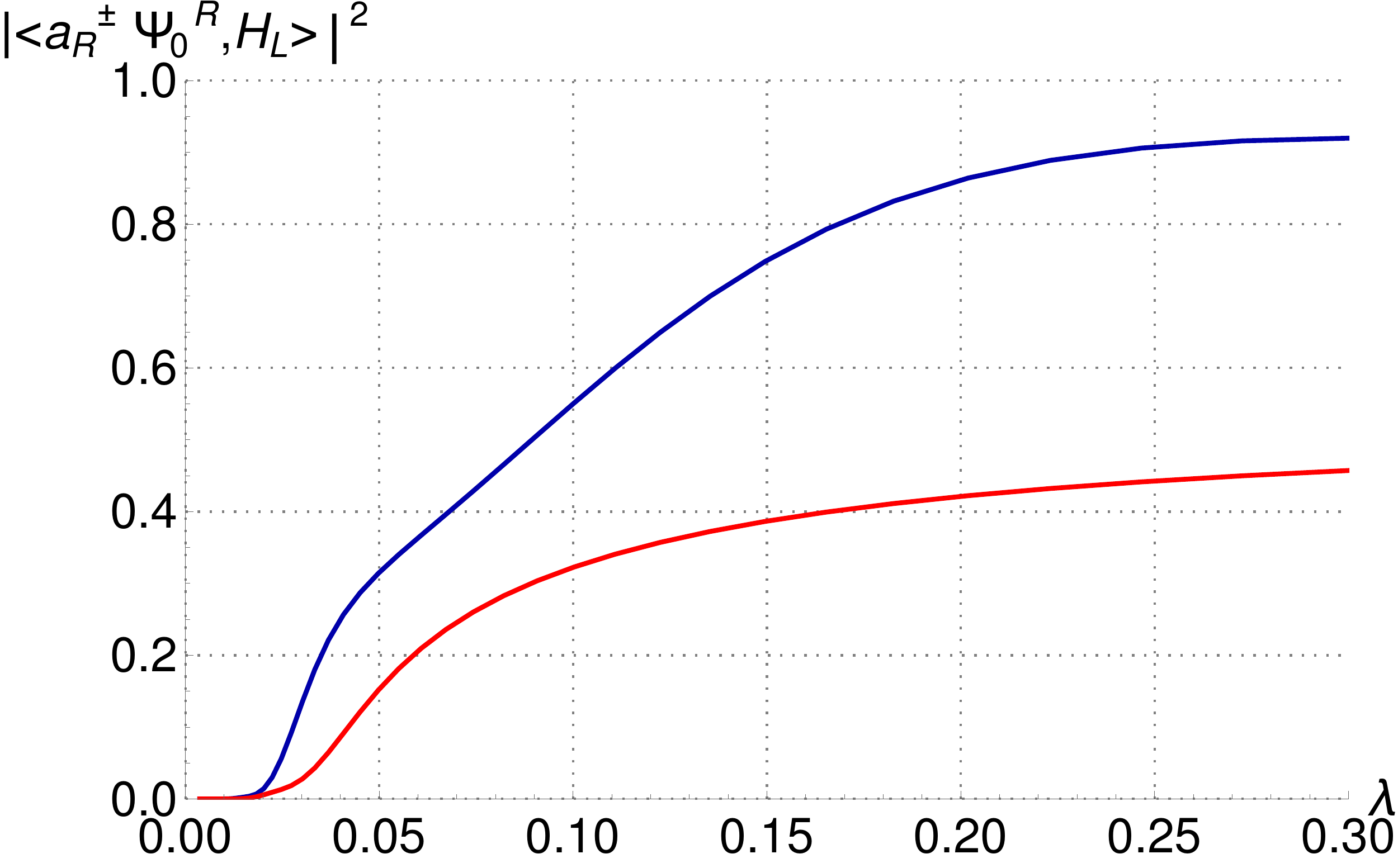}
	\qquad
	\includegraphics[width=0.45\textwidth]{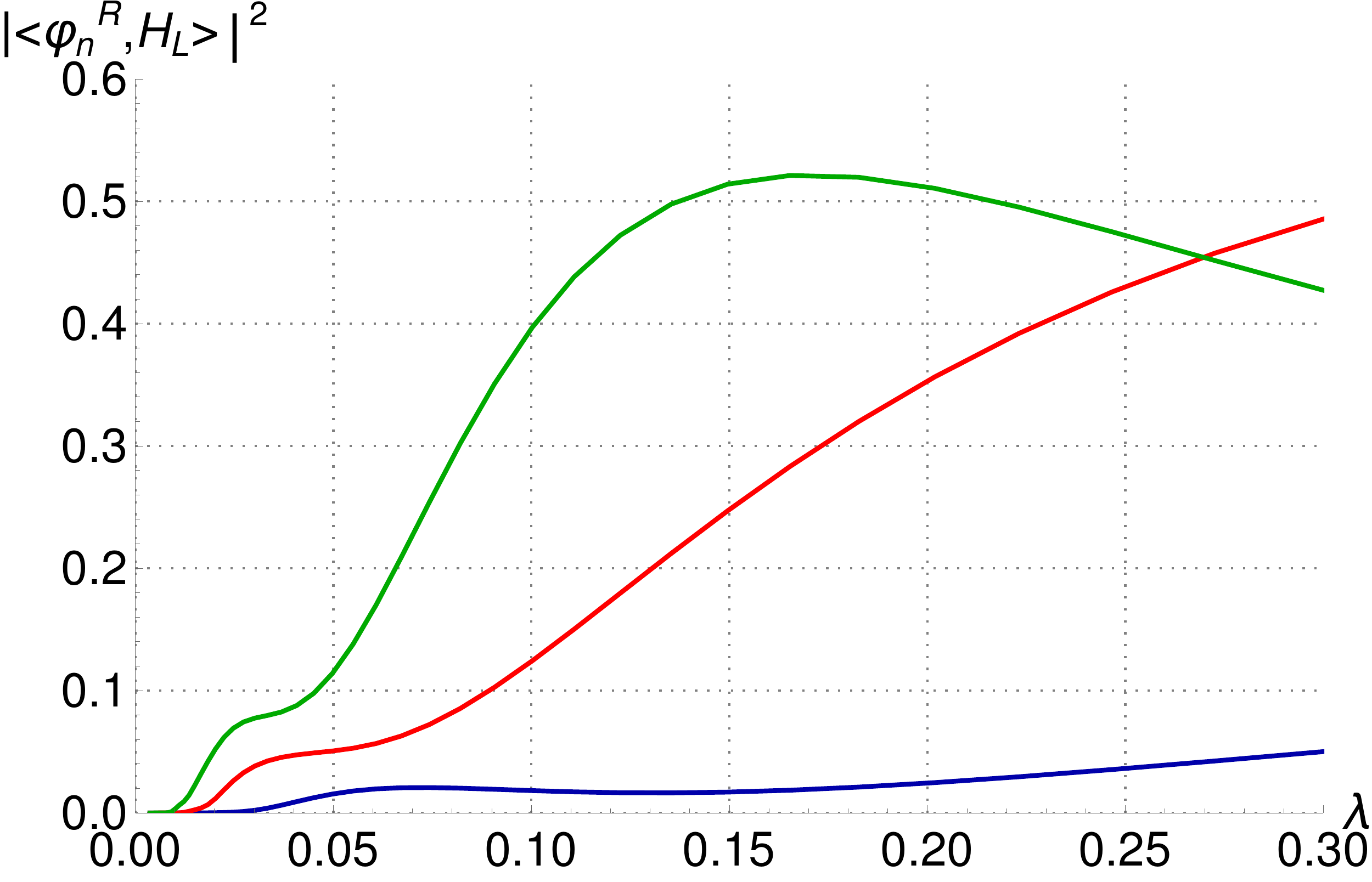}
	\centering
	\caption{Left: A measure of an overlap of the states $a_R \Psi_0^R$ (blue) and $a_R^+ \Psi_0^R$ (red) with the left Hilbert space $\mathcal{H}_L$ as function of the coupling $\lambda$. Right: A measure of an overlap of the perturbative states $\varphi_n^R$ with $\mathcal{H}_L$ for $n = 0$ (blue), $n = 1$ (red) and $n = 2$ (green). The overlap of a state $\psi \in \mathcal{H}$ with $\mathcal{H}_L$ is defined as $| \< \psi, \mathcal{H}_L \> |^2 = \| P_L \psi \|^2$. For $\lambda \rightarrow 0$ the overlaps approach zero in a characteristic non-perturbative fashion.\label{fig:overlap}}
\end{figure}

All hatted operators such as creation-annihilation operators $\hat{a}_L, \hat{a}_L^+, \hat{a}_R, \hat{a}_R^+$ or the number operator $\hat{N}_A$ \eqref{defhatNA} are by construction perturbative with vanishing off-diagonal terms. Their unhatted versions usually fail to satisfy condition (ii), as indicated by \eqref{aLofaR} or by the firewall in \eqref{firewallA}. On the other hand, one expects that condition (i) holds, since $P_R a_R P_R \varphi_n^R = \sqrt{n} \varphi_{n-1}^R + o(\lambda^\infty)$ or equivalently $P_L a_R P_R = o(\lambda^\infty)$. Figure \ref{fig:overlap} shows the norm of the state $P_L a_R \Psi_0^R$ with a characteristic exponential fall-off around $\lambda = 0$. In this sense hatted and unhatted creation-annihilation operators agree on their corresponding perturbative Hilbert spaces. Schematically, $a_R = \hat{a}_R + o(\lambda^\infty)$ on $\H_R$ and $a_L = \hat{a}_L + o(\lambda^\infty)$ on $\H_L$ together with their conjugates. Only on $\H_L$, the complement of $\H_R$, the operators $a_R$ and $\hat{a}_R$ differ significantly.

The unhatted creation-annihilation operators satisfy commutation relation \eqref{commaLaR}. However, in black hole physics, locality at the level of the effective bulk theory requires that left and right operators commute. This is indeed the case for the hatted operators, where we find
\begin{equation}
[\hat{a}_L, \hat{a}_L^+]  = [\hat{a}_R, \hat{a}_R^+] = 1 + o(\lambda^\infty), \qquad\qquad [\hat{a}_L, \hat{a}_R] = [\hat{a}_L, \hat{a}_R^+] = 0.
\end{equation}
The canonical commutation relations on $\H_L$ and $\H_R$ are altered by a non-perturbative factor, while left and right operators commute\footnote{Slightly different definitions of creation-annihilation operators as indicated in Section \ref{sec:no_firewalls} may result in non-perturbative corrections to the locality condition $[\hat{a}_L, \hat{a}_R] = [\hat{a}_L, \hat{a}_R^+] = o(\lambda^\infty)$. In perturbation theory operators are required to be defined uniquely only up to non-perturbative terms.}. Hence locality is maintained up to non-perturbative effects as predicted by a number of papers, \textit{e.g.}, \cite{Papadodimas:2012aq,Papadodimas:2013jku,Papadodimas:2013wnh,Kabat:2014kfa,Raju:2016vsu}. This, however, comes at the cost of the action of, say, right perturbative operators on the left perturbative states to be either unrelated to the action of unhatted creation-annihilation operators (extremely `non-local' as in \cite{Hamilton:2006az,Hamilton:2006fh}), or simply ill-defined \cite{Kabat:2018pbj}.

In this way we have resolved the problem of `fitting' the left creation-annihilation operators into the full Hilbert space. By identifying the right perturbative Hilbert space as $\H_R$ and the right perturbative operators as $\hat{a}_R$ and $\hat{a}_R^+$ we found the proper set of degrees of freedom and operators associated with a perturbative (asymptotic) observer in the right vacuum. In particular every state in $\H_R$ is perturbative with respect to the right minimum, but non-perturbative from the point of view of the left observer. 

In the context of black holes a number of papers \cite{sanchez,Chamblin:2006xd,Hooft:2016itl,Hooft:2016vug} have suggested to remove `half' of the states by considering an antipodal identification of the spacetime. The total Hilbert space then consists of only parity even or odd wave functions. Such an approach would remove a degeneracy within perturbation theory leading to each black hole having a single microstate. In our model we do not find a support for such an identification, but we also do not find any obstacles in its implementation. From the point of view of a single observer, the two situations are indistinguishable as long as perturbative processes of small energies are considered. Only at higher energies one would be able to notice `missing' states in the total Hilbert space. 

Finally let us point out that the original Hamiltonian $H$ is a perturbative operator. The off-diagonal elements $P_R H P_L = o(\lambda^\infty)$ and $P_L H P_R = o(\lambda^\infty)$ are related to the tunneling rate and can be calculated by standard methods within the WKB approximation. With our definition of non-perturbative effects this statement remains true for all energies, even if actual matrix elements become numerically large. We return to time-dependent processes in Section \ref{sec:time}.

\section{Low energy effective theory: state-dependence} \label{sec:4}

In the previous section we have constructed states and operators in the full theory that are the natural perturbative objects from the standpoint of semiclassical (or asymptotic) observers. In this section we relate these operators (observables) in the full theory to operators (observables) in the effective low energy theory. In doing so we will find that the resulting operators in the effective theory are state-dependent.

\subsection{Perturbative observables} \label{sec:state_dep}

In semiclassical black hole physics one usually takes the Hilbert space to be a tensor product $\mathcal{F}_L \otimes \mathcal{F}_R$ of the degrees of left/right freedom on both sides of the horizon. The creation-annihilation operators $b_L, b_L^+$ in $\mathcal{F}_L$ and $b_R, b_R^+$ on $\H_R$ give rise to creation-annihilation operators in the tensor product: $\id \otimes b_R, \id \otimes b_R^+$, \textit{etc}. The bulk of the Hermitian operators are of the form $A_{LL} \otimes \id + \id \otimes A_{RR}$. The black hole microstate is not a vacuum state, but rather an excited state.

Due to the large potential barrier, low occupancy states in our toy-model exhibit an approximate tensor product structure $\H_L \otimes \H_R$. Indeed, states $\Psi_{mn} = \Psi_m^L + \Psi_n^R$ for $m,n \ll N$ resemble elements of the tensor product, as they are combinations of two, nearly decoupled states centered around, respectively, the left and right perturbative vacuum, with a number of excitations given by $\< \Psi_{mn} | \hat{N}_A | \Psi_{mn} \> = m + n + O(\sqrt{\lambda})$ as expected.

Beyond this, however, the two systems differ. A linear structure of the direct sum $\H_L \oplus \H_R$ and the tensor product $\H_L \otimes \H_R$ is vastly different and we cannot hope to reproduce all perturbative operators on the tensor product in the full theory. In particular the action of linear operators in the tensor product is different and there are significantly `more' linear operators on the tensor product.

Consider now an operator $B = A_{LL} \otimes \id + \id \otimes A_{RR}$ on $\H_L \otimes \H_R$ and assume $\chi_m^L$ and $\chi_n^R$ form a set of normalized eigenfunctions of $A_{LL}$ and $A_{RR}$ respectively, with eigenvalues $\lambda^L_m$ and $\lambda^R_n$. This means that
\begin{equation} \label{Aeigen}
B (\chi_m^L \otimes \chi_n^R) = (\lambda^L_m + \lambda^R_n) (\chi_m^L \otimes \chi_n^R).
\end{equation}
We can use the identification of the direct sum with the direct product $\H_L \oplus \H_R \cong \H_L \times \H_R$ to define an operator $A$ on $\H_L \oplus \H_R$ as
\begin{equation} \label{Bdef}
A(\chi_m^L, \chi_n^R) = (\lambda^L_m + \lambda^R_n) (\chi_m^L, \chi_n^R).
\end{equation}
The operator $A$ acts on a state $\chi_m^L + \chi_n^R$ in the same way as $B$ acts on $\chi_m^L \otimes \chi_n^R$. Furthermore the expectation values match,
\begin{equation}
\< \chi_m^L \otimes \chi_n^R | B | \chi_m^L \otimes \chi_n^R \> = \lambda_m^L + \lambda_n^R = \< (\chi_m^L, \chi_n^R) | A | (\chi_m^L, \chi_n^R) \>.
\end{equation}
In this sense $A$ realizes the same relations on $\H_L \oplus \H_R$ as $B$ on $\H_L \otimes \H_R$. 

But the operator $A$ is not linear. It is not even very clear how to extend it to the entire Hilbert space $\H_L \oplus \H_R$. For instance, we may consider projections $A_L$ and $A_R$ of $A$ on $\H_L$ and $\H_R$, so that $A = A_L \oplus A_R$. Operators $A_L$ and $A_R$ can be extended separately to bilinear operators by linearity in each argument. For example, we find that for any $\psi_R \in \H_R$
\begin{equation} \label{BR}
A_R(\chi_m^L, \psi_R) = P_R A(\chi_m^L, \psi_R) = (\lambda^L_m \id + A_{RR}) \psi_R.
\end{equation}
This is clearly a linear operator with respect to the right portion of the state, $\psi_R \in \H_R$. The appearance of $\lambda_m^L$. however, amounts to a form of `state-dependence': The value of the right operator $A_R$ depends on the `hidden' left portion of the state. In this state-dependent sense we can think about $A_R$ as an operator on $\H_R$ only, and  we can write $A_R^{\psi_L}(\psi_R) = A_R(\psi_L, \psi_R)$. That is, we can regard $A_R^{\psi_L}$ as an observable of the right observer. Its value on $\psi_R$, however, depends on the `black hole microstate' $\psi_L$, \textit{i.e.}, on the shape of the wave function on the other side of the potential barrier.

In Section \ref{sec:global_to_eff} we relax our assumption that the action of the operator $A$ takes the form \eqref{Bdef}. Starting from the weaker requirement that expectation values of operators in the full theory agree with those of the corresponding operators in the effective theory, up to non-perturbative terms, we will show there that the corresponding global operators again cannot be defined in a `state-independent' manner. However, the `state-dependence' then no longer pertains to the entire portion of the `hidden' wave function $\psi_L$, but instead is restricted to a choice of perturbative vacuum state $\mu \in \M$. More generally, regardless of the details, we find one cannot realize perturbative observables in the full theory as linear operators. Mathematically this is a consequence of the incompatibility of the linear structure of the direct sum and that of the tensor product. Physically, this means that perturbative interpretations of various operators depend on microstates of the system, as argued in \cite{Papadodimas:2013jku,Papadodimas:2013wnh}.

\subsection{The tensor product: loss of information} \label{sec:tensor}

In the previous section we have argued that it is impossible to represent all perturbative observables on the tensor product $\H_L \otimes \H_R$ by linear operators in the full theory $\H_L \oplus \H_R$. On the other hand, the low energy physics in our model does effectively take place on the tensor product. One therefore expects that an effective theory based on the tensor product should capture the low energy physics, up to non-perturbative effects.

Let $s : \H_L \times \H_R \rightarrow \H_L \otimes \H_R$ denote the canonical bilinear map $s(\chi_L, \chi_R) = \chi_L \otimes \chi_R$. Since $\H_L \oplus \H_R \cong \H_L \times \H_R$ we can assign to any state $\chi = \chi_L + \chi_R$, with $\chi_L \in \H_L$ and $\chi_R \in \H_R$ an effective state $\chi^{\dbl}$ as
\begin{equation} \label{dbl}
\chi = \chi_L + \chi_R \ \longmapsto \ \chi^{\dbl} = \mathcal{N} s(\chi) = \mathcal{N} \: \chi_L \otimes \chi_R,
\end{equation}
where $\mathcal{N}$ is a normalization. If $s(\chi) \neq 0$, we choose
\begin{equation}
\mathcal{N}^2 = \frac{\| \chi_L + \chi_R \|^2}{\| \chi_L \otimes \chi_R \|^2} = \frac{\| \chi_L \|^2 + \| \chi_R \|^2}{\| \chi_L \|^2 \| \chi_R \|^2},
\end{equation}
so that the norms of $\chi$ and $\chi^{\dbl}$ are equal.

First notice that if either $\chi_L = 0$ or $\chi_R = 0$, then $\chi^{\dbl} = 0$. In other words if a state $\chi \in \H$ is perturbative with respect to any minimum, then $\chi^{\dbl} = 0$. We will say that a state is \emph{typical} if it is represented by a non-vanishing effective state in the effective theory. Clearly, every typical state $\chi = \chi_L + \chi_R$ is generic in the sense that upon `random choice' of $\chi_L$ and $\chi_R$ it is unlikely to end up with an atypical state. Only typical states can be represented in the effective theory. 

The image of the map \eqref{dbl} consists of a set $\mathcal{H}_{\text{pert}}$ of all simple tensors $\psi_L \otimes \psi_R \in \H_L \otimes \H_R$. A pre-image of a given simple tensor $\psi_L \otimes \psi_R$, however, is not unique. Every normed state $\psi^{\dbl} \in \H_{\text{pert}}$ can be written as $\psi^{\dbl} = e^{\I \theta} \psi_L \otimes \psi_R$ with $\| \psi_L \| = \| \psi_R \| = 1$ and an irrelevant phase $\theta$. The most general form of $\psi \in \H$ mapped onto $\psi^{\dbl}$ is
\begin{equation} \label{inv_dbl}
\psi = \alpha_L \psi_L + \alpha_R \psi_R, \text{ with } | \alpha_L |^2 + | \alpha_R |^2 = 1,
\end{equation}
where $\arg \alpha_L + \arg \alpha_R = \theta + 2 \pi n$. We will refer to any effective state $\psi^{\dbl}$ as a \emph{macrostate} whereas all corresponding $\psi$ are \emph{microstates}. Many different states $\psi$ are mapped onto the same macrostate in the effective theory.

In our toy-model, all possible microstates corresponding to a given macrostate $\psi^{\dbl}$ are parametrized by a unit vector $(\alpha_L, \alpha_R) \in \C^2$. Equivalently, this ambiguity amounts to a choice of a normed perturbative vacuum $\mu \in \M$. Physically this means that knowledge of the left and right portions $\psi_L$ and $\psi_R$ of the wave function is not sufficient to reconstruct the full wave function $\alpha_L \psi_L + \alpha_R \psi_R$. Instead a prescription for the continuation of the wave functions through the potential barrier is needed. This is clearly a non-perturbative effect, and hence invisible in the effective theory.

Given a simple tensor $\psi_L \otimes \psi_R \in \mathcal{H}_{\text{pert}}$, one can define its projection on, say, $\H_R$ as $\psi_L \otimes \psi_R \mapsto \psi_R$. From the point of view of the right vacuum these are pure states since, when traced over $\H_L$, they lead to the pure state density matrix $| \psi_R \> \< \psi_R |$. Every other state $\psi^{\dbl} \in \H_L \otimes \H_R$ that does not belong to $\H_{\text{pert}}$ can be regarded as a mixed state with the density matrix $\rho_R = \Tr_L | \psi^{\dbl} \> \< \psi^{\dbl} |$.

Note also that from the point of view of the right asymptotic observer, only states that lie in $\H_R \subseteq \mathcal{F}_R$ are perturbative. States beyond $\H_R$ lack any perturbative interpretation. Hence states in $\mathcal{F}_L \otimes \mathcal{F}_R$ that do not belong to $\mathcal{H}_L \otimes \mathcal{H}_R$ lack a perturbative description from the standpoint of both asymptotic regions, even as mixed states.

\subsection{Operators in the effective theory} \label{sec:global_to_eff}

Given an operator $A$ in the full theory, we would also like to construct an operator $A^{\dbl}$ in the effective theory such that 
\begin{equation} \label{equil}
\< \psi | A | \psi \> = \frac{1}{Z} \< \psi^{\dbl} | A^{\dbl} | \psi^{\dbl} \> + o(\lambda^\infty),
\end{equation}
assuming $\psi$ and $\psi^{\dbl}$ are normalized to one. The proportionality factor $Z$ should correspond to the number of microstates $\psi$ represented by an identical macrostate $\psi^{\dbl}$. The relation \eqref{equil} is known as the equilibrium condition \cite{Harlow:2014yoa}. It is in this sense that the full theory is realized by the effective theory up to non-perturbative terms.

Note that in order for \eqref{equil} to hold, one must have $\psi^{\dbl} \neq 0$ if $\psi$ is non-zero, \textit{i.e.}, the state $\psi$ must be typical. Secondly, off-diagonal elements of $A$ must be non-perturbatively small, \textit{i.e.}, the operator $A$ must be perturbative. Given a perturbative operator $A$ a natural guess for its effective counterpart $A^{\dbl}$ would be
\begin{equation} \label{Atry}
B = A_{LL} \otimes \id + \id \otimes A_{RR}.
\end{equation}
Many operators considered in the effective theory are expected to be of this form. Unfortunately, $B$ does not satisfy relation \eqref{equil}. Indeed, consider a normed state $\psi = \alpha_L \psi_L + \alpha_R \psi_R$ with $\| \psi_L \| = \| \psi_R \| = \| \psi \| = 1$. The effective state is $\psi^{\dbl} = e^{i \theta} \psi_L \otimes \psi_R$ with an irrelevant overall phase $\theta$, which drops out from the expectation value. The left hand side of \eqref{equil} then reads
\begin{equation} \label{psiAlhs}
\< \psi | A | \psi \> = | \alpha_L |^2 \< \psi_L | A_{LL} | \psi_L \> + | \alpha_R |^2 \< \psi_R | A_{RR} | \psi_R \> + o(\lambda^\infty)
\end{equation}
while the right hand side is
\begin{equation} \label{psiArhs}
\< \psi^{\dbl} | B | \psi^{\dbl} \> = \< \psi_L | A_{LL} | \psi_L \> + \< \psi_R | A_{RR} | \psi_R \>.
\end{equation}
The mismatch is not surprising: the correlator $\< \psi | A | \psi \>$ clearly distinguishes specific microstates of the system, whereas $\< \psi^{\dbl} | B | \psi^{\dbl} \>$ depends on the overall macrostate only.

In order for the condition \eqref{equil} to hold one possibility is for the system to be in a special `equilibrium' state, and with an operator $A$ that does not distinguish between microstates. Mathematically, if $\Theta \psi = \pm \psi$ and $A_{LL} = \Theta A_{RR} \Theta$, then indeed $A^{\dbl} = B$ and we find
\begin{equation}
\< \psi | A | \psi \> = \frac{1}{2} \< \psi^{\dbl} | A^{\dbl} | \psi^{\dbl} \> + o(\lambda^\infty).
\end{equation}
The $Z$ factor accounts for the degeneracy of the macrostate, $Z = e^{S_{B}} = 2$, where $S_B$ is the Boltzmann entropy \eqref{SBoltz}. 

What are equilibrium states in our model? Notice that the condition $\Theta \psi = \pm \psi$ means $\alpha_R = \pm \alpha_L = 2^{-1/2}$. Hence every energy eigenstate $\Psi_n^{\pm}$ is an equilibrium state. Furthermore every state of fixed parity, even or odd, is also an equilibrium state. Operators satisfying $A_{LL} = \Theta A_{RR} \Theta$ are those that act on both sides of the potential well `in the same way' regardless of a specific microstate. This is closely related to the definition of operators satisfying the Eigenstate Thermalization Hypothesis, \textit{e.g.}, \cite{Srednicki:1995pt,Lashkari:2016vgj,Lashkari:2017hwq}. In particular the total number operator and the Hamiltonian are of this form.

By contrast, if the microstate $\psi$ is not an equilibrium state, then \eqref{psiAlhs} depends on the specific values of $\alpha_L$ and $\alpha_R$. Given a global operator $A$, we can construct a class of effective operators $A^{\dbl}_\mu$, which depend on these parameters. Indeed, by comparing \eqref{psiAlhs} with \eqref{psiArhs} we see that we need
\begin{equation} \label{Aeff}
A^{\dbl}_{\mu} = A_{LL} \otimes | \alpha_R |^2 \id + | \alpha_L |^2 \id \otimes A_{RR}.
\end{equation}
In this case $Z = 1$, as the right hand side of \eqref{equil} produces the expectation value of $A$ within a single, given microstate $\psi$. The family of the operators $A^{\dbl}_{\mu}$ is parametrized by a unit vector $(\alpha_L, \alpha_R) \in \C^2 \cong \M$, or equivalently, by a perturbative vacuum state $\mu \in \M$. In particular every state of the form
\begin{equation}
\sum_{n=0}^{\infty} \left( \gamma_n^L \frac{(\hat{a}_L^+)^n}{\sqrt{n!}} + \gamma_n^R \frac{(\hat{a}_R^+)^n}{\sqrt{n!}} \right) \mu, \qquad \sum_{n=0}^{\infty} | \gamma_n^L |^2 = \sum_{n=0}^{\infty} | \gamma_n^R |^2 = 1
\end{equation}
is characterized by the same vector $(\alpha_L, \alpha_R)$ with $\mu = \alpha_L \Psi_0^L + \alpha_R \Psi_0^R$.

It is a defining property of an effective theory that it should give an approximate description of the full theory within its region of validity. In the context of quantum theory this means that the correlation functions calculated within the effective theory should approximate those in the full theory. Given an operator $A$ in the full theory, the operators $A_\mu^{\dbl}$ defined in \eqref{Aeff} satisfy this condition. This is a family of operators together with a `fake' type of state-dependence as described in \cite{Harlow:2014yoa}. The parameter $\mu$ can be thought of as parametrizing degenerated perturbative vacua. Indeed, each $A_\mu^{\dbl}$ is a perfectly well-defined linear operator on $\H_L \otimes \H_R$, since the numbers $\alpha_L$ and $\alpha_R$ are fixed parameters from the point of view of the effective theory.

One can however revert this last construction. Starting from an operator $B$ as given in \eqref{Atry}, one can try to construct a global operator $A$ such that \eqref{equil} is satisfied. This is a weaker condition than what we have considered in Section \ref{sec:state_dep}, since here we only demand the agreement between the expectation values of the operators (up to non-perturbative effects). It is obvious that a definition of $A$ must depend on the specific microstate of the full system. We can construct a family of operators $A_\mu$ in the full theory such that their action on a microstate $\psi = \alpha_L \psi_L + \alpha_R \psi_R$ with $\psi_L \in \H_L$, $\psi_R \in \H_R$ and $\| \psi \| = \| \psi_L \| = \| \psi_R \| = 1$ reads
\begin{equation} \label{Amu}
A_{\mu} \psi = \left( \frac{A_{LL}}{| \alpha_L |^2} + \frac{A_{RR}}{|\alpha_R|^2} \right) \psi.
\end{equation}
With this definition \eqref{equil} holds with $Z = 1$. The functions $A_\mu$ are now non-linear, \textit{i.e.}, `state-dependent', as they implicitly depend on the parameters $\alpha_L$ and $\alpha_R$ of the state $\psi = \alpha_L \psi_L + \alpha_R \psi_R$ they act on. In other words they depend on the choice of the microstate within a given macrostate $\psi^{\dbl} = e^{\I \theta} \psi_L \otimes \psi_R$.

This is an explicit construction of state-dependent operators in the theory in the spirit of \cite{Papadodimas:2013jku,Papadodimas:2013wnh,Papadodimas:2015jra,Papadodimas:2015xma}. Operators of the form \eqref{Atry} corresponding to naive perturbative observables become non-linear functions or, equivalently, state-dependent operators. In the context of \cite{Marolf:2015dia} their matrix elements represent certain conditional probabilities. Furthermore, since the time evolution generically mixes various microstates, comparison of their matrix elements at different times seems problematic.

\subsection{Global time evolution}

It is common in the context of quantum field theory to calculate time-dependent field operator correlation functions such as $\< 0 | \phi(t_1) \phi(t_2) \ldots \phi(t_n) | 0 \>$. To consider such correlators in our toy-model, we must decide what are the corresponding state $|0\>$, the field operator $\phi$, and the Hamiltonian driving the time evolution. 

From the point of view of the full theory, $|0\>$ is the vacuum state $\Psi_0^+$, $\phi = x$ is the field operator, and time evolution is governed by the full Hamiltonian $H$. In the context of black hole physics, however, one typically considers correlation functions in the perturbation theory corresponding to the viewpoint of a single, say right, asymptotic observer. In perturbation theory $|0\>$ corresponds to the right perturbative vacuum $\Psi_0^R$, and $\phi$ corresponds to the right perturbative field operator $\hat{y}_R = P_R y_R P_R = (\hat{a}_R + \hat{a}_R^+)/\sqrt{2}$.

What remains to be analyzed is the time evolution governed by the unitary operator $U(t) = e^{\I t H}$. Its effective version $U^{\dbl}(t)$ should be such that the evolution of the states and the operators in the full theory matches with that in the effective theory. Assume that $U(t)$ satisfies the condition \eqref{np-small}, \textit{i.e.}, $U_{LR}(t) = o(\lambda^\infty)$ and $U_{RL}(t) = o(\lambda^\infty)$, and that the diagonal elements $U_{LL}(t)$ and $U_{RR}(t)$ remain unitary, at least up to non-perturbative effects. Since the Hamiltonian $H$ satisfies \eqref{np-small}, this is the case for times $t$ sufficiently small for the tunneling effects to be insignificant. Now assign
\begin{equation} \label{Udbl}
U(t) = U_{LL}(t) + U_{RR}(t) \ \longmapsto \ U^{\dbl}(t) = U_{LL}(t) \otimes U_{RR}(t).
\end{equation}
Since $U_{LL}$ and $U_{RR}$ are unitary, they preserve the norm of the states and hence for any $\psi \in \H$ we have
\begin{equation} \label{Udbl_props}
\left[ U(t) \psi \right]^{\dbl} = U^{\dbl}(t) \psi^{\dbl}, \qquad\qquad (U^+)^{\dbl}(t) = (U^{\dbl})^+(t).
\end{equation}
This means that the states evolve in the same way both in the full and in the effective theory. In particular
\begin{equation}
A_{\mu}^{\dbl}(t) = \left[ U^+(t) A_{\mu} U(t) \right]^{\dbl} = (U^{\dbl})^+(t) A_{\mu}^{\dbl} U^{\dbl}(t).
\end{equation}
with $A_{\mu}^{\dbl}$ defined as in the previous section. Therefore the condition \eqref{equil} holds for time-dependent operators as well,
\begin{equation} \label{equil_t}
\< \psi | A(t) | \psi \>  = \frac{1}{Z} \< \psi^{\dbl} | A_{\mu}^{\dbl}(t) | \psi^{\dbl} \> + o(\lambda^\infty)
\end{equation}
With \eqref{Udbl} the effective theory correctly describes the expectation values of time-dependent operators as long as the perturbation theory remains valid.

\subsection{Time reversal and the `wrong sign' commutation relations}

In the previous section we have defined a notion of time evolution in the effective theory that is specified by the full theory, based in particular on the global time $t$ inherited from the full Hamiltonian $H$ in the Hilbert space $\H$. From the perspective of a single observer (say the right one), however, the common practice is to consider the doubled Hilbert space built out of the right Hilbert space $\H_R$. In this case, the effective theory as constructed by the right observer lives on $\H_R \otimes \H_R = (\Theta \H_L) \otimes \H_R$, with effective states defined as
\begin{equation} \label{dbl_conj}
\chi = \chi_L + \chi_R \ \longmapsto \ \tilde{\chi}^{\dbl} = \mathcal{N} \: (\Theta \chi_L) \otimes \chi_R \in \H_R \otimes \H_R
\end{equation}
where $\mathcal{N}$ is the same normalization as in \eqref{dbl}. This does not change the analysis of previous sections in any significant way. Simply, every operator acting on $\H_L$ must now be accompanied by a conjugation by $\Theta$. For example, equation \eqref{Aeff} would read
\begin{equation}
\tilde{A}^{\dbl}_{\mu} = \Theta A_{LL} \Theta \otimes | \alpha_R |^2 \id + | \alpha_L |^2 \id \otimes A_{RR}
\end{equation}
and so on. For the Hamiltonian $H$ of the full theory we have $H_{LL} = \Theta H_{RR} \Theta$ and hence we find its effective version
\begin{equation} \label{Hcdbl}
\tilde{H}^{\dbl}_{\mu} = H_{RR} \otimes | \alpha_R |^2 \id + | \alpha_L |^2 \id \otimes H_{RR}.
\end{equation}

Consider now the time evolution operator $U(t) = e^{\I t H}$. Due to the additional factor of $\I$ we have $\Theta U_{LL}(t) \Theta = U_{RR}(-t)$. Hence, in order to maintain \eqref{Udbl_props}, we have to include an additional conjugation by $\Theta$ in \eqref{Udbl}, \textit{i.e.}, to define
\begin{equation}
\tilde{U}^{\dbl}(t) = \Theta U_{LL}(t) \Theta \otimes U_{RR}(t) = U_{RR}(-t) \otimes U_{RR}(t).
\end{equation}
From the point of view of the right observer, in the effective theory on $\H_R \otimes \H_R$ time directions are opposite in both component spaces $\H_R$. In particular the Hermitian generator $\tilde{h}$ of $\tilde{U}^{\dbl}(t)$ satisfying $\tilde{U}^{\dbl}(t) = e^{\I t \tilde{h}}$ is
\begin{equation} \label{htilde}
\tilde{h} = - H_{RR} \otimes \id + \id \otimes H_{RR}.
\end{equation}
This is a Hermitian generator of time translations in the effective theory on $\H_R \otimes \H_R$, but it is not bounded from below. Furthermore, it is not proportional to any $\tilde{H}^{\dbl}_\mu$, the family of effective operators corresponding to the full Hamiltonian. Finally, $\tilde{h}$ exhibits the famous `wrong' commutation relations \cite{Almheiri:2012rt,Papadodimas:2012aq,Papadodimas:2013jku,Papadodimas:2013wnh,Marolf:2013dba} with left creation-annihilation operators satisfying,
\begin{equation}
[ \tilde{h}, \hat{a}_L \otimes \id] = +\hat{a}_L \otimes \id + O(\sqrt{\lambda}), \qquad\qquad [ \tilde{h}, \hat{a}_L^+ \otimes \id] = - \hat{a}_L^+ \otimes \id + O(\sqrt{\lambda}).
\end{equation}
This is in no contradiction with unitarity or any other property of a well-defined quantum theory. The effective theory is merely designed to mimic the full theory within the regime of its validity. It recognizes the full Hamiltonian as $\tilde{H}^{\dbl}_\mu$ defined in \eqref{Hcdbl}. The time evolution of the effective theory is however driven by $\tilde{h}$ in such a way that \eqref{equil_t} holds. This is different from the effective theory on $\H_L \otimes \H_R$ (or $\H_L \otimes \H_L$ as perceived by the left observer). Different effective viewpoints realize observables in different ways, despite describing the same theory. However expectation values of corresponding operators in corresponding states are equal in all effective theories we have constructed here.

\section{Dynamics} \label{sec:time}

With all elements of our toy-model in place we now turn to a number of dynamical processes that are both tractable and have a clear dual interpretation in terms of black hole physics. Before we proceed, let us point out the obvious: our toy-model with Hamiltonian \eqref{H} is unitary. Since it also has a unique vacuum state, $\Psi_0^+$, it would seem that our model violates Hawking's theorem, as stated in \cite{Mathur:2009hf}. However, the vacuum state in our effective theory is doubly degenerate, $\Psi_0^{\pm}$, and, since the theorem describes the semiclassical situation, its assumptions are not satisfied.

First we investigate tunneling through the potential barrier. Depending on the energy range, this can be viewed either as the decay of one of the perturbative vacua, the Hawking radiation process, or scattering of waves off of the black hole. Then we consider the evolution of a classical particle. We identify signatures of the chaotic behavior and scrambling.

\subsection{Tunneling and Hawking radiation} \label{sec:tunnel}

The evolution operator $e^{i t H}$ restricted to the space of perturbative vacua $\M$ slowly mixes the left and right states $\Psi_0^L$ and $\Psi_0^R$, as
\begin{equation} \label{evolvePsi}
e^{i t H} \left(\begin{array}{c} \Psi_0^L \\ \Psi_0^R \end{array} \right) = e^{\I t E_0} \mat{\cos \left( \tfrac{1}{2} t \Delta E_0 \right)}{- i \sin \left( \tfrac{1}{2} t \Delta E_0 \right)}{- i \sin \left( \tfrac{1}{2} t \Delta E_0 \right)}{\cos \left( \tfrac{1}{2} t \Delta E_0 \right)} \left(\begin{array}{c} \Psi_0^L \\ \Psi_0^R \end{array} \right)
\end{equation}
where
\begin{equation}
E_0 = \frac{1}{2} (E_0^+ + E_0^-), \qquad\qquad \Delta E_0 = E_0^+ - E_0^-.
\end{equation}
Assume that at $t = 0$ the system is in the state $\Psi_0^R$. As the system evolves, the wave function slowly leaks into the left minimum, evolving into `more typical' black hole microstates. This is reminiscent of known instabilities of perturbative vacua \cite{Ooguri:2016pdq,Freivogel:2016qwc} evolving into black holes and envisioned \textit{e.g.}, in the formation of fuzzballs \cite{Mathur:2008kg,Hertog:2017vod}.

On the other hand, as the wave functions builds up on the left, the right observer perceives the inflow of highly excited particles. Indeed, according to the results of Section \ref{sec:fire}, when decomposed in terms of semiclassical modes $\varphi_n^R$, $\Psi_0^L$ is a highly-excited state. We can interpret these particles as the Hawking radiation. The same remains true for any low energy black hole state $\psi$. We can identify some features of the Hawking radiation in our model by tracing back these high occupancy modes through the potential barrier. 

\begin{figure}[ht]
	\includegraphics[width=0.45\textwidth]{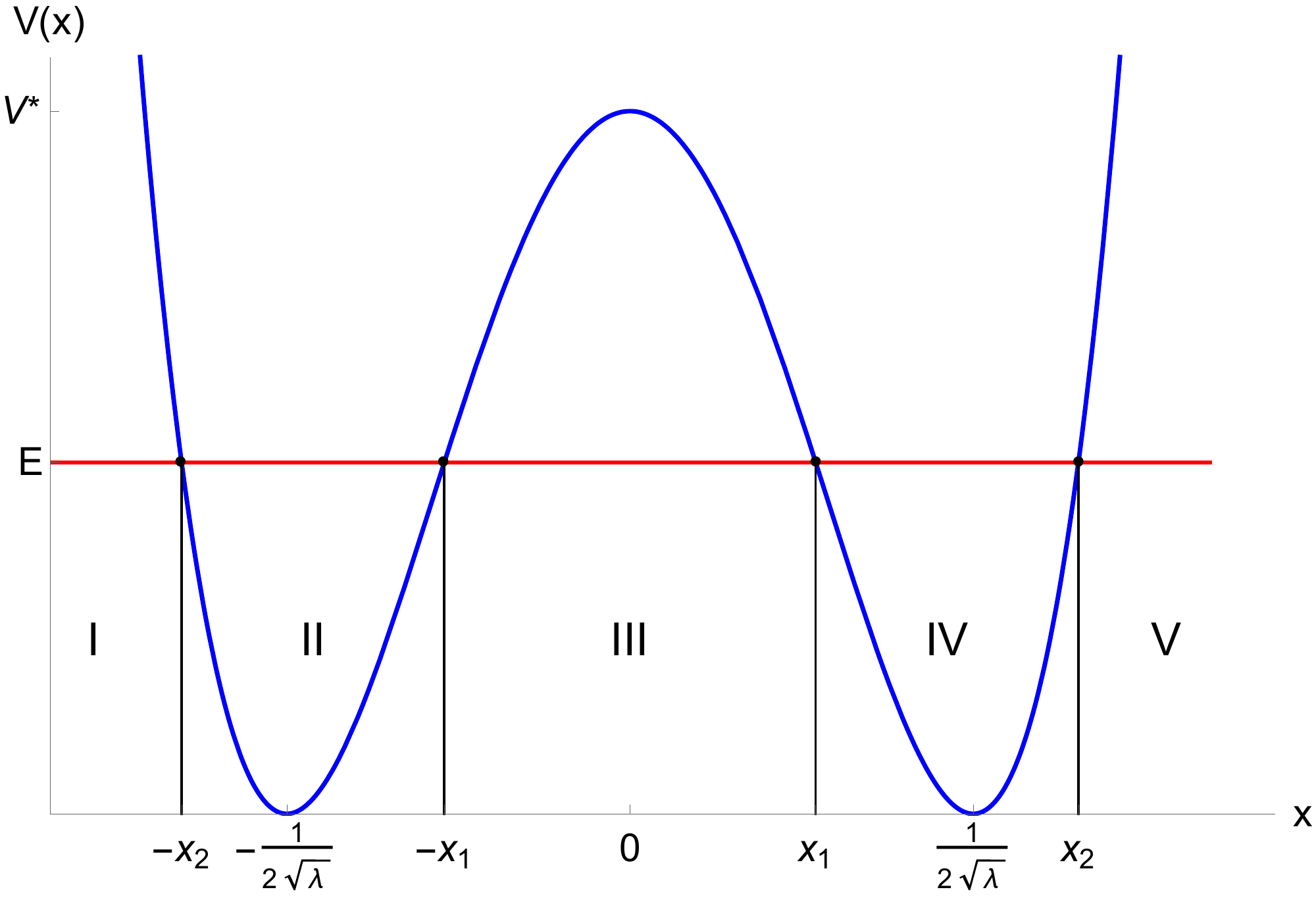}
	\qquad
	\includegraphics[width=0.45\textwidth]{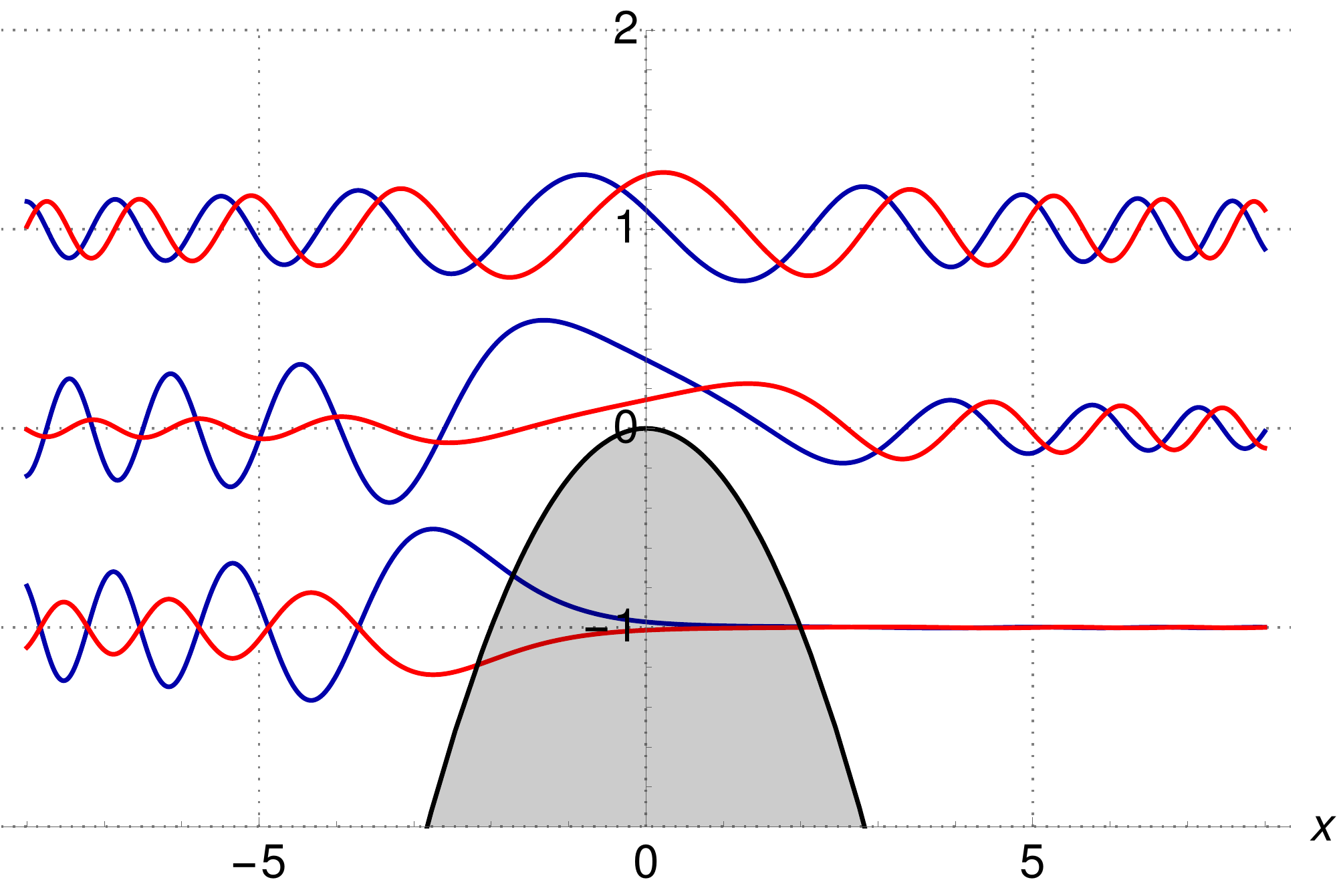}
	\centering
	\caption{Left: The tunneling probability is computed in region III in the WKB approximation. Right: The shape of the real part (blue) and imaginaty part (red) of the (generalized) energy eigenfunctions representing waves incident from the left and scattered by the potential of the inverted harmonic oscillator. The functions correspond to $\nu^2 = 1/2$ and energies $\Delta = -1, 0, 1$.\label{fig:tunelinv}}
\end{figure}

In the WKB approximation the tunneling probability $\Lambda$ for a particle of energy $E$ in a double well potential $V$ is given by
\begin{align} \label{Lambda}
\Lambda &= \int_{-x_1}^{x_1}\sqrt{2(V(x)-E)}dx \ ,
\end{align}
where $\pm x_1$ are the turning points as illustrated in Figure \ref{fig:tunelinv}, and $E < V_{\ast} = V(0) = 1/(32 \lambda)$. The tunneling time $\tau$ and the tunneling rate $\Gamma$ are then $\tau \sim \Gamma^{-1/2} \sim e^{\Lambda}$. In our toy-model the tunneling probability \eqref{Lambda} within the WKB approximation can be expressed in terms of complete elliptic integrals,
\begin{align} \label{Tresult}
\Lambda&= \frac{1}{6\lambda} \sqrt{1+\sqrt{\e}} \left[
\EllipticE\left( \sqrt{\frac{1 - \sqrt{\e}}{1 + \sqrt{\e}}} \right) - \sqrt{\e} \EllipticK\left(\sqrt{\frac{1-\sqrt{\e}}{1+\sqrt{\e}}} \right)
\right],
\end{align}
where $\e = E/V_{\ast} = 32\lambda E$ with $0 < \e < 1$, and $\EllipticK$ denotes the complete elliptic integral of the first kind. 

For energies $E$ close to $V_{\ast}$ we have $\e \sim 1$ and then \eqref{Tresult} reduces to
\begin{equation}
\Lambda(\lambda, \delta) = \sqrt2 \pi\delta +3\sqrt2 \pi \lambda \delta^2 + O(\delta^3), \quad \delta = V_{\ast}-E. \label{LambdaE0}
\end{equation}
The first term in  \eqref{LambdaE0} corresponds to tunneling in an inverted harmonic oscillator with the potential $V_{\text{iho}}=-\omega^2 \frac{x^2}{4}$, for a particle of energy $E = \omega \delta$. The tunneling rate becomes
\begin{align}
\Gamma_{\text{iho}}(E=\omega \delta)=\exp \left(-2\sqrt2 \pi \delta - 6\sqrt2 \pi \lambda \delta^2 + O(\delta^3) \right).
\end{align}
This resonates with \cite{Parikh:1999mf} where it was shown that Hawking radiation can be viewed as a tunneling effect between regions near both sides of the horizon. Furthermore, specific deviations from an exact thermal spectrum were found in such a process, in a range of different black hole backgrounds. A comparison with \cite{Parikh:1999mf} shows that the Hawking radiation in our toy-model corresponds roughly to a single pair of modes of a field of frequency $\omega \sim \sqrt{\lambda} \delta$ in a black hole state of mass $M \sim 1/\sqrt{\lambda} = N$.

Building on the analysis of \cite{Parikh:1999mf} it has been shown in \cite{Betzios:2016yaq} (see also \cite{Horowitz:2009wm}) that the scattering matrix of an inverted harmonic oscillator also appears in the investigation of scattering of shock waves from in the black hole background. Indeed, this is precisely the conclusion from our toy-model. Around $x = 0$ energy eigenfunctions are approximated by wave functions of the inverted harmonic oscillator with energy $E$. The approximate Hamiltonian reads
\begin{align}\label{H-IHO}
H_{\text{iho}} = \frac{p^2}{2}-\frac12 \nu^2 x^2
\end{align}
and in our specific toy-model $\nu^2 = \omega^2/2 = 1/2$. To analyze quantum scattering, we consider oscillating wave functions $\psi$ satisfying the Schr\"{o}dinger equation with the Hamiltonian \eqref{H-IHO},
	\begin{align}\label{SchrIHO}
	\frac{\partial^2}{\partial \xi^2}\psi(\xi)+\left( \frac14\xi^2+\frac{\EE}{\nu} \right)\psi(\xi)=0
	\end{align}
where $\EE = E-V_{\ast}$ and $\xi = \sqrt{2 \nu} x$. A pair of solutions exists for each energy value $\Delta$,
which can be expressed in terms of hypergeometric functions, see \cite{Bermudez:2012fe}. With an appropriate normalization the two solutions represent waves incident from the left and from the right, as shown in Figure \ref{fig:tunelinv}.  By expanding the wave functions in the asymptotic regions transmission and reflection coefficients can be found,
\begin{align} \label{TandR}
T&=\frac{e^{\pi \EE/2\nu}}{\sqrt{2\pi}}\Gamma\left(\frac12-\frac{i \EE}{\nu}\right)
\ , \qquad
R=-i\frac{e^{-\pi \EE/2\nu}}{\sqrt{2\pi}}\Gamma\left(\frac12-\frac{i \EE}{\nu}\right)\ ,
\end{align}
which satisfy $|T|^2+|R|^2=1$. Loosely speaking these coefficients describe the behavior in our toy-model of excitations near a black hole horizon. It is remarkable that the same form of the transmission and reflection coefficients \eqref{TandR} was recovered there from the analysis of the dynamics of shock waves traveling in the black hole background in \cite{Betzios:2016yaq}. To leading order the result follows simply from the repulsive nature of the inverted harmonic oscillator and depends on a single parameter $\nu$ in \eqref{H-IHO}.

\subsection{Classical evolution and chaos} \label{sec:classical}

Consider now a classical particle of energy $E \geq 0$. Such a particle will stay on a closed orbit. If $E < V_{\ast} = V(0)$, then the orbit remains `outside the black hole', \textit{i.e.}, it does not cross the maximum of the potential at $x = 0$. The period of the classical motion is then
\begin{equation}
T_{\text{trapped}} = \frac{8}{\sqrt{1 - \sqrt{\e}} + \sqrt{1 + \sqrt{\e}}} \EllipticK \left( \sqrt{ \frac{1 - \sqrt{1 - \e}}{1 + \sqrt{1 - \e}}} \right),
\end{equation}
When expanded around $\e = 0$ one finds as expected, $T_{\text{trapped}} = 2 \pi + O(\e)$. On the other hand the period diverges logarithmically when $E \rightarrow V_{\ast}$,
\begin{equation}
T_{\text{trapped}} = \sqrt{2} \log \left( \frac{2}{\lambda \delta} \right) + O(\epsilon), \qquad \delta = V_{\ast} - E.
\end{equation}

\begin{figure}[t]
	\includegraphics[width=0.45\textwidth]{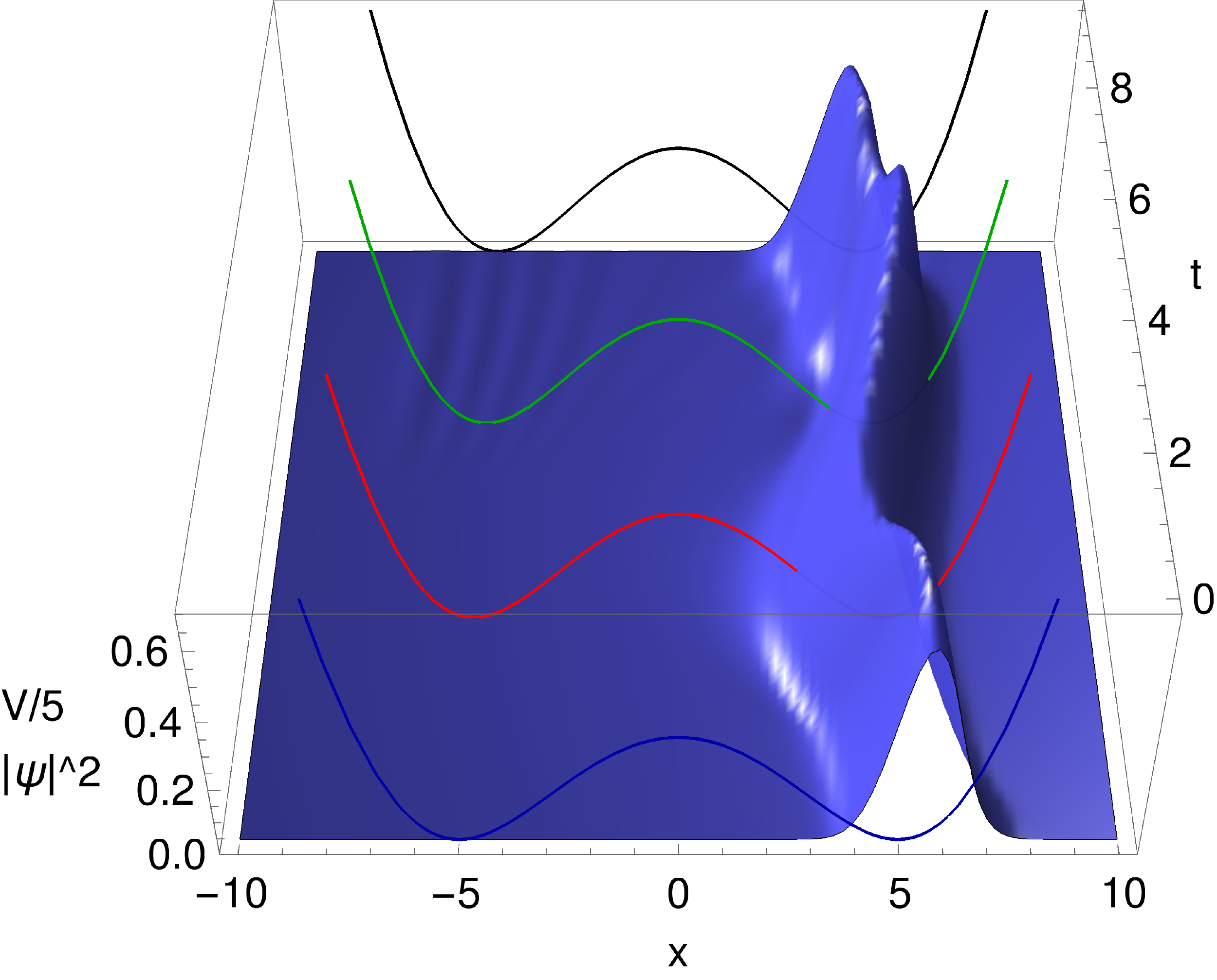}
	\qquad
	\includegraphics[width=0.45\textwidth]{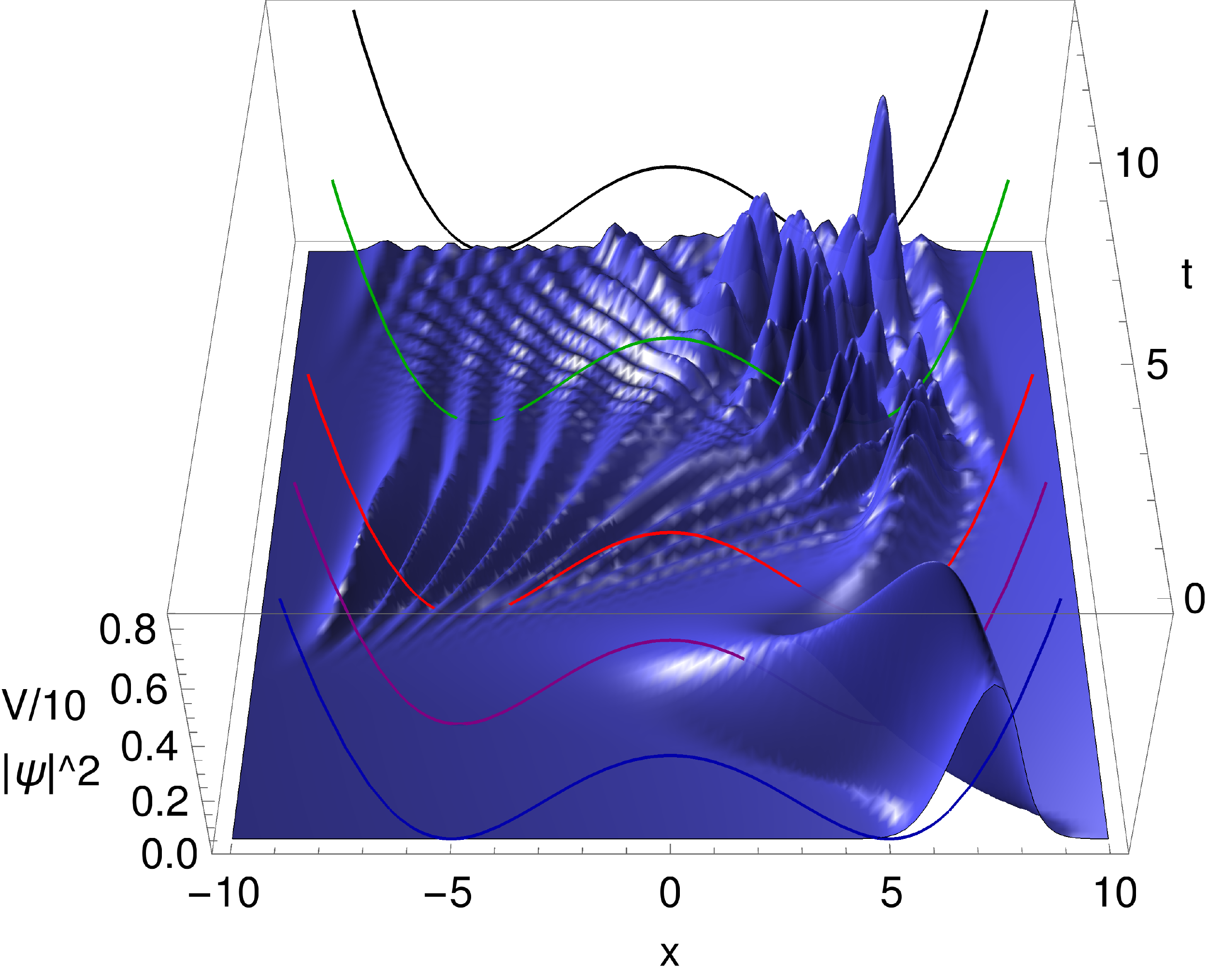} \\
	\includegraphics[width=0.45\textwidth]{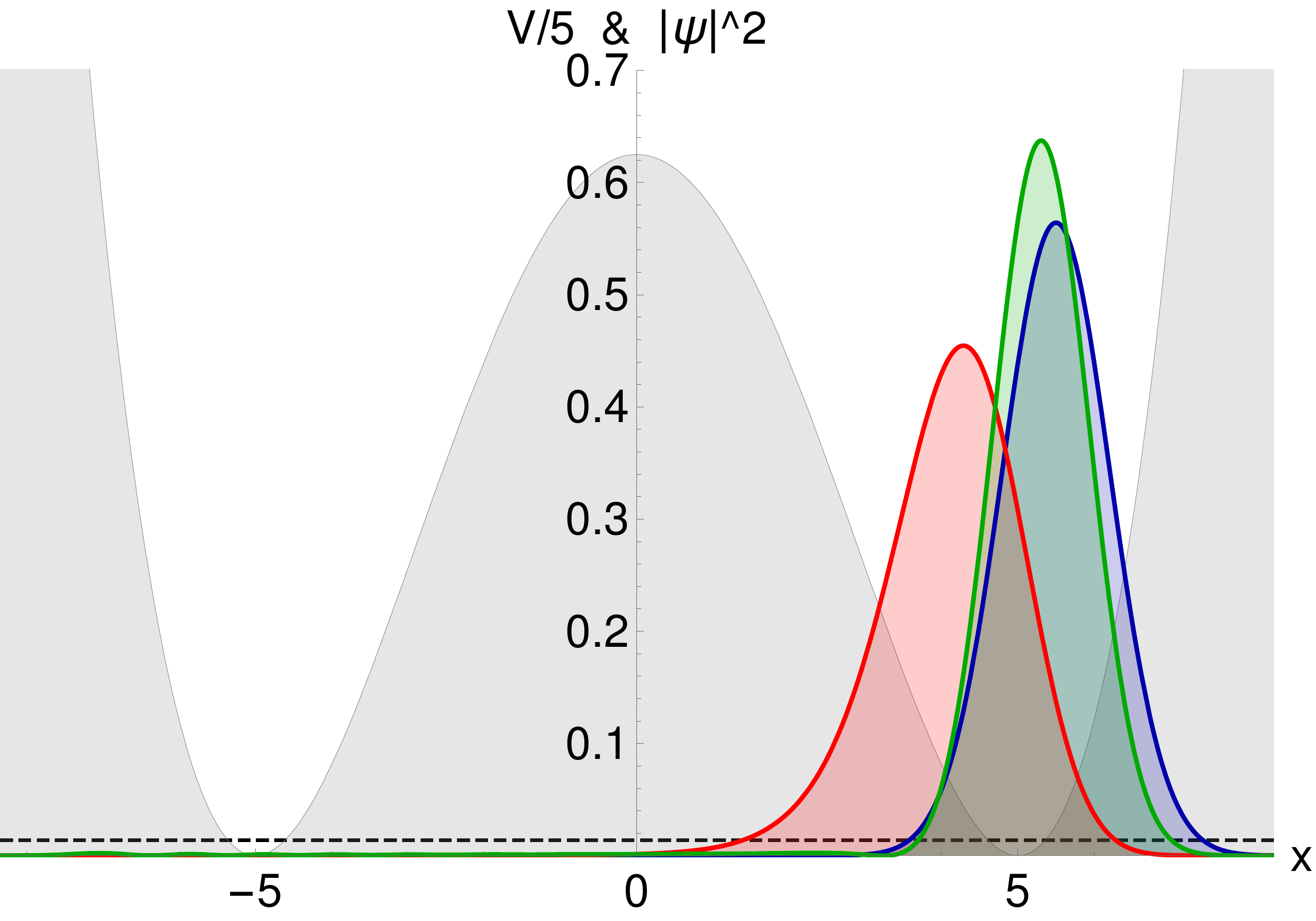}
	\qquad
	\includegraphics[width=0.45\textwidth]{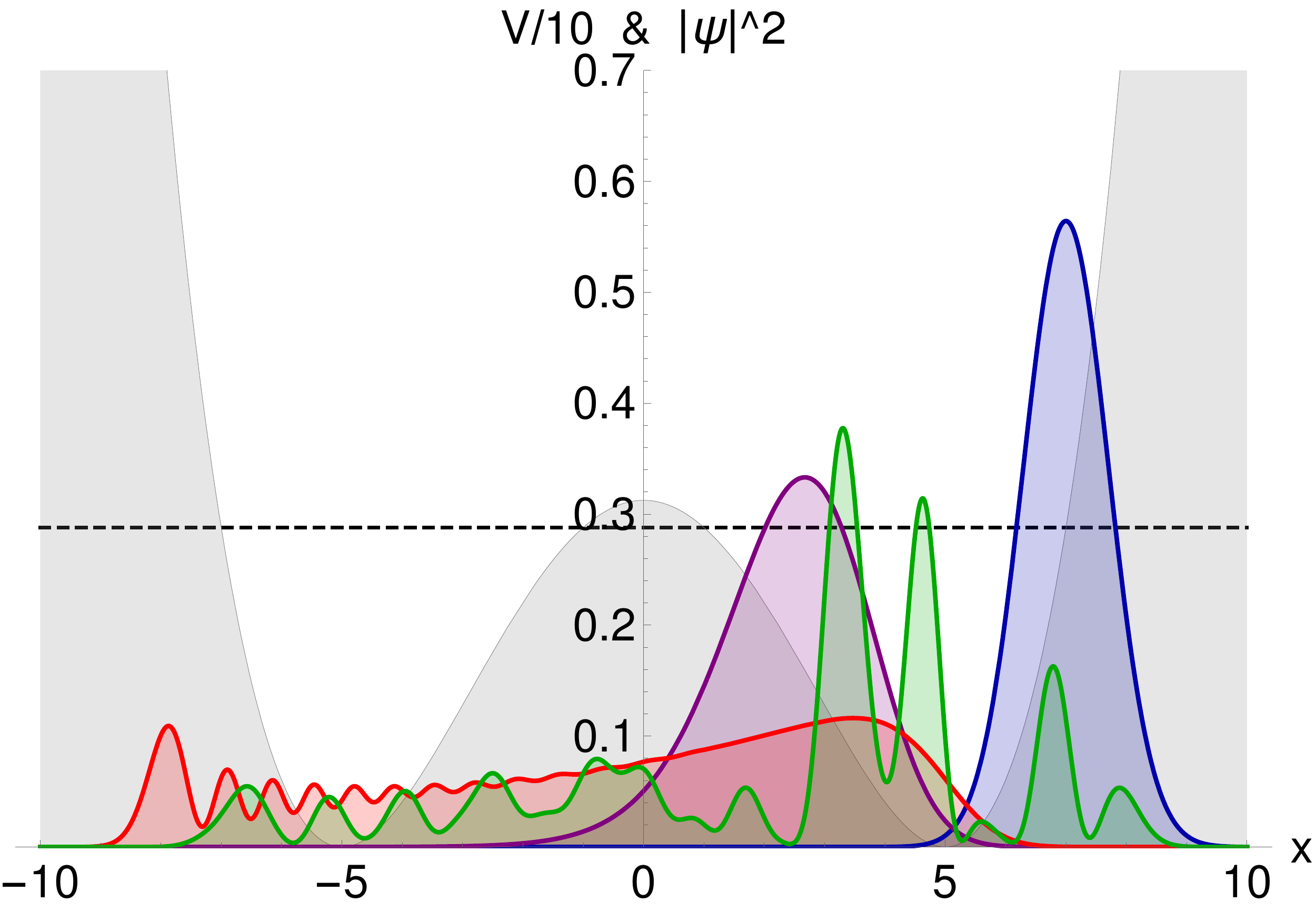}
	\centering
	\caption{Evolution of the density $|\psi|^2$ of an initially classical system represented as a coherent state. Plots on the left show the evolution of a low energy wave packet on a classical path starting at $x = 5.5$ (classical energy $E = 0.14$). Plots on the right present the evolution of a wave packet of energy just below the energy $V_{\ast}$ of the tip of the potential ($E = 2.88$, $V_{\ast} = 3.125$), with the initial position at $x = 7$. In both cases $\lambda = 1/100$. The outlines of the potential in the 3D plots are placed at times: $t = 0$ (blue), quarter of the classical period (purple, only on the right), half of the period (red), and the full classical period (green). The 2D plots show the shapes of the wavefunctions at the corresponding times. The dashed lines indicate the classical energy of the wave packet.\label{fig:coherent}}
\end{figure}

This is a sign of a critical behavior. When analyzed from the point of view of the time-dependent position $x(t)$ of the particle, it is a manifestation of chaos. To see it explicitly, assume that $\delta$ is small enough for the particle to be located close to the tip of the potential for a long time. Hence we can neglect interaction terms and consider the potential of the inverted harmonic oscillator only. The solution to the equations of motion is then simply $x(t) = x_0 \cosh(\nu t) + v_0/\nu \sinh(\nu t)$, where $\nu$ is the `frequency' of the inverted harmonic oscillator as defined in \eqref{H-IHO}. Parameters $x_0$ and $v_0$ are initial position and velocity at $t = 0$ and under their variation $\delta x(t) \sim e^{\nu t} (\delta x_0 + \delta v_0 / \nu)$ for large times. This is by definition chaotic behavior with the Lapunov exponent $\nu = \omega/\sqrt{2} = 1/\sqrt{2}$.

As in the previous section the chaotic behavior is driven by the inverted harmonic oscillator and parametrized by a single parameter $\nu$. The same conclusion arises from the analysis of shock waves in a black hole background in \cite{Shenker:2013pqa,Shenker:2013yza,Maldacena:2017axo,vanBreukelen:2017dul}. Perhaps unsurprisingly, our analysis shows a connection between the classical chaotic behavior and quantum Hawking radiation from the point of view of the tunneling process as analyzed in the previous section. Both processes are described by the same parameter $\nu$ characterizing the `frequency' of the inverted harmonic oscillator \eqref{H-IHO}.

One can also analyze the quantum evolution of the system numerically and show how the classical evolution becomes inaccurate when $E$ approaches $V_{\ast}$. To do so, we consider the evolution of suitable initially coherent states. The resulting evolution is illustrated in Figure \ref{fig:coherent}. As we can see, at low energies the original coherent state (blue curve on the left) bounces back and retains its Gaussian shape after a full period (green). A small change in the shape is caused by interactions of order $\sqrt{\lambda}$. The wave function leaks slightly into the left minimum, creating barely visible ripples in the density $|\psi|^2$. On the other hand a coherent state of energy only slightly lower than $V_{\ast}$ rapidly turns into a highly quantum oscillating wave. The original classical particle (blue curve on the right) starts moving towards the maximum of the potential. At the quarter of the period the peak widens (purple curve). By the time it bounces back at half the period it already breaks into a highly oscillating quantum wave (red curve). After the full classical period passes, it becomes completely scrambled and spread out over the entire domain (green curve). It cannot be viewed as a classical, localized state any more.

\section{Summary and conclusions}\label{con}

Motivated by the holographic description of certain classes of black holes in AdS we have put forward a new and solvable dual toy-model of black holes in terms of a quantum mechanical particle in a double well potential. The effective low energy description involves the tensor product of two decoupled harmonic oscillators representing the degrees of freedom on both sides of the horizon, or in both asymptotic regions. At this level our model captures many of the usual paradoxes of semiclassical black hole physics expressed here in quantum field theoretical language without explicit reference to geometry. 

The effective low energy description of the system as a pair of decoupled oscillators is altered drastically by non-perturbative interactions. We have carefully explored how states and operators in the effective theory emerge from and relate to corresponding quantities in the full model. This elucidates how holographic black hole models involving non-perturbative interactions between two decoupled low energy theories resolve some of the paradoxes of semiclassical black hole physics. Our key findings are the following:

\begin{itemize}
\item {\it Firewalls:} Black hole states in our model are represented by wave functions with significant support in both minima\footnote{This resonates with the model in \cite{Hollowood:2014hta} which was also argued to remove the firewall.}. At first sight the low energy theory predicts a firewall. This is because the expectation value in black hole states of the naive number operator \eqref{defNA} is large. This number operator has the same form in the decoupling limit and hence it is usually assumed that it should represent the number operator for a small coupling $\lambda$ as well. However, by carefully identifying and disentangling the perturbative degrees of freedom we have shown that this is incorrect. We have constructed a different number operator \eqref{defhatNA} that is well-defined in the full model and perturbative in a precise sense. In the decoupling limit this operator correctly reproduces \eqref{defNA}. We found this does not predict a firewall. Hence our model satisfies all four postulates of \cite{Almheiri:2012rt}.
\item {\it State-dependence:} Our toy-model describes both sides of the horizon in terms of a single well-defined, local, unitary quantum theory. Hence it does not require any `state-dependent' operators to describe behind the horizon physics. However we have shown that a clear notion of `state-dependence' emerges when one relates perturbative operators in the full theory to observables in the effective low energy model on the tensor product. Mathematically this is because the linear structure of the direct sum is very different from that of the tensor product. Physically, the state-dependence accounts for the dependence of perturbative operators in the full theory on the black hole microstate, \textit{i.e.}, on the shape of the portion of the wave function behind the barrier that is inaccessible to a given asymptotic observer. 
\item {\it Vacuum structure:}  Our model circumvents Mathur's no-go theorem \cite{Mathur:2008wi}, based on Hawking's original calculation, that states that the information paradox cannot be resolved by exponentially small corrections to correlation functions. This is because this theorem assumes there is a unique vacuum. In our toy-model the perturbative vacuum is degenerate, and the degeneracy is only broken by non-perturbative effects leading to a unique ground state $\Psi_0^+$ in the full theory. In our model the vacuum of an infalling observer is not represented by a semiclassical vacuum, but rather by a superposition of two semiclassical vacua in line with \textit{e.g.}, \cite{Hollowood:2014hta,Bao:2017rnv}. Phrased differently, one could say our model takes seriously the doubled copy of the system and in fact realizes an ensemble of states in the full interacting theory that are to some extent similar to the thermofield double state.
\item {\it Time evolution:} `Wrong sign' commutation relations naturally emerge in the effective description of the system from the standpoint of an observer in one of the asymptotic regions. This is not in contradiction with unitarity of the full model since the Hamiltonian of the full theory is represented by a different operator \eqref{Hcdbl} than the generator \eqref{htilde} of time translations in the effective theory around one of the perturbative vacua.
\end{itemize}

A major advantage of our toy-model is that it is solvable. This has enabled us to analyze the role of non-perturbative interactions in a number of interesting dynamical processes. The appearance of an inverted harmonic oscillator potential separating both wells as a toy-model for non-perturbative interactions leads to features, such as chaotic behavior, which have a natural analog or `dual interpretation' in black hole backgrounds where similar behavior was obtained \textit{e.g.}, in the analysis of shock waves. Moreover, the breakdown of classical evolution of initially coherent states in our model shows that the evolution of an infalling object becomes highly quantum from the viewpoint of an external observer, in line with the principle of black hole complementarity \cite{Susskind:1993if}. It would be interesting to investigate the role of non-perturbative effects and the emergence of the effective theory in more complex models. These could include matrix and tensor models \cite{Banks:1996vh,Iizuka:2008eb,Iizuka:2008hg,Ferrari:2016bvq,Maldacena:2016hyu,Polchinski:2016xgd}, where many features discussed in this paper emerge. Another direction includes the CFT analysis in the context of holography \cite{Fitzpatrick:2015dlt,Fitzpatrick:2016ive,Chen:2017yze}, where non-perturbative effects also become essential in the understanding of unitarity and locality.

\section*{Acknowledgments}

It is a pleasure to thank Ramy Brustein, Jan De Boer, and Kyriakos Papadodimas for helpful discussions. This work is supported in part by the European Research Council grant ERC-2013-CoG 616732 HoloQosmos, the C16/16/005 grant of the KU Leuven, and by the National Science Foundation of Belgium (FWO) grant G092617N.  AB is supported by the CEA Enhanced Eurotalents Fellowship. The work of AG is supported	by a Marie Sk\l odowska-Curie Individual Fellowship of the  European	Commission Horizon 2020 Program under contract number 702548 \emph{GaugedBH}.

\appendix

\providecommand{\href}[2]{#2}\begingroup\raggedright\endgroup


\begin{thebibliography}{10}
	
	\bibitem{Hawking:1976ra}
	S.~W. Hawking, \emph{{Breakdown of Predictability in Gravitational Collapse}},
	\href{http://dx.doi.org/10.1103/PhysRevD.14.2460}{\emph{Phys. Rev.} {\bf D14}
		(1976) 2460--2473}.
	
	\bibitem{Almheiri:2012rt}
	A.~Almheiri, D.~Marolf, J.~Polchinski and J.~Sully, \emph{{Black Holes:
			Complementarity or Firewalls?}},
	\href{http://dx.doi.org/10.1007/JHEP02(2013)062}{\emph{JHEP} {\bf 02} (2013)
		062}, [\href{http://arxiv.org/abs/1207.3123}{{\tt 1207.3123}}].
	
	\bibitem{Almheiri:2013hfa}
	A.~Almheiri, D.~Marolf, J.~Polchinski, D.~Stanford and J.~Sully, \emph{{An
			Apologia for Firewalls}},
	\href{http://dx.doi.org/10.1007/JHEP09(2013)018}{\emph{JHEP} {\bf 09} (2013)
		018}, [\href{http://arxiv.org/abs/1304.6483}{{\tt 1304.6483}}].
	
	\bibitem{Bousso:2012as}
	R.~Bousso, \emph{{Complementarity Is Not Enough}},
	\href{http://dx.doi.org/10.1103/PhysRevD.87.124023}{\emph{Phys. Rev.} {\bf
			D87} (2013) 124023}, [\href{http://arxiv.org/abs/1207.5192}{{\tt
			1207.5192}}].
	
	\bibitem{Marolf:2013dba}
	D.~Marolf and J.~Polchinski, \emph{{Gauge/Gravity Duality and the Black Hole
			Interior}},
	\href{http://dx.doi.org/10.1103/PhysRevLett.111.171301}{\emph{Phys. Rev.
			Lett.} {\bf 111} (2013) 171301}, [\href{http://arxiv.org/abs/1307.4706}{{\tt
			1307.4706}}].
	
	\bibitem{Papadodimas:2013jku}
	K.~Papadodimas and S.~Raju, \emph{{State-Dependent Bulk-Boundary Maps and Black
			Hole Complementarity}},
	\href{http://dx.doi.org/10.1103/PhysRevD.89.086010}{\emph{Phys. Rev.} {\bf
			D89} (2014) 086010}, [\href{http://arxiv.org/abs/1310.6335}{{\tt
			1310.6335}}].
	
	\bibitem{Papadodimas:2013wnh}
	K.~Papadodimas and S.~Raju, \emph{{Black Hole Interior in the Holographic
			Correspondence and the Information Paradox}},
	\href{http://dx.doi.org/10.1103/PhysRevLett.112.051301}{\emph{Phys. Rev.
			Lett.} {\bf 112} (2014) 051301}, [\href{http://arxiv.org/abs/1310.6334}{{\tt
			1310.6334}}].
	
	\bibitem{Papadodimas:2015jra}
	K.~Papadodimas and S.~Raju, \emph{{Remarks on the necessity and implications of
			state-dependence in the black hole interior}},
	\href{http://dx.doi.org/10.1103/PhysRevD.93.084049}{\emph{Phys. Rev.} {\bf
			D93} (2016) 084049}, [\href{http://arxiv.org/abs/1503.08825}{{\tt
			1503.08825}}].
	
	\bibitem{Papadodimas:2015xma}
	K.~Papadodimas and S.~Raju, \emph{{Local Operators in the Eternal Black Hole}},
	\href{http://dx.doi.org/10.1103/PhysRevLett.115.211601}{\emph{Phys. Rev.
			Lett.} {\bf 115} (2015) 211601}, [\href{http://arxiv.org/abs/1502.06692}{{\tt
			1502.06692}}].
	
	\bibitem{Mathur:2008wi}
	S.~D. Mathur, \emph{{What Exactly is the Information Paradox?}},
	\href{http://dx.doi.org/10.1007/978-3-540-88460-6_1}{\emph{Lect. Notes Phys.}
		{\bf 769} (2009) 3--48}, [\href{http://arxiv.org/abs/0803.2030}{{\tt
			0803.2030}}].
	
	\bibitem{Mathur:2009hf}
	S.~D. Mathur, \emph{{The Information paradox: A Pedagogical introduction}},
	\href{http://dx.doi.org/10.1088/0264-9381/26/22/224001}{\emph{Class. Quant.
			Grav.} {\bf 26} (2009) 224001}, [\href{http://arxiv.org/abs/0909.1038}{{\tt
			0909.1038}}].
	
	\bibitem{Hawking:2016msc}
	S.~W. Hawking, M.~J. Perry and A.~Strominger, \emph{{Soft Hair on Black
			Holes}}, \href{http://dx.doi.org/10.1103/PhysRevLett.116.231301}{\emph{Phys.
			Rev. Lett.} {\bf 116} (2016) 231301},
	[\href{http://arxiv.org/abs/1601.00921}{{\tt 1601.00921}}].
	
	\bibitem{Hertog:2004dr}
	T.~Hertog and K.~Maeda, \emph{{Black holes with scalar hair and asymptotics in
			N = 8 supergravity}},
	\href{http://dx.doi.org/10.1088/1126-6708/2004/07/051}{\emph{JHEP} {\bf 07}
		(2004) 051}, [\href{http://arxiv.org/abs/hep-th/0404261}{{\tt
			hep-th/0404261}}].
	
	\bibitem{Hertog:2004ns}
	T.~Hertog and G.~T. Horowitz, \emph{{Designer gravity and field theory
			effective potentials}},
	\href{http://dx.doi.org/10.1103/PhysRevLett.94.221301}{\emph{Phys. Rev.
			Lett.} {\bf 94} (2005) 221301},
	[\href{http://arxiv.org/abs/hep-th/0412169}{{\tt hep-th/0412169}}].
	
	\bibitem{Hertog:2005hu}
	T.~Hertog and G.~T. Horowitz, \emph{{Holographic description of AdS
			cosmologies}},
	\href{http://dx.doi.org/10.1088/1126-6708/2005/04/005}{\emph{JHEP} {\bf 04}
		(2005) 005}, [\href{http://arxiv.org/abs/hep-th/0503071}{{\tt
			hep-th/0503071}}].
	
	\bibitem{Hertog:2006rr}
	T.~Hertog, \emph{{Towards a Novel no-hair Theorem for Black Holes}},
	\href{http://dx.doi.org/10.1103/PhysRevD.74.084008}{\emph{Phys. Rev.} {\bf
			D74} (2006) 084008}, [\href{http://arxiv.org/abs/gr-qc/0608075}{{\tt
			gr-qc/0608075}}].
	
	\bibitem{Giddings:2012gc}
	S.~B. Giddings, \emph{{Nonviolent nonlocality}},
	\href{http://dx.doi.org/10.1103/PhysRevD.88.064023}{\emph{Phys. Rev.} {\bf
			D88} (2013) 064023}, [\href{http://arxiv.org/abs/1211.7070}{{\tt
			1211.7070}}].
	
	\bibitem{Giddings:2017mym}
	S.~B. Giddings, \emph{{Nonviolent unitarization: basic postulates to soft
			quantum structure of black holes}},
	\href{http://dx.doi.org/10.1007/JHEP12(2017)047}{\emph{JHEP} {\bf 12} (2017)
		047}, [\href{http://arxiv.org/abs/1701.08765}{{\tt 1701.08765}}].
	
	\bibitem{Chowdhury:2013tza}
	B.~D. Chowdhury, \emph{{Black holes versus firewalls and thermo-field
			dynamics}}, \href{http://dx.doi.org/10.1142/S021827181342011X}{\emph{Int. J.
			Mod. Phys.} {\bf D22} (2013) 1342011},
	[\href{http://arxiv.org/abs/1305.6343}{{\tt 1305.6343}}].
	
	\bibitem{Maldacena:2013xja}
	J.~Maldacena and L.~Susskind, \emph{{Cool horizons for entangled black holes}},
	\href{http://dx.doi.org/10.1002/prop.201300020}{\emph{Fortsch. Phys.} {\bf
			61} (2013) 781--811}, [\href{http://arxiv.org/abs/1306.0533}{{\tt
			1306.0533}}].
	
	\bibitem{Susskind:2014moa}
	L.~Susskind, \emph{{Entanglement is not enough}},
	\href{http://dx.doi.org/10.1002/prop.201500095}{\emph{Fortsch. Phys.} {\bf
			64} (2016) 49--71}, [\href{http://arxiv.org/abs/1411.0690}{{\tt 1411.0690}}].
	
	\bibitem{Balasubramanian:2014gla}
	V.~Balasubramanian, M.~Berkooz, S.~F. Ross and J.~Simon, \emph{{Black Holes,
			Entanglement and Random Matrices}},
	\href{http://dx.doi.org/10.1088/0264-9381/31/18/185009}{\emph{Class. Quant.
			Grav.} {\bf 31} (2014) 185009}, [\href{http://arxiv.org/abs/1404.6198}{{\tt
			1404.6198}}].
	
	\bibitem{Shenker:2013pqa}
	S.~H. Shenker and D.~Stanford, \emph{{Black holes and the butterfly effect}},
	\href{http://dx.doi.org/10.1007/JHEP03(2014)067}{\emph{JHEP} {\bf 03} (2014)
		067}, [\href{http://arxiv.org/abs/1306.0622}{{\tt 1306.0622}}].
	
	\bibitem{Shenker:2013yza}
	S.~H. Shenker and D.~Stanford, \emph{{Multiple Shocks}},
	\href{http://dx.doi.org/10.1007/JHEP12(2014)046}{\emph{JHEP} {\bf 12} (2014)
		046}, [\href{http://arxiv.org/abs/1312.3296}{{\tt 1312.3296}}].
	
	\bibitem{Gao:2016bin}
	P.~Gao, D.~L. Jafferis and A.~Wall, \emph{{Traversable Wormholes via a Double
			Trace Deformation}},
	\href{http://dx.doi.org/10.1007/JHEP12(2017)151}{\emph{JHEP} {\bf 12} (2017)
		151}, [\href{http://arxiv.org/abs/1608.05687}{{\tt 1608.05687}}].
	
	\bibitem{Maldacena:2017axo}
	J.~Maldacena, D.~Stanford and Z.~Yang, \emph{{Diving into traversable
			wormholes}}, \href{http://dx.doi.org/10.1002/prop.201700034}{\emph{Fortsch.
			Phys.} {\bf 65} (2017) 1700034}, [\href{http://arxiv.org/abs/1704.05333}{{\tt
			1704.05333}}].
	
	\bibitem{vanBreukelen:2017dul}
	R.~van Breukelen and K.~Papadodimas, \emph{{Quantum teleportation through
			time-shifted AdS wormholes}},  \href{http://arxiv.org/abs/1708.09370}{{\tt
			1708.09370}}.
	
	\bibitem{Hamilton:2005ju}
	A.~Hamilton, D.~N. Kabat, G.~Lifschytz and D.~A. Lowe, \emph{{Local bulk
			operators in AdS/CFT: A Boundary view of horizons and locality}},
	\href{http://dx.doi.org/10.1103/PhysRevD.73.086003}{\emph{Phys. Rev.} {\bf
			D73} (2006) 086003}, [\href{http://arxiv.org/abs/hep-th/0506118}{{\tt
			hep-th/0506118}}].
	
	\bibitem{Hamilton:2006az}
	A.~Hamilton, D.~N. Kabat, G.~Lifschytz and D.~A. Lowe, \emph{{Holographic
			representation of local bulk operators}},
	\href{http://dx.doi.org/10.1103/PhysRevD.74.066009}{\emph{Phys. Rev.} {\bf
			D74} (2006) 066009}, [\href{http://arxiv.org/abs/hep-th/0606141}{{\tt
			hep-th/0606141}}].
	
	\bibitem{Fitzpatrick:2016ive}
	A.~L. Fitzpatrick, J.~Kaplan, D.~Li and J.~Wang, \emph{{On information loss in
			AdS$_{3}$/CFT$_{2}$}},
	\href{http://dx.doi.org/10.1007/JHEP05(2016)109}{\emph{JHEP} {\bf 05} (2016)
		109}, [\href{http://arxiv.org/abs/1603.08925}{{\tt 1603.08925}}].
	
	\bibitem{Martinez:2004nb}
	C.~Martinez, R.~Troncoso and J.~Zanelli, \emph{{Exact black hole solution with
			a minimally coupled scalar field}},
	\href{http://dx.doi.org/10.1103/PhysRevD.70.084035}{\emph{Phys. Rev.} {\bf
			D70} (2004) 084035}, [\href{http://arxiv.org/abs/hep-th/0406111}{{\tt
			hep-th/0406111}}].
	
	\bibitem{Acena:2013jya}
	A.~Ace{\~n}a, A.~Anabal{\'o}n, D.~Astefanesei and R.~Mann, \emph{{Hairy planar
			black holes in higher dimensions}},
	\href{http://dx.doi.org/10.1007/JHEP01(2014)153}{\emph{JHEP} {\bf 01} (2014)
		153}, [\href{http://arxiv.org/abs/1311.6065}{{\tt 1311.6065}}].
	
	\bibitem{Anabalon:2012ta}
	A.~Anabalon, \emph{{Exact Black Holes and Universality in the Backreaction of
			non-linear Sigma Models with a potential in (A)dS4}},
	\href{http://dx.doi.org/10.1007/JHEP06(2012)127}{\emph{JHEP} {\bf 06} (2012)
		127}, [\href{http://arxiv.org/abs/1204.2720}{{\tt 1204.2720}}].
	
	\bibitem{Faedo:2017veq}
	F.~Faedo, D.~Klemm and M.~Nozawa, \emph{{Hairy black holes in $N = 2$ gauged
			supergravity}},  in \emph{{Proceedings, 14th Marcel Grossmann Meeting on
			Recent Developments in Theoretical and Experimental General Relativity,
			Astrophysics, and Relativistic Field Theories (MG14) (In 4 Volumes): Rome,
			Italy, July 12-18, 2015}}, vol.~4, pp.~4204--4207, 2017.
	\newblock \href{http://dx.doi.org/10.1142/9789813226609_0562}{DOI}.
	
	\bibitem{Anabalon:2017yhv}
	A.~Anabalon, D.~Astefanesei, A.~Gallerati and M.~Trigiante, \emph{{Hairy Black
			Holes and Duality in an Extended Supergravity Model}},
	\href{http://arxiv.org/abs/1712.06971}{{\tt 1712.06971}}.
	
	\bibitem{Henneaux:2004zi}
	M.~Henneaux, C.~Martinez, R.~Troncoso and J.~Zanelli, \emph{{Asymptotically
			anti-de Sitter spacetimes and scalar fields with a logarithmic branch}},
	\href{http://dx.doi.org/10.1103/PhysRevD.70.044034}{\emph{Phys. Rev.} {\bf
			D70} (2004) 044034}, [\href{http://arxiv.org/abs/hep-th/0404236}{{\tt
			hep-th/0404236}}].
	
	\bibitem{Hertog:2004bb}
	T.~Hertog and K.~Maeda, \emph{{Stability and thermodynamics of AdS black holes
			with scalar hair}},
	\href{http://dx.doi.org/10.1103/PhysRevD.71.024001}{\emph{Phys. Rev.} {\bf
			D71} (2005) 024001}, [\href{http://arxiv.org/abs/hep-th/0409314}{{\tt
			hep-th/0409314}}].
	
	\bibitem{Papadimitriou:2007sj}
	I.~Papadimitriou, \emph{{Multi-Trace Deformations in AdS/CFT: Exploring the
			Vacuum Structure of the Deformed CFT}},
	\href{http://dx.doi.org/10.1088/1126-6708/2007/05/075}{\emph{JHEP} {\bf 05}
		(2007) 075}, [\href{http://arxiv.org/abs/hep-th/0703152}{{\tt
			hep-th/0703152}}].
	
	\bibitem{Witten:2001ua}
	E.~Witten, \emph{{Multitrace operators, boundary conditions, and AdS / CFT
			correspondence}},  \href{http://arxiv.org/abs/hep-th/0112258}{{\tt
			hep-th/0112258}}.
	
	\bibitem{Berkooz:2002ug}
	M.~Berkooz, A.~Sever and A.~Shomer, \emph{{'Double trace' deformations,
			boundary conditions and space-time singularities}},
	\href{http://dx.doi.org/10.1088/1126-6708/2002/05/034}{\emph{JHEP} {\bf 05}
		(2002) 034}, [\href{http://arxiv.org/abs/hep-th/0112264}{{\tt
			hep-th/0112264}}].
	
	\bibitem{Hertog:2005hm}
	T.~Hertog and S.~Hollands, \emph{{Stability in designer gravity}},
	\href{http://dx.doi.org/10.1088/0264-9381/22/24/007}{\emph{Class. Quant.
			Grav.} {\bf 22} (2005) 5323--5342},
	[\href{http://arxiv.org/abs/hep-th/0508181}{{\tt hep-th/0508181}}].
	
	\bibitem{Harlow:2014yoa}
	D.~Harlow, \emph{{Aspects of the Papadodimas-Raju Proposal for the Black Hole
			Interior}}, \href{http://dx.doi.org/10.1007/JHEP11(2014)055}{\emph{JHEP} {\bf
			11} (2014) 055}, [\href{http://arxiv.org/abs/1405.1995}{{\tt 1405.1995}}].
	
	\bibitem{Harlow:2014yka}
	D.~Harlow, \emph{{Jerusalem Lectures on Black Holes and Quantum Information}},
	\href{http://dx.doi.org/10.1103/RevModPhys.88.015002}{\emph{Rev. Mod. Phys.}
		{\bf 88} (2016) 15002}, [\href{http://arxiv.org/abs/1409.1231}{{\tt
			1409.1231}}].
	
	\bibitem{Reed/Simon}
	M.~Reed and B.~Simon, \emph{Analysis of operators}.
	\newblock Methods of Modern Mathematical Physics. Academic Press, 1978.
	
	\bibitem{Maldacena:2001kr}
	J.~M. Maldacena, \emph{{Eternal black holes in anti-de Sitter}},
	\href{http://dx.doi.org/10.1088/1126-6708/2003/04/021}{\emph{JHEP} {\bf 04}
		(2003) 021}, [\href{http://arxiv.org/abs/hep-th/0106112}{{\tt
			hep-th/0106112}}].
	
	\bibitem{Czech:2012be}
	B.~Czech, J.~L. Karczmarek, F.~Nogueira and M.~Van~Raamsdonk, \emph{{Rindler
			Quantum Gravity}},
	\href{http://dx.doi.org/10.1088/0264-9381/29/23/235025}{\emph{Class. Quant.
			Grav.} {\bf 29} (2012) 235025}, [\href{http://arxiv.org/abs/1206.1323}{{\tt
			1206.1323}}].
	
	\bibitem{VanRaamsdonk:2013sza}
	M.~Van~Raamsdonk, \emph{{Evaporating Firewalls}},
	\href{http://dx.doi.org/10.1007/JHEP11(2014)038}{\emph{JHEP} {\bf 11} (2014)
		038}, [\href{http://arxiv.org/abs/1307.1796}{{\tt 1307.1796}}].
	
	\bibitem{ZinnJustin:2004ib}
	J.~Zinn-Justin and U.~D. Jentschura, \emph{{Multi-instantons and exact results
			I: Conjectures, WKB expansions, and instanton interactions}},
	\href{http://dx.doi.org/10.1016/j.aop.2004.04.004}{\emph{Annals Phys.} {\bf
			313} (2004) 197--267}, [\href{http://arxiv.org/abs/quant-ph/0501136}{{\tt
			quant-ph/0501136}}].
	
	\bibitem{Papadodimas:2012aq}
	K.~Papadodimas and S.~Raju, \emph{{An Infalling Observer in AdS/CFT}},
	\href{http://dx.doi.org/10.1007/JHEP10(2013)212}{\emph{JHEP} {\bf 10} (2013)
		212}, [\href{http://arxiv.org/abs/1211.6767}{{\tt 1211.6767}}].
	
	\bibitem{Hartle:2015bna}
	J.~Hartle and T.~Hertog, \emph{{Quantum transitions between classical
			histories}}, \href{http://dx.doi.org/10.1103/PhysRevD.92.063509}{\emph{Phys.
			Rev.} {\bf D92} (2015) 063509}, [\href{http://arxiv.org/abs/1502.06770}{{\tt
			1502.06770}}].
	
	\bibitem{Hertog:2017vod}
	T.~Hertog and J.~Hartle, \emph{{Observational Implications of Fuzzball
			Formation}},  \href{http://arxiv.org/abs/1704.02123}{{\tt 1704.02123}}.
	
	\bibitem{Susskind:1993if}
	L.~Susskind, L.~Thorlacius and J.~Uglum, \emph{{The Stretched horizon and black
			hole complementarity}},
	\href{http://dx.doi.org/10.1103/PhysRevD.48.3743}{\emph{Phys. Rev.} {\bf D48}
		(1993) 3743--3761}, [\href{http://arxiv.org/abs/hep-th/9306069}{{\tt
			hep-th/9306069}}].
	
	\bibitem{Sekino:2008he}
	Y.~Sekino and L.~Susskind, \emph{{Fast Scramblers}},
	\href{http://dx.doi.org/10.1088/1126-6708/2008/10/065}{\emph{JHEP} {\bf 10}
		(2008) 065}, [\href{http://arxiv.org/abs/0808.2096}{{\tt 0808.2096}}].
	
	\bibitem{Ghosh:2016fvm}
	S.~Ghosh and S.~Raju, \emph{{The Breakdown of String Perturbation Theory for
			Many External Particles}},  \href{http://arxiv.org/abs/1611.08003}{{\tt
			1611.08003}}.
	
	\bibitem{Ghosh:2017pel}
	S.~Ghosh and S.~Raju, \emph{{Loss of locality in gravitational correlators with
			a large number of insertions}},  \href{http://arxiv.org/abs/1706.07424}{{\tt
			1706.07424}}.
	
	\bibitem{Jafferis:2017tiu}
	D.~L. Jafferis, \emph{{Bulk reconstruction and the Hartle-Hawking
			wavefunction}},  \href{http://arxiv.org/abs/1703.01519}{{\tt 1703.01519}}.
	
	\bibitem{Kabat:2014kfa}
	D.~Kabat and G.~Lifschytz, \emph{{Finite N and the failure of bulk locality:
			Black holes in AdS/CFT}},
	\href{http://dx.doi.org/10.1007/JHEP09(2014)077}{\emph{JHEP} {\bf 09} (2014)
		077}, [\href{http://arxiv.org/abs/1405.6394}{{\tt 1405.6394}}].
	
	\bibitem{Raju:2016vsu}
	S.~Raju, \emph{{Smooth Causal Patches for AdS Black Holes}},
	\href{http://dx.doi.org/10.1103/PhysRevD.95.126002}{\emph{Phys. Rev.} {\bf
			D95} (2017) 126002}, [\href{http://arxiv.org/abs/1604.03095}{{\tt
			1604.03095}}].
	
	\bibitem{Hamilton:2006fh}
	A.~Hamilton, D.~N. Kabat, G.~Lifschytz and D.~A. Lowe, \emph{{Local bulk
			operators in AdS/CFT: A Holographic description of the black hole interior}},
	\href{http://dx.doi.org/10.1103/PhysRevD.75.106001,
		10.1103/PhysRevD.75.129902}{\emph{Phys. Rev.} {\bf D75} (2007) 106001},
	[\href{http://arxiv.org/abs/hep-th/0612053}{{\tt hep-th/0612053}}].
	
	\bibitem{Kabat:2018pbj}
	D.~Kabat and G.~Lifschytz, \emph{{Does boundary quantum mechanics imply quantum
			mechanics in the bulk?}},  \href{http://arxiv.org/abs/1801.08101}{{\tt
			1801.08101}}.
	
	\bibitem{sanchez}
	N.~Sanchez and B.~Whiting, \emph{Quantum field theory and the antipodal
		identification of black-holes},
	\href{http://dx.doi.org/http://dx.doi.org/10.1016/0550-3213(87)90289-6}{\emph{Nuclear
			Physics B} {\bf 283} (1987) 605 -- 623}.
	
	\bibitem{Chamblin:2006xd}
	A.~Chamblin and J.~Michelson, \emph{{Alpha-Vacua, Black Holes, and AdS/CFT}},
	\href{http://dx.doi.org/10.1088/0264-9381/24/6/013}{\emph{Class. Quant.
			Grav.} {\bf 24} (2007) 1569--1604},
	[\href{http://arxiv.org/abs/hep-th/0610133}{{\tt hep-th/0610133}}].
	
	\bibitem{Hooft:2016itl}
	G.~'t~Hooft, \emph{{Black hole unitarity and antipodal entanglement}},
	\href{http://dx.doi.org/10.1007/s10701-016-0014-y}{\emph{Found. Phys.} {\bf
			46} (2016) 1185--1198}, [\href{http://arxiv.org/abs/1601.03447}{{\tt
			1601.03447}}].
	
	\bibitem{Hooft:2016vug}
	G.~'t~Hooft, \emph{{The firewall transformation for black holes and some of its
			implications}},  \href{http://arxiv.org/abs/1612.08640}{{\tt 1612.08640}}.
	
	\bibitem{Srednicki:1995pt}
	M.~Srednicki, \emph{{Thermal fluctuations in quantized chaotic systems}},
	\href{http://dx.doi.org/10.1088/0305-4470/29/4/003}{\emph{J. Phys.} {\bf A29}
		(1996) L75--L79}, [\href{http://arxiv.org/abs/chao-dyn/9511001}{{\tt
			chao-dyn/9511001}}].
	
	\bibitem{Lashkari:2016vgj}
	N.~Lashkari, A.~Dymarsky and H.~Liu, \emph{{Eigenstate Thermalization
			Hypothesis in Conformal Field Theory}},
	\href{http://arxiv.org/abs/1610.00302}{{\tt 1610.00302}}.
	
	\bibitem{Lashkari:2017hwq}
	N.~Lashkari, A.~Dymarsky and H.~Liu, \emph{{Universality of Quantum Information
			in Chaotic CFTs}},  \href{http://arxiv.org/abs/1710.10458}{{\tt 1710.10458}}.
	
	\bibitem{Marolf:2015dia}
	D.~Marolf and J.~Polchinski, \emph{{Violations of the Born rule in cool
			state-dependent horizons}},
	\href{http://dx.doi.org/10.1007/JHEP01(2016)008}{\emph{JHEP} {\bf 01} (2016)
		008}, [\href{http://arxiv.org/abs/1506.01337}{{\tt 1506.01337}}].
	
	\bibitem{Ooguri:2016pdq}
	H.~Ooguri and C.~Vafa, \emph{{Non-supersymmetric AdS and the Swampland}},
	\href{http://arxiv.org/abs/1610.01533}{{\tt 1610.01533}}.
	
	\bibitem{Freivogel:2016qwc}
	B.~Freivogel and M.~Kleban, \emph{{Vacua Morghulis}},
	\href{http://arxiv.org/abs/1610.04564}{{\tt 1610.04564}}.
	
	\bibitem{Mathur:2008kg}
	S.~D. Mathur, \emph{{Tunneling into fuzzball states}},
	\href{http://dx.doi.org/10.1007/s10714-009-0837-3}{\emph{Gen. Rel. Grav.}
		{\bf 42} (2010) 113--118}, [\href{http://arxiv.org/abs/0805.3716}{{\tt
			0805.3716}}].
	
	\bibitem{Parikh:1999mf}
	M.~K. Parikh and F.~Wilczek, \emph{{Hawking radiation as tunneling}},
	\href{http://dx.doi.org/10.1103/PhysRevLett.85.5042}{\emph{Phys. Rev. Lett.}
		{\bf 85} (2000) 5042--5045}, [\href{http://arxiv.org/abs/hep-th/9907001}{{\tt
			hep-th/9907001}}].
	
	\bibitem{Betzios:2016yaq}
	P.~Betzios, N.~Gaddam and O.~Papadoulaki, \emph{{The Black Hole S-Matrix from
			Quantum Mechanics}},
	\href{http://dx.doi.org/10.1007/JHEP11(2016)131}{\emph{JHEP} {\bf 11} (2016)
		131}, [\href{http://arxiv.org/abs/1607.07885}{{\tt 1607.07885}}].
	
	\bibitem{Horowitz:2009wm}
	G.~Horowitz, A.~Lawrence and E.~Silverstein, \emph{{Insightful D-branes}},
	\href{http://dx.doi.org/10.1088/1126-6708/2009/07/057}{\emph{JHEP} {\bf 07}
		(2009) 057}, [\href{http://arxiv.org/abs/0904.3922}{{\tt 0904.3922}}].
	
	\bibitem{Bermudez:2012fe}
	D.~Bermudez and D.~J.~F. C, \emph{{Factorization method and new potentials from
			the inverted oscillator}},
	\href{http://dx.doi.org/10.1016/j.aop.2013.02.015}{\emph{Annals Phys.} {\bf
			333} (2013) 290--306}, [\href{http://arxiv.org/abs/1206.4519}{{\tt
			1206.4519}}].
	
	\bibitem{Hollowood:2014hta}
	T.~J. Hollowood, \emph{{Schr{\"o}dinger's cat and the firewall}},
	\href{http://dx.doi.org/10.1142/S0218271814410041}{\emph{Int. J. Mod. Phys.}
		{\bf D23} (2014) 1441004}, [\href{http://arxiv.org/abs/1403.5947}{{\tt
			1403.5947}}].
	
	\bibitem{Bao:2017rnv}
	N.~Bao, S.~M. Carroll and A.~Singh, \emph{{The Hilbert Space of Quantum Gravity
			Is Locally Finite-Dimensional}},
	\href{http://dx.doi.org/10.1142/S0218271817430131}{\emph{Int. J. Mod. Phys.}
		{\bf D26} (2017) 1743013}, [\href{http://arxiv.org/abs/1704.00066}{{\tt
			1704.00066}}].
	
	\bibitem{Banks:1996vh}
	T.~Banks, W.~Fischler, S.~H. Shenker and L.~Susskind, \emph{{M theory as a
			matrix model: A Conjecture}},
	\href{http://dx.doi.org/10.1103/PhysRevD.55.5112}{\emph{Phys. Rev.} {\bf D55}
		(1997) 5112--5128}, [\href{http://arxiv.org/abs/hep-th/9610043}{{\tt
			hep-th/9610043}}].
	
	\bibitem{Iizuka:2008eb}
	N.~Iizuka, T.~Okuda and J.~Polchinski, \emph{{Matrix Models for the Black Hole
			Information Paradox}},
	\href{http://dx.doi.org/10.1007/JHEP02(2010)073}{\emph{JHEP} {\bf 02} (2010)
		073}, [\href{http://arxiv.org/abs/0808.0530}{{\tt 0808.0530}}].
	
	\bibitem{Iizuka:2008hg}
	N.~Iizuka and J.~Polchinski, \emph{{A Matrix Model for Black Hole
			Thermalization}},
	\href{http://dx.doi.org/10.1088/1126-6708/2008/10/028}{\emph{JHEP} {\bf 10}
		(2008) 028}, [\href{http://arxiv.org/abs/0801.3657}{{\tt 0801.3657}}].
	
	\bibitem{Ferrari:2016bvq}
	F.~Ferrari, \emph{{Black Hole Horizons and Bose-Einstein Condensation}},
	\href{http://arxiv.org/abs/1601.08120}{{\tt 1601.08120}}.
	
	\bibitem{Maldacena:2016hyu}
	J.~Maldacena and D.~Stanford, \emph{{Remarks on the Sachdev-Ye-Kitaev model}},
	\href{http://dx.doi.org/10.1103/PhysRevD.94.106002}{\emph{Phys. Rev.} {\bf
			D94} (2016) 106002}, [\href{http://arxiv.org/abs/1604.07818}{{\tt
			1604.07818}}].
	
	\bibitem{Polchinski:2016xgd}
	J.~Polchinski and V.~Rosenhaus, \emph{{The Spectrum in the Sachdev-Ye-Kitaev
			Model}}, \href{http://dx.doi.org/10.1007/JHEP04(2016)001}{\emph{JHEP} {\bf
			04} (2016) 001}, [\href{http://arxiv.org/abs/1601.06768}{{\tt 1601.06768}}].
	
	\bibitem{Fitzpatrick:2015dlt}
	A.~L. Fitzpatrick and J.~Kaplan, \emph{{Conformal Blocks Beyond the
			Semi-Classical Limit}},
	\href{http://dx.doi.org/10.1007/JHEP05(2016)075}{\emph{JHEP} {\bf 05} (2016)
		075}, [\href{http://arxiv.org/abs/1512.03052}{{\tt 1512.03052}}].
	
	\bibitem{Chen:2017yze}
	H.~Chen, C.~Hussong, J.~Kaplan and D.~Li, \emph{{A Numerical Approach to
			Virasoro Blocks and the Information Paradox}},
	\href{http://arxiv.org/abs/1703.09727}{{\tt 1703.09727}}.
	
\end{thebibliography}
\end{document}